\documentclass[a4paper,11pt]{article}
\pdfoutput=1 

\usepackage{jheppub} 

\usepackage[T1]{fontenc} 
\usepackage{graphicx}
\usepackage{epstopdf}
\usepackage{graphics}
\usepackage{hyperref}
\usepackage[small]{subfigure}

\newcommand{\Ord}{\mathcal{O}}
\newcommand{\dd}{\mathrm{d}}

\title{NLO Massive Event-Shape Differential and Cumulative Distributions}

\preprint{\begin{flushright} IFT-UAM/CSIC-19-149, UWThPh-2019-34\end{flushright}\vspace*{-1cm}}

\author[a]{Christopher Lepenik}
\author{and}
\author[b,c]{Vicent Mateu}

\affiliation[a]{University of Vienna, Faculty of Physics, Boltzmanngasse 5, A-1090 Wien, Austria}
\affiliation[b]{Departamento de F\'isica Fundamental e IUFFyM,\\Universidad de Salamanca, E-37008 Salamanca, Spain}
\affiliation[c]{Instituto de F\'isica Te\'orica UAM-CSIC, E-28049 Madrid, Spain}

\emailAdd{christopher.lepenik@univie.ac.at}
\emailAdd{vmateu@usal.es}

\abstract{
We provide a general method to effectively compute differential and cumulative event-shape distributions to $\mathcal{O}(\alpha_s)$ precision for
massive quarks produced primarily at an $e^+e^-$ collider. In particular, we show that at this order, due to the screening of collinear singularities
by the quark mass, for all event shapes linearly sensitive to soft dynamics, there appear only two distributions at threshold:~a Dirac delta function
and a plus distribution.
Furthermore, we show that the coefficient of the latter is universal for any infra-red and collinear safe event shape, and provide an analytic
expression for it. Likewise, we compute a general formula for the coefficient of the Dirac delta function, which depends only
on the event-shape measurement function in the soft limit. Finally, we present an efficient algorithm to compute the differential and cumulative distributions,
which does not rely on Monte Carlo methods, therefore achieving a priory arbitrary precision even in the extreme dijet region. We implement this
algorithm in a numeric code and show that it agrees with 
analytic results on the distribution for 2-jettiness, heavy jet mass and a massive generalization of C-parameter.}

\begin{document}
\maketitle
\flushbottom

\section{Introduction}\label{sec:intro}
Recent years have seen tremendous progress in the understanding and computation of event-shape cross sections for $e^+e^-$ machines such
as LEP or the future linear and circular colliders. This is mainly achieved with the use of factorized expressions (see e.g.\ Refs.~\cite{Schwartz:2007ib,Bauer:2008dt})
derived in the frame of the effective field theory (EFT)\footnote{Factorization formulas can also be derived in the Collins, Soper and Sterman (CSS)
formalism~\cite{Collins:1981uk,Korchemsky:1998ev, Korchemsky:1999kt,Korchemsky:2000kp, Berger:2003iw}.} known as Soft-Collinear Effective
Theory (SCET)~\cite{Bauer:2000ew, Bauer:2000yr, Bauer:2001ct, Bauer:2001yt, Bauer:2002nz}. The state of the art for massless event shapes is
next-to-next-to-leading-log (N$^2$LL)~\cite{Hornig:2009vb,Becher:2012qc,Bell:2018gce} and next-to-next-to-next-to-leading-log
(N$^3$LL)~\cite{Becher:2008cf,Chien:2010kc,Hoang:2014wka,Moult:2018jzp,HJM-theo}.\footnote{Resummation can also be worked out in the coherent
branching formalism~\cite{Catani:1992ua}, which achieves N$^2$LL precision in an automated, numeric way~\cite{Banfi:2004yd,Banfi:2014sua}.}
The computation of the soft function has been fully automatized analytically at ${\cal O}(\alpha_s)$ in Ref.~\cite{Hoang:2014wka} and numerically
at ${\cal O}(\alpha_s^2)$ in Refs.~\cite{Bell:2018vaa,Bell:2018oqa} (in those articles one can also find analytic results for some event shapes such as C-parameter).
Analytic fixed-order perturbative predictions exist at ${\cal O}(\alpha_s)$ for thrust, heavy jet mass (HJM) and C-parameter, while for other event shapes such as
angularities or jet broadening, the cross sections can be expressed as a 1D numerical integral. Numerical results exist at ${\cal O}(\alpha_s^2)$~\cite{Catani:1996vz,Dixon:2018qgp}
and ${\cal O}(\alpha_s^3)$~\cite{GehrmannDeRidder:2007bj, GehrmannDeRidder:2007hr,Weinzierl:2008iv, Ridder:2014wza,
Weinzierl:2009ms,DelDuca:2016ily}. In Ref.~\cite{Mateu:2013gya} the fixed-order cross sections for oriented event shapes have been computed up to
${\cal O}(\alpha_s^2)$. These results have been used, in particular, to determine the strong coupling constant with very high
precision~\cite{Becher:2012qc,Chien:2010kc,Abbate:2010xh,Abbate:2012jh,Gehrmann:2012sc,Hoang:2015hka,HJM-fit}.

The theoretical knowledge for event shapes involving massive quarks is comparatively much poorer. Fixed-order predictions have been obtained numerically
at ${\cal O}(\alpha_s)$ (analytic results at this order exist but are scarce) and ${\cal O}(\alpha_s^2)$~\cite{Nason:1997nw,Bernreuther:1997jn,Rodrigo:1999qg}.
Factorization and resummation for 2-jettiness with massive quarks~\cite{Fleming:2007xt,Fleming:2007qr} has recently achieved N$^3$LL precision through
the computation of matrix elements at two loops~\cite{Jain:2008gb,Gritschacher:2013pha,Pietrulewicz:2014qza,Hoang:2015vua,Hoang:2019fze}. These results
have been used in Ref.~\cite{Butenschoen:2016lpz} to calibrate the {\scshape Pythia}~8.205~\cite{Sjostrand:2007gs} top quark mass parameter in terms of the
short-distance MSR scheme~\cite{Hoang:2008yj,Hoang:2017suc}. In this direction, further theoretical progress has been made, and a $p_T$ cutoff as implemented
in angular ordered parton showers has been included in Ref.~\cite{Hoang:2018zrp}. Event-shapes for massive particles shall play an important role at the future
linear collider, where they will presumably be used to determine the top quark mass from boosted events in a well defined scheme within quantum field theory.
This measurement is complementary to threshold scans with partially orthogonal experimental uncertainties. Furthermore, from the conclusions that can
be drawn from this article, they might also be a useful tool to determine the strong coupling $\alpha_s$.

For a complete description of massive event shapes with N$^2$LL accuracy in the peak region where the SCET and bHQET effective theories can be applied,\footnote{The
logarithmic counting refers to the kinematic limit $p_J^2~\sim m_q^2\sim Q^2 \lambda^2$, with $p^\mu_J$ the jet four-momentum and $\lambda$ de EFT power-counting
parameter. In this limit, the set of terms summed up at leading log in the cumulative cross section have the form $\log^{n+1}(\lambda)\alpha_s^n$ for any integer $n>0$.
At threshold there are no double logs and therefore the meaning of logarithmic accuracy is different.}
but also valid in the tail and far-tail of the distribution, the resummed cross section has to be matched to the fixed-order prediction at $\Ord(\alpha_s)$. In full generality,
the differential cross section for massive particles up to this order can be written as
\begin{align}\label{eq:general-diff}
\frac{1}{\sigma^C_0} \frac{\dd \sigma_C}{\dd e} ={}&R^0_C(\hat m)\,\delta (e - e_{\rm min}) +
C_F \frac{\alpha_s}{\pi} A^C_e({\hat m})\delta (e - e_{\rm min}) \\
& + C_F \frac{\alpha_s}{\pi} B^C_{\rm plus}({\hat m})\biggl[\frac{1}{e - e_{\rm min}} \biggr]_+
+ C_F \frac{\alpha_s}{\pi} F^{\rm NS}_e (e, \hat{m}) + \Ord(\alpha_s^2)\,,\nonumber
\end{align}
with $F^{\rm NS}_e$ a function regular at $e_{\rm min}$, the lower endpoint of the distribution, $\sigma^C_0$ the massless Born-cross section and $R^0_C$ the tree level R-ratio.
The sub- and super-scripts $C$ denote the type of current considered (vector or axial-vector), omitted for $F_e^\mathrm{NS}$ to keep the notation simple.
In this paper we compute the differential and cumulative cross sections for all event shapes, for both vector and axial-vector currents, reaching the same standard
as for massless quarks at this order (where essentially all results are known analytically in terms of relatively simple expressions), by obtaining analytical results for
$A_e$ and $B_{\rm plus}$, and computing $F^{\rm NS}_e$ through 1-dimensional numerical integrals in a way which is almost as precise and stable as for a full analytic
result. Our calculation shows that $A_e$ and $F^{\rm NS}_e$ depend on the specific event-shape variable, and $B_{\rm plus}$ is a universal function of the reduced mass
$\hat m \equiv m/Q$. This matching program has already been carried out in previous work~\cite{Butenschoen:2016lpz} in a less efficient way.
While at the time of the ``NLO revolution'' the $\Ord(\alpha_s)$ results seem completely standard, having a dedicated article on those is
still useful because: a)~known results are mainly numeric and provided as binned distributions, which makes the matching to resummed results unpractical, b)~analytic
results are faster and easier to implement; c)~non-zero quark masses entail that different event-shape schemes can be used, allowing to control the sensitivity to the
mass, to the best of our knowledge a possibility never discussed so far;\footnote{Schemes for event shapes were first introduced in Ref.~\cite{Salam:2001bd} to study
hadron mass effects in hadronization power corrections. Here we extend the analysis to heavy quarks, and study how to gain/loose sensitivity to their mass.}
d)~new strategies to efficiently compute the cross section are presented, which we believe will
be useful for future work; e)~our results are an important input for further studies of massive event shapes, which aim to improve our understanding of heavy quark
mass determinations in general. This article provides all ingredients that are needed for a full N$^2$LL computation in a way which is useful and easy to implement, and
in that sense it will be a reference for many future analysis in this field. In particular our results will help to clarify the top quark Monte Carlo (MC) mass interpretation problem.

This article is organized as follows: in Sec.~\ref{sec:results} we introduce massive schemes, and show how to implement them into massive
event shapes. In Sec.~\ref{sec:derivation} we provide a direct computation of our main analytic result, $A_e$, the coefficient of the delta function,
computing the real and virtual contributions directly, canceling the infrared singularities explicitly. Sec.~\ref{sec:numeric} deals with $F_e^{\rm NS}$,
describing our numerical algorithm, which is applied to differential and cumulative cross sections as well as moments. Analytic results for cross sections
of a few event shapes are shown in Sec.~\ref{sec:analytic}. Our conclusions are summarized in Sec.~\ref{sec:conclusions}. We provide a second analytic
computation of the delta-function coefficient in Appendix~\ref{sec:indirect}, requiring that the integration of the differential cross section over its
entire physical range reproduces the total hadronic cross section. Applications of our master formula for a selection of event shapes, analytic
formulae for the massive total hadronic cross section and some master integrals are provided in the remaining appendices.

\section{Event Shapes for Massive Particles}\label{sec:results}
In this section we introduce generalizations of classic event shapes for massive heavy quarks, for which we present analytic results of the delta-function coefficient in Appendix~\ref{app:anadel}. These generalizations go under the name of \,``mass schemes'', and have no effect on the
partonic cross section for massless particles, but for massive quarks dramatically modify the sensitivity to their mass already at parton level even
at the lowest order. This control on the sensitivity to the mass is of high interest, and in particular sets the effective renormalization scale of the
heavy quark mass. No systematic study of these schemes exists yet for massive quarks, and here we intend to fill this gap.

Mass schemes for event shapes were originally introduced to study non-perturbative power corrections in the context of non-zero hadron
masses~\cite{Salam:2001bd, Mateu:2012nk}. In the case of massless quarks, partonic cross sections are unaffected by scheme changes, but
power corrections substantially depend on the scheme choice. In particular it has been shown that the leading non-perturbative power correction
is universal for the so-called E-scheme. They have been also used to study power corrections in the dressed-gluon
approximation~\cite{Gardi:2003iv} and to determine the mass of heavy quarks in boosted events~\cite{Fleming:2007qr}. Due to increasing
theoretical and experimental precision, it is necessary to include the finite bottom quark mass in high-precision calculations, e.g.\ for extracting
the strong coupling constant~\cite{Abbate:2010xh}. On the other hand, mass effects are dominant and have to be included when extracting the
quark mass itself or when carrying out mass-related studies such as analyzing the properties of the top quark Monte Carlo mass parameter.

When using classical event shapes like thrust, C-parameter, jet broadening etc.\ in presence of massive quarks, the way how one treats energies
and three-momenta magnitudes is important, since (unlike for massless particles) $E_p \neq |\vec p\,|$. To categorize different ways of how to treat
energies and three-momenta, several ``schemes'' can be defined for massive event shapes, which are distinguished exactly by the way how $E_p$
and $|\vec p\,|$ are interpreted. Obviously, all these schemes reduce to the original definition in the massless case.

Starting from the original definition of event shapes, the ``E-scheme'' is defined by the replacement $\vec p \to (E_p/|\vec p\,|)\,\vec p$, while the
``P-scheme'' is defined by the substitution $E_p \to |\vec p\,|$. Some event shapes originally defined in P-scheme are thrust~\cite{Farhi:1977sg},
\begin{equation}
\tau \equiv \tau^P = \frac{1}{Q_P} \sum_i p_{i,\bot} \mathrm e^{-|\eta_i|} = \frac{1}{Q_P}\min_{\hat t}\sum_i(|\vec p_i| - |\hat t\cdot \vec p_i|)\,,
\end{equation}
C-parameter~\cite{Parisi:1978eg,Donoghue:1979vi} (it is often useful to define the reduced C-parameter as $\widetilde C = C/6$)
\begin{equation}
C \equiv C^P = \frac{3}{2Q_P^2}\sum_{i,j}|\vec p_i||\vec p_j|\sin^2\theta_{ij} =
\frac{3}{2}\biggl[1 - \frac{1}{Q_P^2} \sum_{i, j} \frac{(\vec p_i\cdot \vec p_j)^2}{|\vec p_i||\vec p_j|}\biggr]\,,
\end{equation}
and broadening~\cite{Rakow:1981qn}
\begin{equation}
B_T \equiv B_T^P = \frac{1}{2Q_P}\sum_i p_{i,\bot} = \frac{1}{2 Q_P} \sum_i (|\vec p_i| - |\hat t\cdot\vec p_i|)^{1/2}
(|\vec p_i| + |\hat t\cdot\vec p_i|)^{1/2}\,,
\end{equation}
while angularities~\cite{Berger:2003pk} were originally defined in the E-scheme:
\begin{equation}
\tau_a \equiv \tau_a^E = \frac{1}{Q} \sum_i \frac{E_i}{|\vec p_i|} p_{i, \bot}\mathrm e^{-|\eta_i|(1-a)} =
\frac{1}{2 Q} \sum_i \frac{E_i}{|\vec p_i|}(|\vec p_i| - |\hat t\cdot\vec p_i|)^{1-\frac{a}{2}}(|\vec p_i| + |\hat t\cdot\vec p_i|)^{\frac{a}{2}}\,,
\end{equation}
where $\eta$ denotes the pseudo-rapidity, $p_\bot \equiv |\vec p_\bot|$ the transverse momentum measured with respect to the thrust axis, and $m_\bot \equiv \sqrt{p_\bot^2 + m^2}$ is the transverse mass.
\begin{figure*}[t!]
\subfigure[]
{
\includegraphics[width=0.48\textwidth]{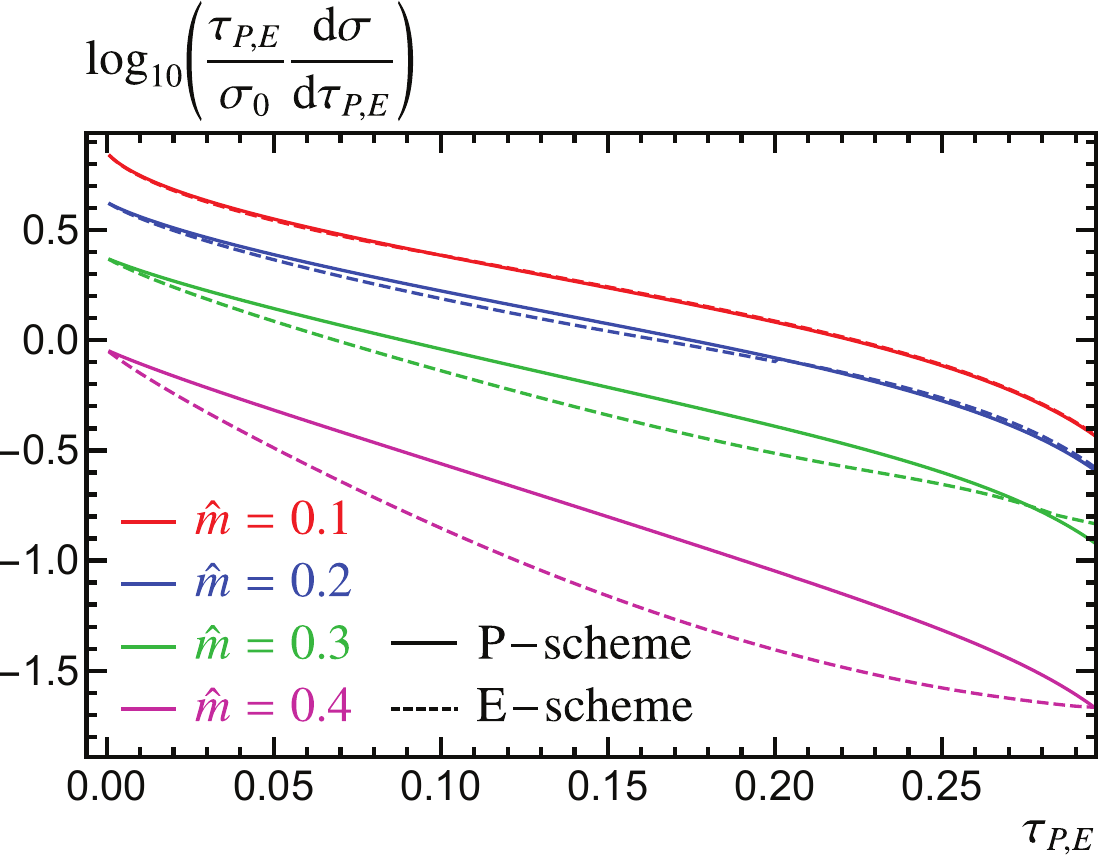}
\label{fig:EPTau}
}
\subfigure[]{
\includegraphics[width=0.47\textwidth]{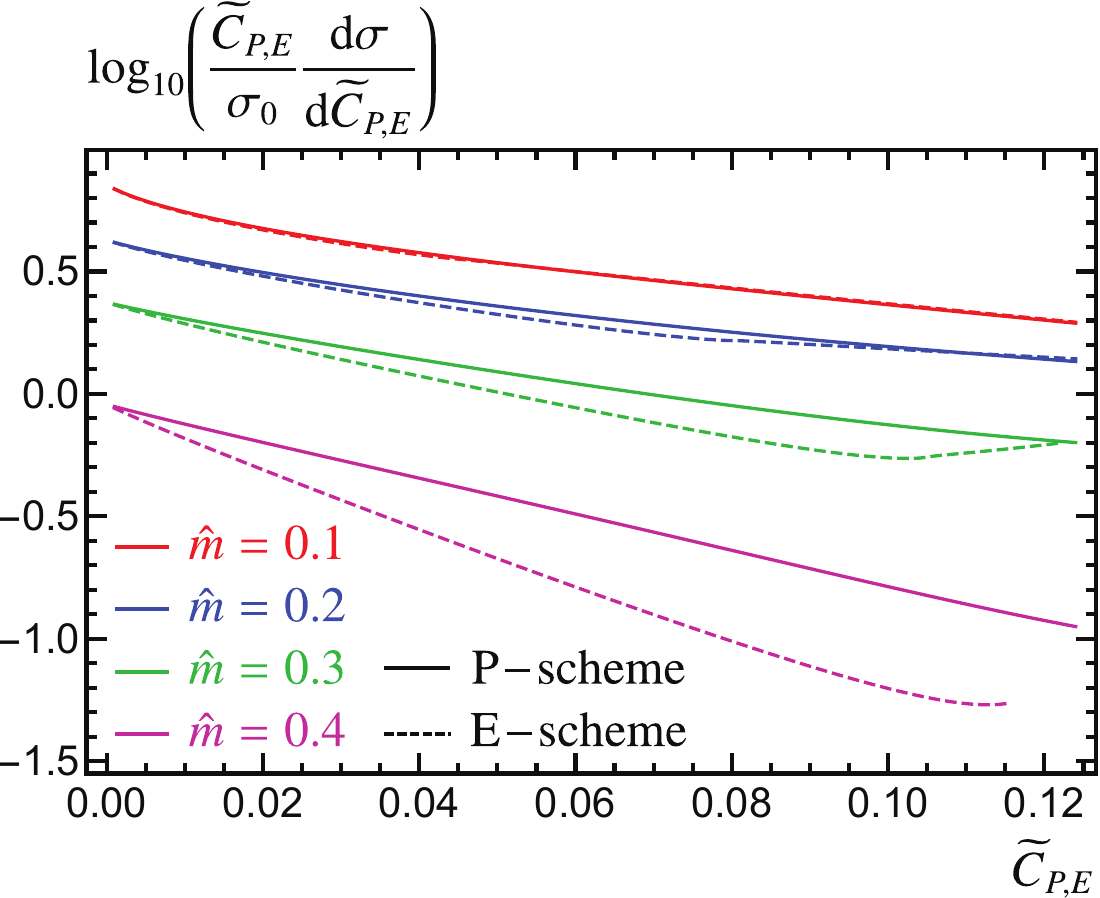}
\label{fig:EPC}
}
\subfigure[]{
\includegraphics[width=0.48\textwidth]{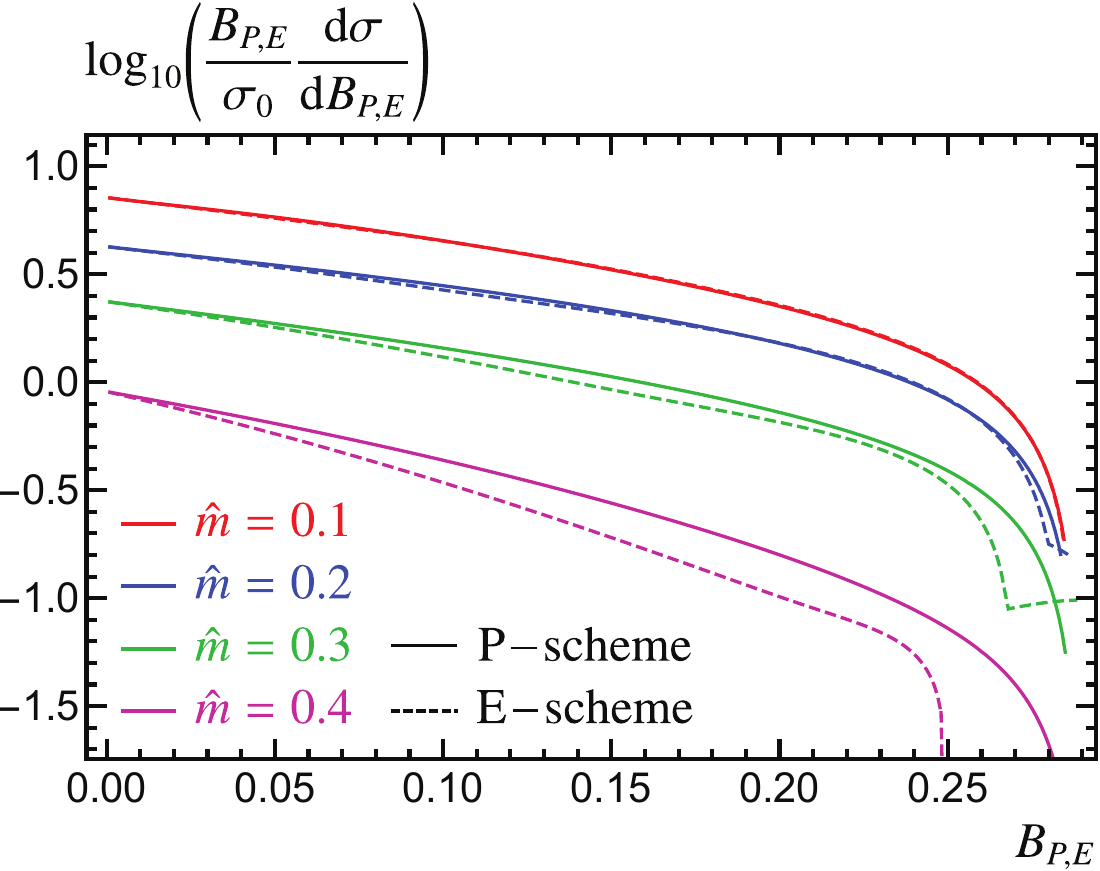}
\label{fig:EPB}
}
\subfigure[]{
\includegraphics[width=0.48\textwidth]{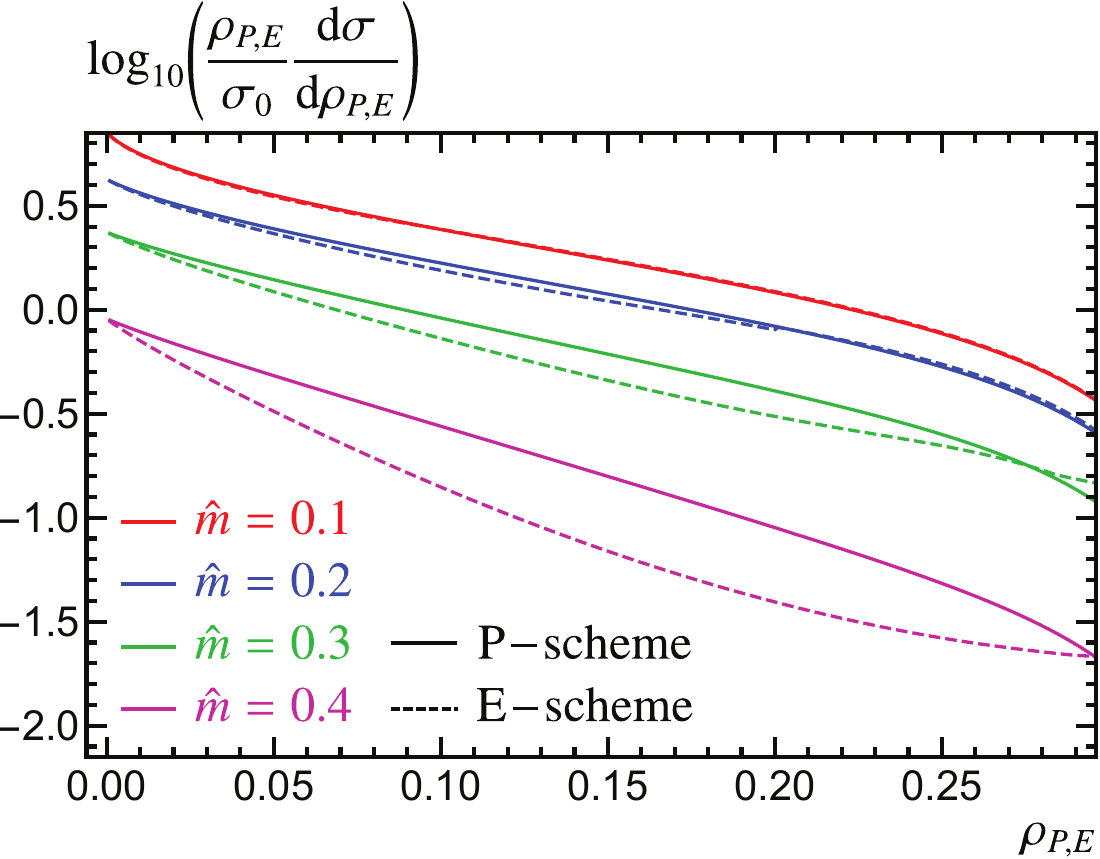}
\label{fig:EPRho}
}
\caption{Event-shape differential distributions for the vector current in the P (solid lines) and E schemes (dashed lines). Panels (a), (b), (c), and (d)
show the cross section for thrust, C-parameter, jet broadening, and heavy jet mass, respectively. All curves are multiplied by $e-e_{\rm min}$, with
$e$ the event-shape value and $e_{\rm min}$ its minimal value, such that the cross section is finite for $e=e_{\rm min}$. Red, blue,
green and magenta show the results for $\hat m=0.1,\,0.2,\,0.3$ and $0.4$, respectively.}
\label{fig:EvsP}
\end{figure*}
The P-scheme version of angularities and E-scheme versions of the other event shapes listed here can be found in Appendix~\ref{app:anadel}. Note
that the substitutions have to be done for the event-shape normalization as well, such that E-scheme event shapes are usually normalized using
$Q = \sum_i E_{p,i}$, while P-scheme event shapes are normalized by $Q_P = \sum_i |\vec p_i|$. The definition of the thrust axis itself does not
change with the scheme, i.e.\ it is always defined with respect to the original P-scheme thrust definition. The minimal value for these event shapes remains
$e_{\rm min} = 0$, meaning that these observables are insensitive to parton masses at leading order, which can be useful in cases where mass
effects are preferred to be small.

Some event shapes are, in their original definitions, neither P- nor E-scheme, sometimes referred to as ``massive scheme'' or ``M-scheme'',
usually containing full momentum information\,\footnote{For more detailed information on how to define consistent substitution rules, see
Ref.~\cite{Preisser:thesis}.}. One of these is heavy jet mass~\cite{Clavelli:1979md, Chandramohan:1980ry, Clavelli:1981yh}, defined as the heavier
of the two hemisphere invariant masses, normalized by $Q^2$
\begin{equation}
\rho = \frac{1}{Q^2}\Biggl(\,\sum_{i \in \text{heavy}}\!\!p_i\!\Biggr)^{\!\!2}\,,
\end{equation}
where the hemispheres are defined to be separated by the plane orthogonal to the thrust axis.
It can be useful to define massive scheme versions of other event shapes as well. Examples include the massive version of thrust
(2-jettiness)~\cite{Stewart:2009yx} and C-parameter (C-jettiness)~\cite{Gardi:2003iv}. 2-jettiness is defined by generalizing the original definition to
\begin{equation}
\tau_J = \frac{1}{Q}\sum_i(|E_i| - |\hat t\cdot \vec p_i|)\,,
\end{equation}
while C-jettiness is based on the Lorentz-invariant form
\begin{equation}
C_J = \frac{3}{2}\biggl[2 - \sum_{i\neq j}\frac{(p_i\cdot p_j)^2}{(p_i\cdot q)(p_j\cdot q)}\biggr],
\end{equation}
introduced in Ref.~\cite{Ellis:1980nc}, with $q = \sum_i p_i$. These event shapes usually have a non-zero minimal value $e_{\rm min} \neq 0$ and are
therefore mass sensitive already at leading order. The increased mass sensitivity can be useful when studying mass related issues, e.g.\ 2-jettiness
was used to calibrate the {\scshape Pythia}~8.205 MC top quark mass~\cite{Butenschoen:2016lpz} and has been proposed to measure the top quark mass
at a future linear collider~\cite{Fleming:2007qr}.

In Appendix~\ref{app:anadel} we present some analytic results for the delta-function coefficients of differential cross sections for the event shapes
listed above in various schemes, together with respective characteristic information on the event shapes.

Differential cross sections in the E- and P-schemes for a selection of event shapes can be seen in Fig.~\ref{fig:EvsP}. The plots have been
generated using the algorithm described in Sec.~\ref{sec:differential}. We do not show
massive-scheme cross sections in this plot since their lower endpoint is different from zero. We have
chosen the plot-range of the P-scheme allowed values, since they are mass-independent, and our $y$ axis is in a logarithmic scale to make the
curves with small values of $\hat m$ visible. In general E-scheme maximal values do depend on the reduced mass (see Appendix~\ref{app:anadel}
for some examples). 
Since the scheme dependence
vanishes for $m=0$, curves are very similar for small values of the reduced mass, resulting in nearly identical red lines in all four panels of the figure.
As the mass increases, the differences grow, and for $\hat m=0.4$ the curves in both schemes are clearly different. We observe that the cross
section is smaller in the E-scheme for most of the spectrum.

\section{Analytic Results for the Distributions at Threshold}\label{sec:derivation}
\begin{figure}[t]\centering
\includegraphics[width=0.15\textwidth]{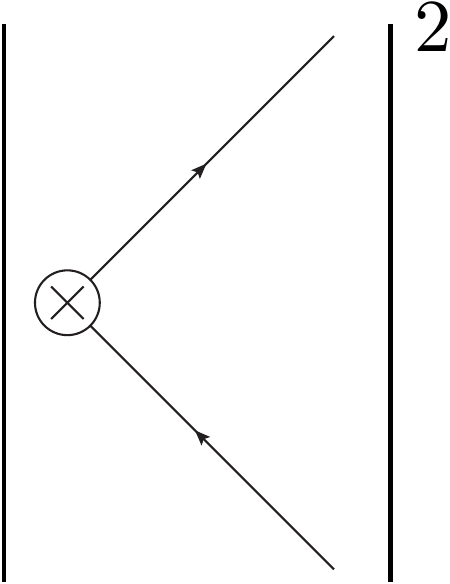}
\caption{Diagrammatic contribution to the $\Ord(\alpha_s^0)$ differential event shape distribution. The diagram has to be squared and
contributes only as a Dirac delta function.\label{fig:Tree}}
\end{figure}
\begin{figure*}[t!]
\subfigure[]
{\includegraphics[width=0.4\textwidth]{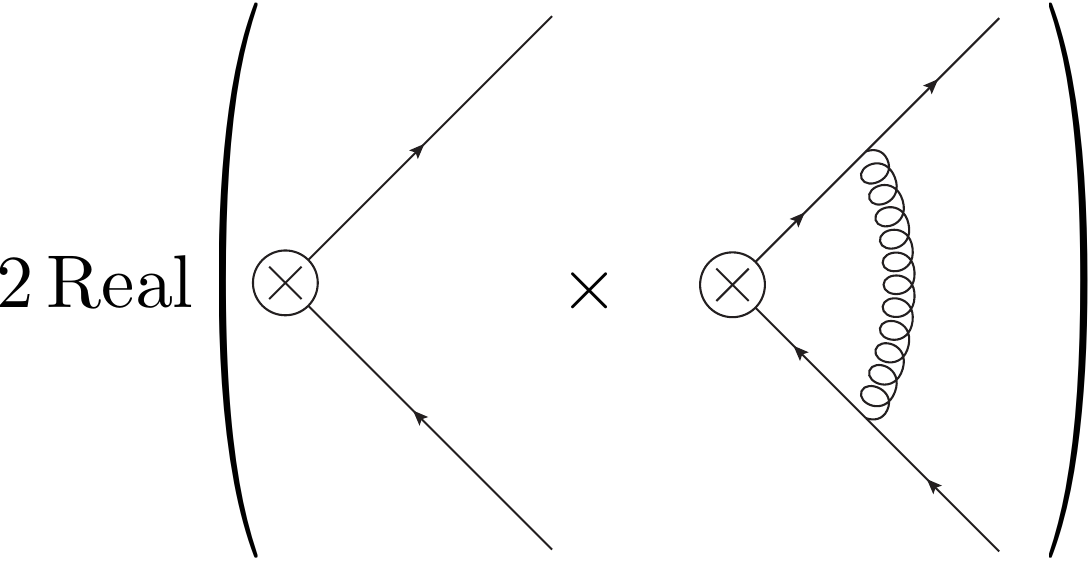}\label{fig:virtual}}~~~~~~~~~~~~
\subfigure[]{\includegraphics[width=0.41\textwidth]{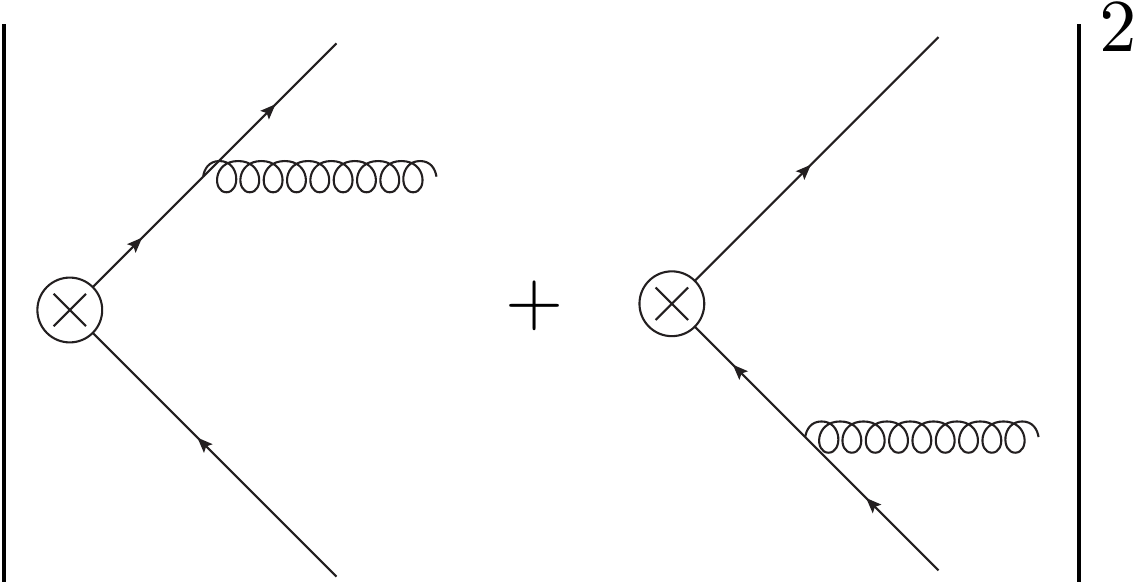}\label{fig:real}}
\caption{Contributions to the $\Ord(\alpha_s)$ differential cross section for event shapes. Panel (a) shows the virtual contribution as twice
the interference of the three-level and one one-loop diagrams, while in panel (b) the two real-radiation diagrams that have to be added and
squared are drawn.}
\label{fig:NLO}
\end{figure*}

In this section we provide the computation of one of our main results, an integral expression for the delta-function coefficient of event-shape
differential cross sections in full QCD. Obviously, the results are different for vector and axial-vector currents, but the computation is analogous
to both processes. Along the computation we show that the coefficient of the plus distribution is the same for any observable linearly sensitive
to soft momentum. The delta-function coefficient receives contributions from the virtual- and real-radiation diagrams, which are separately IR
divergent, although the sum is finite. The Feynman diagrams at LO and NLO are shown in Figs.~\ref{fig:Tree} and \ref{fig:NLO}, respectively.
In the virtual term, the divergence originates from a loop integration, while in the real-radiation it is a consequence of the phase-space integration.
The cancellation can be achieved by computing the two terms explicitly, as in the approach followed in Sec.~\ref{sec:direct} for the differential cross
section, or, in the case of inclusive quantities such as the total hadronic cross section, by taking the imaginary part of the forward scattering amplitude.
In this approach, IR divergences that might appear in individual Feynman diagrams are always a consequence of loop integrals. Furthermore, one never
has to deal with squaring matrix elements. 
We exploit this fact in Appendix~\ref{sec:indirect}, and analytically 
compute the delta-function coefficient by simply imposing that the differential cross
section integrated across the whole spectrum reproduces the total hadronic cross section. 
\subsection[Born Cross Section and \texorpdfstring{$\Ord(\alpha_s^0)$}{Tree Level} Distribution]
{\boldmath Born Cross Section and $\Ord(\alpha_s^0)$ Distribution}\label{sec:Born}
\begin{figure*}[t!]
\subfigure[]
{
\includegraphics[width=0.48\textwidth]{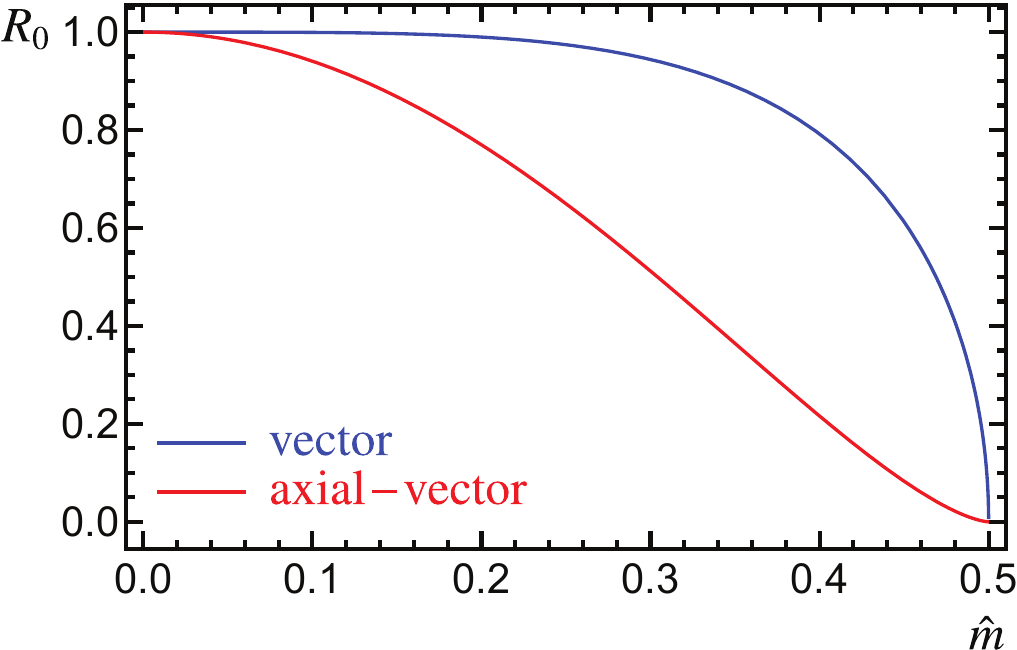}
\label{fig:R0}
}
\subfigure[]{
\includegraphics[width=0.473\textwidth]{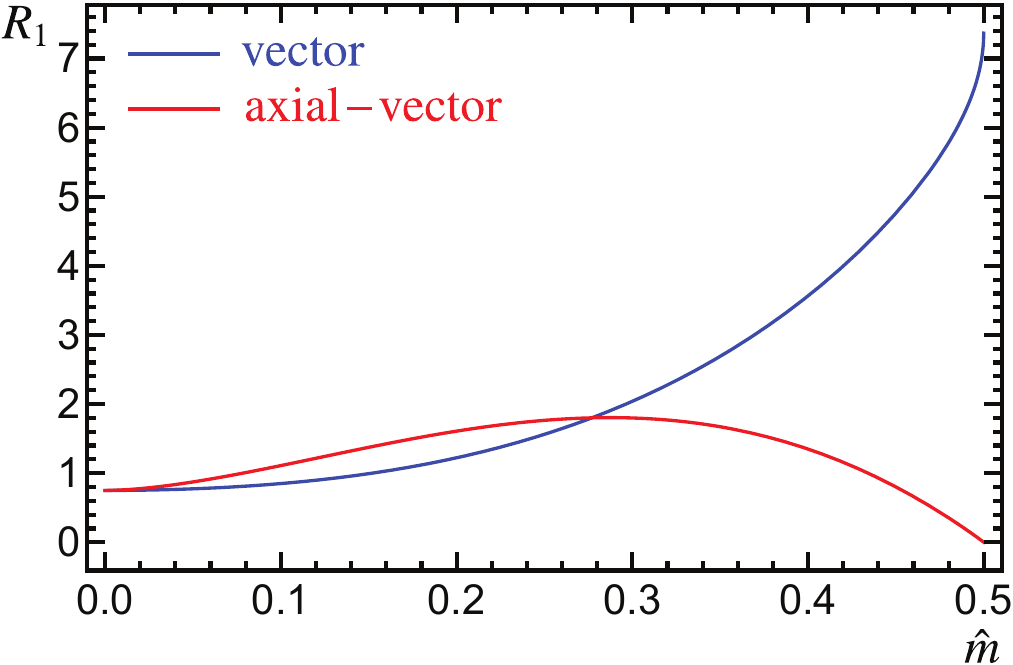}
\label{fig:R1}
}
\caption{Massive total hadronic cross section for vector (blue) and axial-vector (red) currents, at tree-level in panel (a) and
one loop in panel (b).}
\label{fig:Rhad}
\end{figure*}

It is customary to present event-shape distributions normalized to the Born cross section, which is defined as the cross section for massless quarks
at tree-level in four dimensions. The Born cross section is different for vector and and axial-vector currents: the former gets contributions from both
photon and Z-boson exchange, while the letter is mediated by the Z boson only. Taking into account the finite width $\Gamma_Z$ of the Z boson,
one obtains
\begin{align}
\sigma^V_0 & = \frac{N_c}{3} \frac{4 \pi \alpha_{\rm em}^2}{Q^2}
\Biggl[ Q^2_q + \frac{v^2_f (v_e^2 + a_e^2)}{(1 - \hat{m}_Z^2)^2 + \bigl(
\frac{\Gamma_Z}{m_Z} \bigr)^2} + \frac{2\, Q_q v_e v_q (1 - \hat{m}_Z^2)}{(1
- \hat{m}^2_Z)^2 + \bigl( \frac{\Gamma_Z}{m_Z} \bigr)^2} \Biggr],\\
\sigma^A_0 & = \frac{N_c}{3} \frac{4 \pi \alpha_{\rm em}^2}{Q^2}
\Biggl[ \frac{a_q^2 (v_e^2 + a_e^2)}{(1 - \hat{m}_Z^2)^2 + \bigl(
\frac{\Gamma_Z}{m_Z} \bigr)^2} \Biggr],\nonumber
\end{align}
with $ \alpha_{\rm em}$ the electromagnetic coupling, $\hat m_Z=m_Z/Q$ the reduced Z-boson mass, $Q_q$ the quark electric charge, $N_c$ the
number of colors, and $v_e$ and $a_e$ ($v_q$ and $a_q$) the electron (quark) vector and axial-vector couplings to the Z boson. Here and in what
follows, the leptonic trace is always computed in four dimensions. This poses no problem since we are taking the electroweak interactions at leading
order only. For non-zero quark masses, the normalized tree-level cross section is different for vector and axial-vector currents:
\begin{equation}
\frac{\sigma^V_{0, m}}{\sigma^V_0} \equiv R_0^V(\hat m)= \frac{(3-v^2)\,v}{2} \,,\qquad
\frac{\sigma^A_{0, m}}{\sigma^A_0} \equiv R_0^A(\hat m)= v^3\,,
\end{equation}
with $v=\sqrt{1 - 4 \hat{m}^2}$ the velocity of the on-shell massive quarks in the center of mass frame. The functions $R_0^V$ and $R_0^A$ are shown
graphically as a function of $\hat m$ in Fig.~\ref{fig:R0}. In the massless limit $R_0^C(v=1)=1$, while both vanish
at threshold ($v\to0$). 
At $\Ord(\alpha_s^0)$ the differential cross section is obviously
\begin{equation}
\frac{1}{ \sigma^C_0}\frac{\dd \sigma_0}{\dd e} = R_0^C(\hat m)\,\delta (e - e_{\rm min})\,,
\end{equation}
with $C=V,A$ for vector and axial-vector currents, respectively, and $e_{\rm min}$ the minimal value the event shape can take.

\subsection{Phase Space and Kinematic Variables}\label{sec:phase}
\begin{figure}[t]\centering
\includegraphics[width=0.5\textwidth]{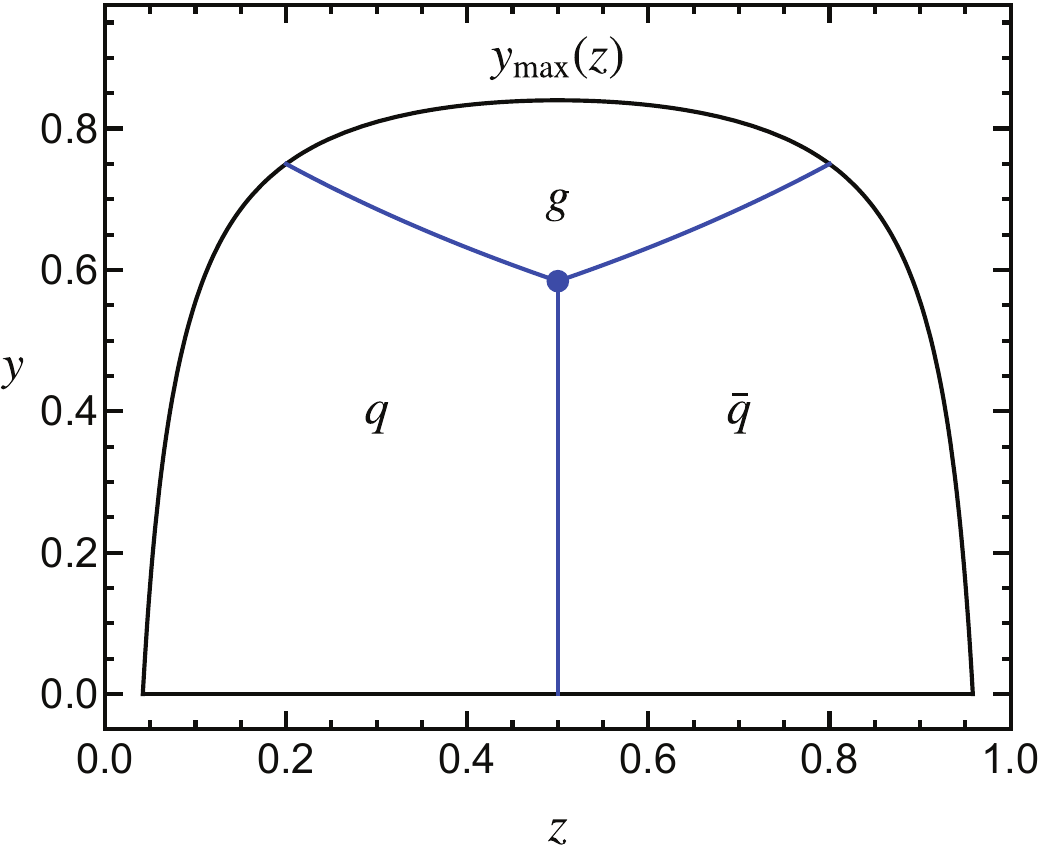}
\caption{Phase space diagram in $(z,y)$ coordinates for two particles with equal mass (quark and anti-quark) and a massless particle (gluon).
The plot is generated with the numerical value $m/Q=0.2$. The available phase space is contained between the $y=0$ and $y=y_{\rm max}(z)$
curves, which intersect at the points $(z=z_{\pm}, y=0)$. Blue lines split the phase space into regions where the thrust axis points into the
direction of the quark, anti-quark or gluon momenta. The three lines meet at the point $(1/2,y_{\rm middle})$, marked by a blue dot.
\label{fig:Phase-space}}
\end{figure}
In this section we introduce some notation and write down the 2- and 3-body phase space in $d=4-2\varepsilon$ dimensions in terms of kinematic
variables that facilitate our computation. All masses appearing in this article are understood in the pole scheme. The phase space for $n$ particles in
$d=4-2\varepsilon$ dimensions is defined as
\begin{equation}
\int\! \dd\Phi_n = (2\pi)^d\! \int\!\Biggl(\,\prod_{i=1}^n\frac{\dd^{d-1} \vec p_i}{2 E_i (2\pi)^{d-1}}\!\Biggr)\, \delta^{(d)}\!
\biggl(\!P^\mu - \sum_{i=1}^n p_i^\mu\biggr)\,,
\end{equation}
with $E_i=\sqrt{|\vec p_i|^2+m_i^2}$, since the particles are on-shell. Let us start with the $2$-body phase space for particles with the same mass $m$:
\begin{equation}
\Phi_2 = (1 - 4 \hat{m}^2)^{\frac{1}{2} - \varepsilon} \,\Phi^{m = 0}_2 \,,\qquad
\Phi^{m = 0}_2 = \frac{\Gamma (1 - \varepsilon)}{2\, \Gamma (2 - 2 \varepsilon)} \frac{Q^{- 2\varepsilon}}{(4 \pi)^{1 - \varepsilon}} \,,
\end{equation}
where, for convenience, we have factored out the $d$-dimensional $2$-body phase space for massless particles.
For three particles, with particle $1$ (quark) and $2$ (anti-quark) having the same mass $m$, while particle $3$
(gluon) is massless, and adding the flux factor one obtains
\begin{align}
\frac{\tilde{\mu}^{2 \varepsilon}}{2 Q^2} P(Q,\varepsilon) \int \!\dd\Phi_3 ={}&
\frac{\Bigl( \frac{4 \pi \tilde{\mu}^2}{Q^2} \Bigr)^{\!\varepsilon}}{256 \pi^3 \Gamma (1 - \varepsilon)}\!
\int \! \dd x_1\, \dd x_2\, \Bigl\{ (1 - x_1) (1 - x_2) (x_1 + x_2 - 1) \\
& - \hat m^2 (2 - x_1 - x_2)^2 \Bigr\}^{- \varepsilon} \,,\nonumber\\
P(Q,\varepsilon) \equiv& \frac{\Phi^{m = 0}_2 |_{\varepsilon \to 0}}{\Phi^{m = 0}_2} \,,\nonumber
\end{align}
with $x_i=2\,E_i/Q$, and $E_i$ the energy of the $i$-th particle such that energy conservation implies $x_1+x_2+x_3=2$.
For convenience we have multiplied by the ratio of the $d=4$ and $d=4-2\varepsilon$ massless $2$-body phase-space factors. This
helps taking the $\varepsilon\to 0$ limit after canceling the IR divergences. Next we implement the following variable
transformation: $x_1 = 1 - (1 - z) \,y$, $x_2 = 1 - z \,y$, such that $y=2E_g/Q$, which makes the soft limit $y \rightarrow 0$ manifest:
\begin{equation}
\frac{\tilde{\mu}^{2 \varepsilon}}{2 Q^2} P(Q,\varepsilon) \!\int \!\dd\Phi_3 =
\frac{\Bigl( \frac{4 \pi \tilde{\mu}^2}{Q^2} \Bigr)^{\!\varepsilon}}{256 \pi^3 \Gamma (1 - \varepsilon)}
\int \!\dd y\, \dd z\, y^{1 - 2 \varepsilon} \, [\,z (1 - z) (1 - y) -{\hat m}^2 \,]^{- \varepsilon}\,.
\end{equation}
This result shows how collinear singularities, that would be located at $z = 0, 1$ for massless particles, are screened by the finite quark mass.
In these coordinates the Dalitz region is parametrized as
\begin{align}
&0 \leq y \leq y_{\rm max} (z) \equiv 1 - \frac{\hat{m}^2}{z (1 - z)}\,,
\qquad z_- \leq z \leq z_+\,, \\
& z_{\pm} \equiv \frac{ 1 \pm v}{2}\,.\nonumber
\end{align}
In Fig.~\ref{fig:Phase-space} we show the
phase-space boundaries for the numerical value $\hat m = 0.2$. The phase-space limits in the $z$ variable satisfy $z_+ z_- = \hat{m}^2$
and $z_+ + z_- = 1$, while the upper limit in the $y$ variable has its maximum at $ y_{\rm max}(1/2)= v^2$. Another useful relation is given by
$(z-z_-)(z_+-z)=z(1-z) - \hat m^2$, a positive quantity inside the Dalitz region. In the massless limit the Dalitz plot is simply a square:
$0 \leq y \leq 1$, $| z | \leq 1$. The phase space (as well as all matrix elements and event-shape measurement functions) are invariant under the
change $z \to 1 - z$ (mirror symmetry with respect to the $z=1/2$ vertical line), since all results remain the same when exchanging quark
and anti-quark.

For later use, it is convenient to split the $3$-body phase space into regions where the thrust axis points into the direction of the quark, anti-quark or
gluon momenta. To that end we define
\begin{equation}\label{eq:psregbound}
y_{\tau} (\hat{m}, z) = \frac{\sqrt{1 - 4 \hat{m}^2 (1 - z^2)} - z}{1 - z^2}\,,
\end{equation}
such that these three regions are given by
\begin{align}\label{eq:regions}
0 &\leq z \leq \frac{1}{2}\,, &0 \leq y &\leq y_{\tau} (\hat{m}, z), & {\rm quark}\,, \nonumber\\
\frac{1}{2} &\leq z \leq 1\,, &0 \leq y &\leq y_{\tau} (\hat{m}, 1 - z), & {\rm anti}{\text -}{\rm quark}\,, \\
0 &\leq z \leq 1\,, &{\rm max}\, [\,y_{\tau} (\hat{m}, z), y_{\tau} (\hat{m}, 1 - z)\,] \leq y &\leq y_{\rm max} (z), & {\rm gluon} \,,
\nonumber
\end{align}
and the lines separating the three regions meet at the point $y_{\rm middle} = 4\bigl(\sqrt{1 - 3 \hat{m}^2} - 1/2 \bigr)/3$, $z = 1 / 2$.
The quark [anti-quark] boundary meets the phase-space boundary at $z = \hat{m}$ [$z = 1 - \hat{m}$], $y = (1 - 2 \hat{m}) / (1 - \hat{m})$.

\begin{figure}[t]\centering
\includegraphics[width=0.5\textwidth]{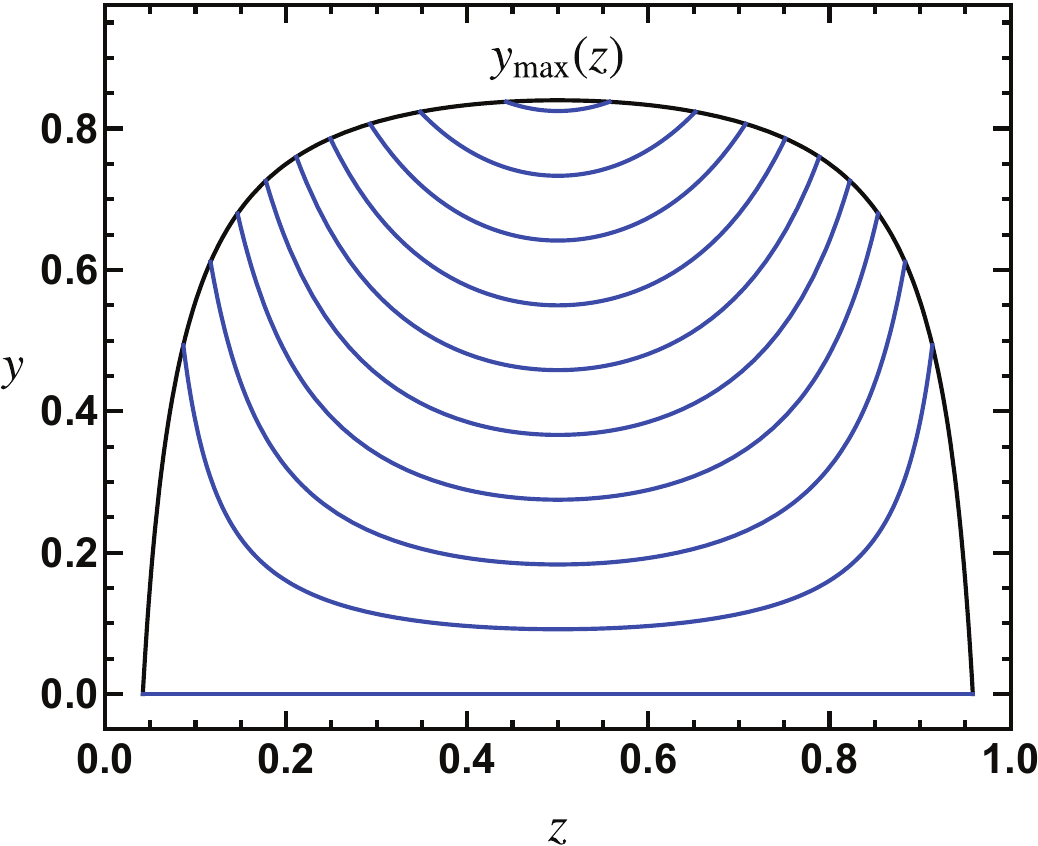}
\caption{Phase space diagram in $(z,y)$ coordinates (black lines) showing curves with constant value of the C-parameter measurement function
in the soft limit $\overline C$ (blue lines) with $\hat m = 0.2$. The lines correspond to $10$ equally spaced values of $\overline C$ between $0$
and $0.681$. For the event shape maximal value $\overline C_{\rm max} = v^2 f_C(1/2)$, the corresponding contour line intersects with the phase
space boundary at one point only, the maximum $(1/2,v^2)$. \label{fig:C-soft}}
\end{figure}

The value of any event-shape variable for events with three particles in the final state (two quarks and a gluon) can be expressed as a function
of the reduced mass $\hat m $ and the $z$ and $y$ phase-space variables. This function, which is not always smooth or continuous, will be
referred to as the measurement function $\hat e(y,z)$ (for simplicity we will omit its mass dependence). Massive event-shape
measurement functions $\hat e(z,y)$ take their minimal value if $y=0$, regardless of the value of $z$, i.e.\ $\hat e(z,0)=e_{\rm min}$, and in the
soft limit $y\to 0$ the measurement function can be expanded as follows:\,\footnote{We consider only the usual case of event shapes linearly
sensitive to soft momentum, that is with $f_e(z)\neq0$. For event shapes with quadratic (or higher) sensitivity to soft momenta, that is with
\begin{equation}
\frac{\dd^n\hat e(y,z)}{\dd y^n}\bigg|_{y=0} \neq 0\,,
\end{equation}
only for some $n>1$, one finds that the differential distribution contains up to the $(n-1)$-th derivative of delta and plus distributions. Since those
event shapes are scarce and of little interest, we do not show any explicit results for them.}

\begin{equation}
\hat e (z, y) = e_{\rm min} + y f_e (z) + \Ord (y^2) \equiv {\bar e}(y,z) + \Ord (y^2) \,,
\end{equation}
where we have defined the soft event-shape variable ${\bar e}$ associated to $e$,\footnote{See Appendix~\ref{app:anadel}
for some event-shape specific expressions for $f_e(z)$.} with measurement function
${\bar e}(y,z)=e_{\rm min} + y f_e (z)$. The soft event shape has the same minimal value
\mbox{$\bar e_{\rm min}=e_{\rm min}$} as the original one, but has a different maximal value, generally larger, that is attained at the highest point of
the Dalitz plot, $(z,y)=(1/2,v^2)$, as can be seen in Fig.~\ref{fig:C-soft}:
\begin{equation}
\bar{e}_{\rm max} = e_{\rm min} + v^2 f_e (1 / 2) \,.
\end{equation}

\subsection[Direct Computation of \texorpdfstring{$\Ord(\alpha_s)$}{One-Loop} Results]
{\boldmath Direct Computation of $\Ord(\alpha_s)$ Results}
\label{sec:direct}
In this approach we directly compute the differential distribution adding up real- and virtual-radiation diagrams. Our computation reproduces
the few known results (either analytic or numeric), but is more general. In this approach for carrying out the calculation we explicitly show
how IR singularities cancel in the sum for IR safe observables already at the differential level. We regulate them using $d=4-2\varepsilon$
dimensions (dimreg) and use plus-distribution identities to keep the computations as general and simple as possible. We have also written the
$3$-body phase space in a way in which IR singularities look as close as possible to UV ones.

\subsubsection{Virtual Radiation}\label{sec:virtual}

We start with the virtual radiation diagrams, which at this order have only two particles in the final state and are common to all event shapes.
The contribution to the differential
cross section at $\Ord(\alpha_s)$ comes from the interference of the tree-level and one-loop diagrams (which have IR divergences treated
in dimensional regularization), and since momenta are fully
constrained by energy-momentum conservation, it only contributes to the delta-function coefficient. The general form of the vector and
axial-vector massive form factors up to one loop take the form\,\footnote{Here we already include the wave-function renormalization in the
OS scheme $Z^{\rm{OS}}_q$ and the term coming from pole mass renormalization.}
\begin{align}
V^{\mu} & = \biggl[ 1 + C_F \frac{\alpha_s}{\pi} A (\hat{m}) \biggr]
\gamma^{\mu} + C_F \frac{\alpha_s}{\pi} \frac{B (\hat{m})}{2 m} (p_1 -p_2)^{\mu} \,, \\
A^{\mu} & = \biggl[ 1 + C_F \frac{\alpha_s}{\pi} C (\hat{m}) \biggr]
\gamma^{\mu} \gamma_5 + C_F \frac{\alpha_s}{\pi} \frac{D (\hat{m})}{2 m}\, \gamma_5\, q^{\mu}\,, \nonumber
\end{align}
with $q = p_1 + p_2$ the photon or Z-boson momentum, and $p_i$ the quark and anti-quark momenta. The vector form factor satisfies the Ward
identity $q_\mu V^\mu = 0$, while the longitudinal part of the axial form factor does not contribute to the cross section.
Their real parts take the form~\cite{Jersak:1981sp, Harris:2001sx}
\begin{align}
{\rm Re} [C (\hat{m})] & = {\rm Re} [A (\hat{m})] + \frac{4\hat{m}^2 }{v}L_v \,,\nonumber\\
{\rm Re} [D (\hat{m})] &= 2\, \hat{m}^2 \biggl[ 1 - \frac{2+v^2}{v} L_v \biggr],\nonumber\\
{\rm Re} [B (\hat{m})] & = \frac{2\, \hat{m}^2}{v}L_v\,,\\
{\rm Re} [A (\hat{m})] &= \biggl( \frac{1+v^2}{2v}L_v -\frac{1}{2} \biggr)\biggl[\frac{1}{\varepsilon} - 2\log\biggl( \frac{m}{\mu} \biggr)\biggr]+
A_{\rm reg}(\hat{m})\,,\nonumber\\
A_{\rm reg}(\hat{m}) & =\frac{3}{2} \,v\, L_v - 1 + \, \frac{1+v^2}{4v} \biggl[ \pi^2 - 2
L_v^2 - 2\, {\rm Li}_2\biggl( \frac{2 v}{1 + v} \biggr) \biggr]\nonumber,
\end{align}
with
\begin{equation}
L_v \equiv \log\biggl( \frac{1 + v}{2 \hat{m}} \biggr).
\end{equation}
IR singularities look the same for both currents, and are fully contained in the transverse form factors $A$ and $C$ (note that due to current
conservation the form factors are UV finite, and therefore all singularities left after carrying out the QCD renormalization program are IR).
The results are singular in the $\hat m\to 0$ limit since collinear singularities are regulated by the finite quark mass. Furthermore in this limit the
longitudinal form factors $B$ and $D$ vanish, and the transverse form factors become identical for
the two currents.

The form factor contribution at $\Ord(\alpha_s)$ is
\begin{align}\label{eq:virtual}
&\frac{1}{ \sigma^C_0}\frac{\dd \sigma^V_{2,\rm virt}}{\dd e} = \frac{\alpha_s}{\pi} C_F R_C^{\rm virt}(\hat m)\,\delta (e - e_{\rm min})\,,\\
&
R_V^{\rm virt}(\hat m)
= 
P(Q,\varepsilon) \, v^{1 - 2\varepsilon}\Bigl\{ 2\, {\rm Re} [A (\hat{m})] (1 + 2 \hat{m}^2
- \varepsilon) - v^2 {\rm Re} [B (\hat{m})] \Bigl\}\nonumber\\
&\quad\quad\quad\;\,\;= 
P(Q,\varepsilon) v \biggl\{ (3-v^2) {\rm Re} [A] - [1 + 2\log (v)]
\biggl( \frac{1+v^2}{v} L_v - 1 \biggr) - v^2\, {\rm Re} [B] + \Ord(\varepsilon)\biggl\}\,,\nonumber\\
&
R_A^{\rm virt}(\hat m) = 2\,
P(Q,\varepsilon) v^{3 - 2\varepsilon} (1 - \varepsilon)\, {\rm Re} [C (\hat{m})]\nonumber\\
&\quad\quad\;\;\,\quad= 2\,
P(Q,\varepsilon) v^3 \biggl\{{\rm Re} [C]- [1 + 2\log (v)]
\biggl( \frac{1+v^2}{v} L_v - \frac{1}{2} \biggr) + \Ord(\varepsilon)\biggl\}\,.\nonumber
\end{align}
When adding these results to the real radiation contributions the $1/\varepsilon$ term present in the $A$ and $C$ form factors cancel,
along with the associated $\mu$ dependence.

\subsubsection{Real Radiation}\label{sec:real}
\begin{figure}[t]\centering
\includegraphics[width=0.5\textwidth]{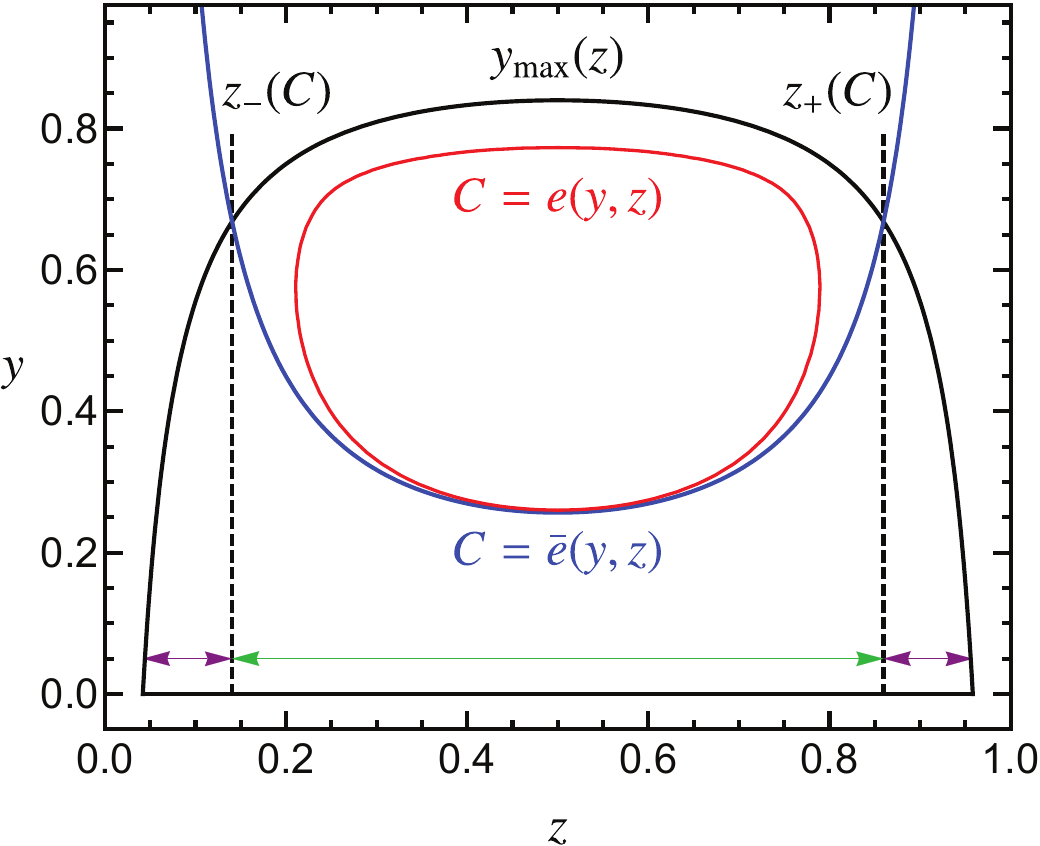}
\caption{Phase space diagram in $(z,y)$ coordinates (black solid lines) showing a curve with constant value of the C-parameter event
shape measurement function (red line) and its soft limit (blue line), using $\hat m = 0.2$, $C=0.42$. The latter corresponds to Eq.~\eqref{eq:iso-soft}, which cuts
the phase-space boundary at the points $z_\pm(C)$. The dashed lines divide the $y=0$ axis into three segments, marked with
double-pointed arrows.
\label{fig:C-intersections}}
\end{figure}

Real radiation diagrams exhibit IR singularities in phase-space integrals, originating from the zero gluon-momentum limit.
The matrix element squared, summed over the polarization of final-state particles and averaged over the lepton spins can be written as
\begin{align}
\frac{|\sum_{\rm spin} \mathcal{M}_C|^2}{4\sigma^C_0}&=\frac{256 \pi^2 \alpha_s \tilde{\mu}^{2 \varepsilon} C_F}{y^2}M_C (y, z, \hat{m}, \varepsilon)\,,\\
M_C (y, z, \hat{m}, \varepsilon)& =M^0_C (z, \hat{m}) + \varepsilon M^1_C (z, \hat{m}) + y M_C^{\rm{hard}}(y, z) +\Ord(\varepsilon^2)\,,\nonumber\\
M^0_V (z, \hat{m}) & = - (1 + 2{\hat m}^2) M^1_V (z, \hat{m})\,,\quad M^1_V (z, \hat{m}) = - \frac{(1 - z) z - \hat{m}^2 }{(1 - z)^2 z^2 }\,,\nonumber\\
M^0_A (z, \hat{m}) & = - M^1_A (z, \hat{m}) = -v^2 M^1_V (z, \hat{m})\,,\nonumber
\end{align}
with $C=V,A$ labeling the current type and $\mu^2 \equiv 4 \pi \tilde{\mu}^2 \mathrm e^{- \gamma_E}$. We denote the pieces which vanish for
$y\to 0$ as the ``hard matrix elements'' $M^{\rm{hard}}_C$. Our results agree with those in
Refs.~\cite{Kramer:1979pg, Ioffe:1978dc, Nilles:1980ic} (note that there is a known sign error in Ref.~\cite{Kramer:1979pg}). The hard matrix
elements can be further split as $M^{\rm{hard}}_C (y, z, \hat{m}) = M^2_C (z, \hat{m}) + y M^3_C (z, \hat{m})$, with
\begin{align}
M^2_C (z, \hat{m}) & = - \binom{1 + 2 \hat{m}^2}{v^2} \frac{1}{z (1 - z)}\,,\\
M^3_C (z, \hat{m}) & = \frac{1}{2 z (1 - z)} \binom{1}{1 + 2 \hat{m}^2} -1\,,\nonumber
\end{align}
with the upper (lower) part of the expression in parentheses belonging to the vector (axial-vector) current.
The radiative one-loop contribution to the differential distribution, partially expanded around $\varepsilon=0$, reads
\begin{align}
&\frac{1}{\sigma^C_0} \frac{\dd \sigma^{\rm real}_C}{\dd e} = \\
& P(Q,\varepsilon) C_F \frac{\alpha_s}{\pi} \frac{\Bigl( \frac{4 \pi \tilde{\mu}^2}{Q^2}
\Bigr)^{\!\varepsilon}}{\Gamma (1 - \varepsilon)} \int\! \frac{\dd y\, \dd z}{y^{1 + 2 \varepsilon}} \, [\,z (1 - z) (1 - y) - {\hat m}^2 \,]^{-\varepsilon} \,
\delta [e - \hat e (y, z)]\, M_C (y, z, \hat{m}, \varepsilon) \nonumber \\
& = P(Q,\varepsilon) C_F\frac{\alpha_s}{\pi} \biggl\{ - \frac{\delta (e - e_{\rm min})}{2}\! \int\!\dd z\, \biggl[ M^1_C (z, \hat{m}) + M^0_C (z, \hat{m})
\biggl( \frac{1}{\varepsilon} + 2 \log \Bigl( \frac{\mu}{Q} \Bigr) \nonumber\\
&\quad - \log\bigl[z (1 - z) - \hat{m}^2 \bigr] \biggr) \biggr] + \!\int\! \dd z\, \dd y\, \biggl[ M^0_C (z,\hat{m}) \biggl[\frac{1}{y} \biggr]_+
+ M_C^{\rm{hard}} (y, z) \biggr]
\delta [e - \hat e(y, z)] \biggl\} . \nonumber
\end{align}
To get to the second line we have collected powers in $y$ and used the identity
\begin{equation}
y^{- 1 - 2 \varepsilon} = - \frac{1}{2 \varepsilon} \delta (y) + \biggl[\frac{1}{y} \biggr]_+ +\Ord (\varepsilon)\,,
\end{equation}
in the terms containing $M^{0,1}_C$. The coefficient of the Dirac delta function is not yet fully explicit, as the integral over the plus function still hides
singular terms. The distributional structure is completely determined in the $y\to0$ limit, therefore we add and subtract the following term to the last
integrand
\begin{align}
M^0_C (z, \hat{m}) \biggl[\frac{1}{y} \biggr]_+ \delta[e - \overline{e} (y, z)]\,,
\end{align}
such that in the sum of the original and subtracted terms the plus prescription can be dropped (this can be done because the integrand goes to zero
linearly with $y$). This strategy is similar to subtraction algorithms used in NLO and NNLO parton-level Monte Carlos to achieve cancellation of IR
singularities between real- and virtual-radiation contributions. In our case, the subtraction helps isolating the distributional structure of the cross
section. For the added term we proceed as follows
\begin{align}\label{eq:soft-delta}
&\int\!\dd z\, \dd y\, M^0_C (z, \hat{m}) \biggl[\frac{1}{y} \biggr]_+ \delta[e - \overline{e} (y, z)]\,\Theta[y_{\rm max}(z)-y]=\nonumber\\
&\int\! \dd z \,\frac{M^0_C (z, \hat{m})}{f_e (z)} \biggl[\frac{f_e (z)}{e - e_{\rm min}} \biggr]_+ \Theta \bigl[ y_{\rm max} (z) - h(e,z) \bigr] =\\
&-\! \delta (e - e_{\rm min}) \!\int\! \dd z\, M^0_C (z, \hat{m}) \log [f_e (z)] + \biggl[\frac{1}{e - e_{\rm min}} \biggr]_+ \int \!\dd z\, M^0_C (z, \hat{m})\,
\Theta \bigl[ y_{\rm max} (z) - h(e,z) \bigr]\,,\nonumber
\end{align}
with
\begin{equation}\label{eq:iso-soft}
h(e,z) \equiv \frac{e - e_{\rm min}}{f_e (z)} \,,
\end{equation}
representing a curve in phase space defined by the condition $\bar e(y,z)=e$ (that is, a contour line with constant value of the soft
event-shape measurement function). This line has the important property of always intersecting with the phase-space boundary $y_{\rm max}$ at two
symmetric points, which will be denoted by $z_{\pm}(e)$, as shown in Fig.~\ref{fig:C-intersections} for the C-parameter event shape. To
get to the last line of Eq.~\eqref{eq:soft-delta} we have used the rescaling identity
\begin{equation}
\biggl[\frac{\log^n (b x)}{b x} \biggr]_+ = \frac{1}{b} \biggl\{ \frac{\log^{n + 1} (b)}{n + 1}\, \delta (x) + \sum_{i = 0}^n \binom{n}{i}
\log^{n - i} (b) \biggl[\frac{\log^n (x)}{x} \biggr]_+ \biggr\}\,,
\end{equation}
and the fact that if $e = e_{\rm min}$ then $h(e_{\rm min},z)=0$ and the constraint imposed by the Heaviside function $\Theta[y_{\rm max}(z)]$
is automatically satisfied [\,since $y_{\rm max}(z)>0$\, for $z_-\leq z\leq z_+$]. The second term in the last line of Eq.~\eqref{eq:soft-delta}
is not a pure distribution yet, but can be converted to such using the relation
\begin{equation}
f(x)\, \biggl[\frac{1}{x} \biggr]_+ = f(0)\, \biggl[\frac{1}{x} \biggr]_+ +\frac{f (x) - f (0)}{x}\, .
\end{equation}
Finally we arrive at
\begin{align}
&\int\!\dd z\, \dd y\, M^0_C (z, \hat{m})\, \biggl[\frac{1}{y} \biggr]_+\! \delta[e - \overline{e} (y, z)]=
-\delta (e - e_{\rm min}) \!\int\! \dd z\, M^0_C (z, \hat{m}) \log [f_e (z)]\\
&+\biggl[\frac{1}{e - e_{\rm min}} \biggr]_+ \int \!\dd z\, M^0_C (z, \hat{m})
- \!\int\! \dd z \,M^0_C (z, \hat{m})\, \frac{\Theta [e - e_{\rm min} - y_{\rm max}(z) f_e (z)]}{e - e_{\rm min}} \,,\nonumber
\end{align}
where we have used the identity $\Theta (x) + \Theta (- x) = 1$ in the last term. The Heaviside theta function in the last integral
requires $h(e,z)>y_{\rm max}(z)$, and therefore restricts the $z$ integration to the two disconnected segments shown in
Fig.~\ref{fig:C-intersections} as purple double-pointed arrows: \mbox{$z_- \leq z \leq z_- (e)$}
and $z_+ (e) \leq z \leq z_+$, where $z_{\pm} (e)$ are the two solutions of the equation $e = \bar e[z,y_{\rm max}(z)]$ 
that lay on the original integration path $z_- \leq z \leq z_+$. The points $z_{\pm} (e)$ depend on 
$e$ and fulfill $z_-<z_-(e)<z_+(e)<z_+$, since $f_e (z)$ is positive (given that by definition $e \geq e_{\min}$)\,. It is useful to write
the Heaviside theta in the last term as an integral over a Dirac delta function:
\begin{align}
\frac{\Theta [e - e_{\rm min} - y_{\rm max}(z) f_e (z)]}{e - e_{\rm min}} =
&\int\! \dd y\, \frac{\Theta [y - y_{\max}(z)]}{y} \,\delta [e - e_{\min} - y f_e (z)]\,.
\end{align}
These results provide the contribution of the real-radiation diagrams to the differential cross section shown in Eq.~\eqref{eq:general-diff}:
\begin{align}\label{eq:real}
A_e(\hat m) ={}& A_e^{\rm real}({\hat m}) + R_C^{\rm virt}({\hat m})\,,\\
A_e^{\rm real}({\hat m}) ={}&-\!\frac{P(Q,\varepsilon)}{2} \!\!
\int\!\!\dd z \biggl\{\! M^1_C (z, \hat{m}) + M^0_C (z, \hat{m})\! \biggl[ \frac{1}{\varepsilon} + 2 \log \Bigl( \frac{\mu}{Q} \Bigr)\!
-\! \log\biggl(\!\frac{z (1 - z) - \hat{m}^2}{[f_e (z)]^2} \biggr)\! \biggr]\! \biggl\},\nonumber\\
B_{\rm plus}({\hat m}) ={}& \! \int \!\dd z\, M^0_C (z, \hat{m})\,,\nonumber\\
F^{\rm NS} ={}& \! \int\! \dd z\, \dd y\, \biggl\{ M_C^{\rm{hard}} (y, z) \delta [e - \hat e(y,z)] +
\frac{ M_C^0(z, \hat{m})}{y} \biggl[\delta [e - \hat e(y,z)]\nonumber\\
& - \Theta [y - y_{\max}(z)] \,\delta [e - \bar e(y,z)] - \delta [e - \bar e(y,z)]\biggr]\biggl\}
\equiv F^{\rm NS}_{\rm hard} + F^{\rm NS}_{\rm soft}\,.\nonumber
\end{align}
In $F^{\rm NS}$ (which can only be computed analytically for some 
simple event shapes), in those terms where no explicit Heaviside function is
shown, a $\Theta[y_{\rm max}(z)-y]$ is understood. $F^{\rm NS}_{\rm hard}$ and
$F^{\rm NS}_{\rm soft}$ correspond to the terms containing $M_C^{\rm hard}$ and $M^0_C$, respectively. This intermediate result already
shows that the coefficient of the plus distribution is identical for all event shapes linearly sensitive to soft momentum. The non-singular term
contains no distributions, and there is no singularity in the integration domain: the hard function tends linearly to zero for $y\to0$, while the soft
term contains one piece which is the difference of two delta functions with the same $y\to0$ limit (therefore again going linearly to zero in the soft
limit), and a theta function such that small values of $y$ are left out.
\subsubsection{Final Result for the Direct Computation}\label{sec:final-direct}
We first give an analytic expression for the $B_{\rm plus}$ coefficient in Eq.~\eqref{eq:real}. The $z$ integration is carried out
using the first line in Eq.~\eqref{eq:real-integrals}, yielding
\begin{equation}\label{eq:plus-coef}
B_{\rm plus}(\hat{m}) = \binom{3-v^2}{2\,v^2} \bigl[(1 +v^2) L_v - v \bigr]\,,
\end{equation}
where again the first and second line of the expression in big parentheses correspond to vector and axial-vector currents, respectively.
The result exhibits a log-type singularity for $m\to 0$, since in that limit the log-plus distribution associated to collinear singularities is
no longer screened by the heavy quark mass.

To obtain the coefficient of the delta term we need to perform two integrals analytically, which are given
in Eq.~\eqref{eq:real-integrals}. Adding the results in Eqs.~\eqref{eq:virtual} and \eqref{eq:real} we cancel the $1/\varepsilon$ singularity along with
the $\mu$ dependence. Therefore taking the limit $\varepsilon\to0$ amounts to setting $P(Q,\varepsilon)\to1$, and we get
\begin{align}\label{eq:main-result}
A^V_e (\hat{m}) ={}& (1 + 2 \hat{m}^2) \biggl\{ (1 - 2 \hat{m}^2) \biggl[
{\rm Li}_2 \biggl( \!- \frac{v (1 + v)}{2 \hat{m}^2} \biggr) - 3 \,{\rm Li}_2
\biggl( \frac{v (1 - v)}{2 \hat{m}^2} \biggr) + 2 \log^2(\hat{m}) + \pi^2 \nonumber\\
& - 2 \log^2 \biggl( \frac{1+v}{2} \biggr) \biggr] + 2 v
\bigl[ \log (\hat{m}) - 1 \bigr] - 2 I_e (\hat{m}) \biggl\} + (4+v^2 - 16 \hat{m}^4) L_v \,, \nonumber\\
A^A_e (\hat{m}) ={}& v^2 \biggl\{ (4+v^2) L_v+ 2 v \bigl[ \log (\hat{m}) - 1 \bigr] - 2 I_e
(\hat{m}) + (1 - 2 \hat{m}^2) \\
& \times \! \biggl[ {\rm Li}_2 \biggl( \!- \frac{v (1 + v)}{2\hat{m}^2} \biggr) - 3\, {\rm Li}_2 \biggl( \frac{v (1 - v)}{2 \hat{m}^2}\biggr)
+ \pi^2 + 2 \log^2 (\hat{m}) - 2 \log^2 \biggl( \frac{1 + v}{2}\biggr) \biggr] \biggl\} , \nonumber
\end{align}
where the only event shape dependent piece is the integral
\begin{equation}\label{eq:ES-dependent}
I_e (\hat{m}) 
= \frac{1}{2} \int_{z_-}^{z_+}\! \dd z\, \frac{(1-z)z - \hat m^2}{(1-z)^2 z^2} \log [f_e (z)]
= \int_{z_-}^{1/2}\! \dd z\, \frac{(1-z)z - \hat m^2}{(1-z)^2 z^2} \log [f_e (z)]\,,
\end{equation}
where we have used the $z\leftrightarrow(1-z)$ symmetry to simplify the integration range.

\section{Numerical Algorithms}\label{sec:numeric}
Before we describe the algorithms to compute the differential and cumulative event-shape cross sections, we show how to write down the
four-momenta of the three-particle phase space in terms of the $(z,y)$ coordinates. This is very useful to figure out an analytic expression
for the event-shape measurement function. These expressions can in turn be used to compute the values of $e_{\rm min}$ and $e_{\rm max}$, and can
be expanded around $y=0$ to obtain $f_e(z)$. Since we are not dealing with oriented event shapes, without any loss of generality we can
choose the three particles contained in the $x-y$ plane, with the gluon 3-momenta pointing into the positive $z$ direction. With the notation
$p=[E,\vec{p}\,]=[E,p_x,p_y,p_z]$ one has:\,\footnote{One can generate vectors with non-zero $x$ component by taking vector products,
e.g.\ when computing jet broadening.}
\begin{align}\label{eq:vectors}
p_g &= \frac{y}{2}\,[1,0,0,1]\,,\nonumber\\
p_q &= \biggl[\frac{1 - y (1 - z)}{2}, 0, \sqrt{(1 - y) (1 - z) z - \hat m^2}, \frac{1 - y (1 - z) - 2 z}{2}\biggr]\,,\\
p_{\bar q} &= \biggl[\frac{1 - y z}{2}, 0, -\sqrt{(1 - y) (1 - z) z - \hat m^2}, \frac{2 z-1-y\,z}{2}\biggr]\,.\nonumber
\end{align}
The magnitude of the (anti-)quark three-momentum reads
\begin{equation}
|\vec{p}_{\bar q}| = \frac{1}{2} \sqrt{(1-y\,z)^2-4\hat m^2}\,,\qquad |\vec{p}_{q}| = |\vec{p}_{\bar q}|_{z\to1-z}\,,
\end{equation}
such that E-scheme 4-momenta are obtained by multiplying the spatial components in Eq.~\eqref{eq:vectors} by $E/|\vec{p}\,|$, while P-scheme
4-momenta require replacing the temporal component by $|\vec{p}\,|$, see Sec.~\ref{sec:results}. The thrust axis is simply $\hat{z}$ when pointing
in the gluon direction [\,see Eq.~\eqref{eq:regions}\,], while it equals $\vec{p}_q/|\vec{p}_q|$ and
$\vec{p}_{\bar q}/|\vec{p}_{\bar q}|$ when pointing to the quark and anti-quark direction, respectively.

Since to compute the radiative tails of the distributions (either differential or cumulative) and the moments of the differential distributions one only
needs the real radiation contribution, in this section we adopt the shorthand notation
\begin{equation}
M_C (y, z) \equiv M_C (y, z,\hat{m}, \varepsilon = 0)\,.
\end{equation}

\subsection{Computation of Moments}\label{sec:moments}
An especially convenient way to compute the $n$\,-th moment of the distribution is expressing it in terms of the total hadronic cross section and
displaced moments:
\begin{equation}
\langle (e - e_{\min})^n \rangle \equiv \frac{1}{\sigma^C_0}\!
\int^{e_{\max}}_{e_{\min}}\! \dd e\, (e - e_{\min})^n \frac{\dd
\sigma}{\dd e}\,, \qquad \langle (e - e_{\min})^0 \rangle = R (\hat{m})\,.
\end{equation}
The reduced moments at $\Ord(\alpha_s)$ and for $n>0$ can be computed directly using a numerical 2D integration, which is
convergent due to the insertion of the displaced measurement function
\begin{equation}
\langle (e - e_{\min})^n \rangle = C_F \frac{\alpha_s}{\pi} \!\int\!\dd y \,\dd z \, [\,\hat e (y, z) - e_{\min}\,]^n\,
\frac{M_C (y, z)}{y}+\Ord(\alpha_s^2)\,.
\end{equation}
The integration can be carried out with a MC procedure, but it is faster and more efficient to use a deterministic
integrator in 2D. For our numerical checks we have used the \texttt{dblquad} routine included in the \texttt{scipy.integrate}~\cite{Virtanen:2019joe}
python module. Finally, an efficient way of numerically computing regular moments is
\begin{equation}
\langle e \rangle^n = e^n_{\rm min}\, R (\hat{m}) + C_F \frac{\alpha_s}{\pi}\!
\int\! \dd y\, \dd z \, [\hat e(y, z)^n - e^n_{\rm min}]\, \frac{M_C (y, z)}{y}+\Ord(\alpha_s^2)\,,
\end{equation}
such that the numerical integral is convergent and can be directly computed in a standard way.

\subsection{Computation of Cross Sections Using a MC}\label{sec:MC}
Before we describe our novel numerical algorithm to directly compute the differential and cumulative cross sections,
we briefly review how this is done using MC methods. The MC can only access the radiative tail of the distribution,
and therefore one needs to consider only real radiation diagrams. To obtain the differential distribution one needs
to integrate over a Dirac delta function of the event-shape measurement function. Since there is no known way of doing
this in the MC approach, one instead bins the distribution, such that the delta gets replaced by the
difference of Heaviside functions. More specifically, by integrating first over the event-shape bin we obtain
\begin{align}\label{eq:bins}
&\Sigma (e_2) - \Sigma (e_1) = 
\! \int \!\frac{\dd y\, \dd z}{y}\, \dd e \, \delta [e-\hat e(y, z)] M_C(y, z)\,\Theta(e-e_1)\, \Theta(e_2-e) =\\
& \int \!\frac{\dd y\, \dd z}{y} \, \Theta[\hat e(y, z)-e_1]\, \Theta[e_2-\hat e(y, z)] M_C(y, z)\,.\nonumber
\end{align}
The advantage of the MC method is that one can compute the binned distribution for all event shapes in a single run. In practice one
chooses a set of bins for each event-shape variable ahead of time. In our implementation of the MC algorithm, we match the $(y,z)$
phase space into the unit square with the following change of variables:
\begin{align}
&y = v^2\,t_1\,,\qquad z = \frac{1}{2} + \Bigl(t_2 - \frac{1}{2}\Bigr)\sqrt{1-\frac{4\hat{m}^2}{1-y}}\,,\\
&\dd y\,\dd z = v^2\sqrt{1-\frac{4\hat{m}^2}{1-y}}\,\dd t_1\,\dd t_2\,.\nonumber
\end{align}
Now Eq.~\eqref{eq:bins} has a nice interpretation in terms of MC's: a)~generate a sample of points in the $(t_1,t_2)$ unit square,
compute for each one of them $(y,z)$ and from that get numerical values for all event-shape variables, matrix elements and
the Jacobian; b)~for each random point and for every event shape, figure out which bin it corresponds to; c)~add the numerical
values of the matrix element (times Jacobian) for each bin; d)~normalize each bin to the total number of points in the random sample.
Statistical uncertainties can be obtained in the usual way, and several independent runs can be combined. The method can of course
be refined using importance sampling, and in our numerical code we use the python implementation of
VEGAS~\cite{Lepage:1980dq}. The advantage of the MC is that a single run can be used
to compute the full distribution for all event shapes at once, and even to compute other quantities such as moments of the distribution or
the cumulative cross-section. 

\subsection{Direct Computation of the Differential Cross Section}\label{sec:differential}
\begin{figure}[t]\centering
\includegraphics[width=0.5\textwidth]{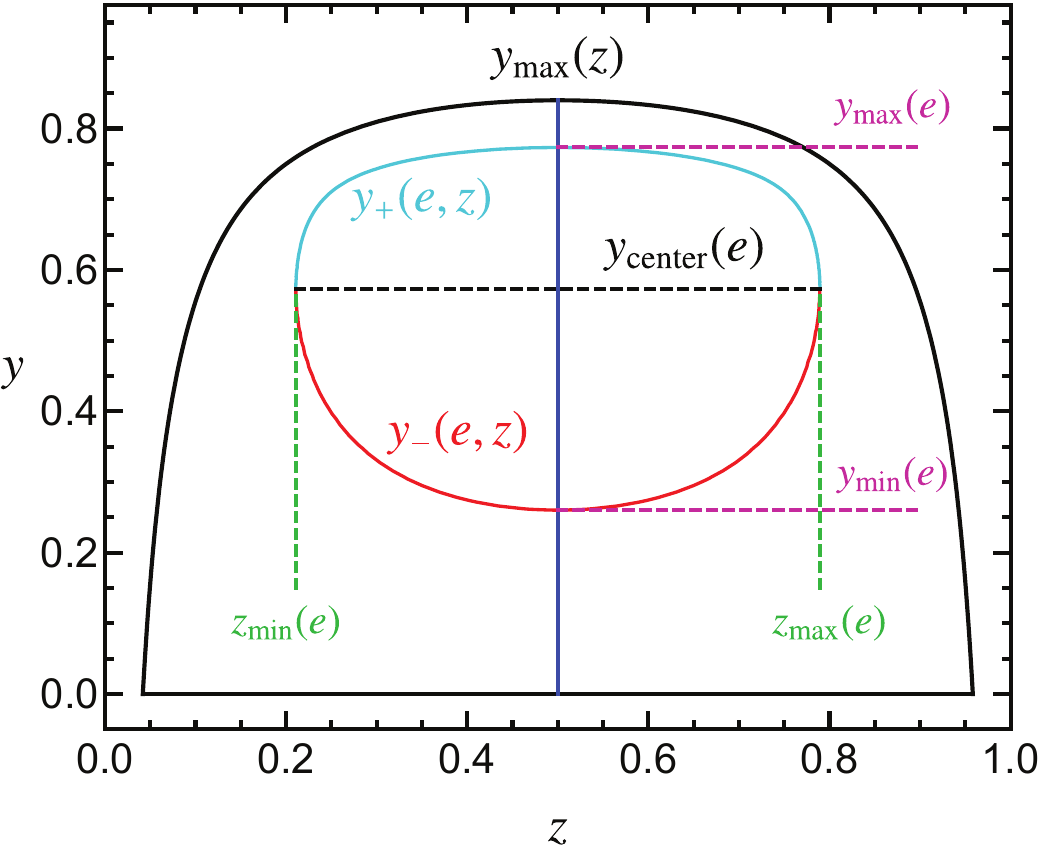}
\caption{Phase space diagram in $(z,y)$ coordinates (black lines) showing in red and cyan the contour line with
constant value of C-parameter, corresponding to the functions $y_+(e,z)$ and $y_-(e,z)$, respectively. The two curves
meet at the points $[z_{\min,\max}(e),y_{\rm center}(e)]$, joined by the black, dashed line. These curves meet the
$z=1/2$ vertical blue line for $y=y_{\min,\max}(e)$, respectively.
Dashed magenta [green] lines mark the points of maximal and minimal values that $y$ [$z$] can take within the red curve.
To generate this plot the values $\hat m = 0.2$ and $C=0.42$ were used.\label{fig:integration}}
\end{figure}
In this section we describe an alternative method which does not have the limitations inherent to a MC (can only compute binned cross sections,
and in general one needs to specify the bins ahead of time). The direct method computes directly the (unbinned) differential cross section, and
since it only uses ``deterministic'' integration methods, it can in principle achieve arbitrary precision in very small run-time. The only
requirement for the method to be applicable is that one can compute the value of the event shape and its first derivative in terms
of the phase-space variables $y$ and $z$. On the other hand, as compared to the MC method, one needs to compute a numerical
integral for each current, each event shape and each point in the spectrum. We denote the radiative tail of the distribution by $F_e(e,\hat m)$,
defined as
\begin{equation}
F_e(e,\hat m)\equiv \frac{B_{\rm plus}({\hat m})}{e - e_{\rm min}} + F_e^{\rm NS}(e,\hat m)\,.
\end{equation}
Since we can compute $F_e$ numerically with high precision and the values of $B_{\rm plus}$ and $e_{\rm min}$ are known analytically,
$F_e^{\rm NS}$ can be readily obtained. This is the last ingredient for the full description of the $\Ord(\alpha_s)$ differential cross
section. It should be noted that even in the massless limit $F_e(e,0)$ is not always analytically known (e.g.\ for angularities or broadening).
On the other hand, in the $m\to 1/2$ limit $F_e(e,1/2) = 0$ and only singular terms survive, which are analytically computed in
Sec.~\ref{sec:derivation}.

Let us first discuss how to obtain the cross section for a toy model: an event shape whose measurement function coincides with the soft
limit of some regular event shape. We proceed by integrating the $y$ variable analytically, followed by a numerical integration of the
$z$ variable. The $z$ integration boundaries are set by $z_\pm(e)$, the intersection of $h(e,z)$ with $y_{\rm max}(z)$, see
Fig.~\ref{fig:C-intersections}. Therefore we get
\begin{align}
\frac{1}{\sigma^C_0} \frac{\dd \sigma_C}{\dd \bar{e}} & = C_F \frac{\alpha_s}{\pi}\!\int\! \dd y\, \dd z
\,\frac{M_C (y,z)}{y}\,\delta [\bar{e} - e_{\rm min} - y f_e(z)] \\ &= \frac{C_F}{\bar{e} - e_{\rm min}}
\frac{\alpha_s}{\pi}\! \int_{z_-(e)}^{z_+(e)}\! \dd z\,M_C\biggl(\frac{\bar{e}-e_{\rm min}}{f_e(z)},z \biggr)\,.\nonumber
\end{align}

\begin{figure}[t]\centering
\includegraphics[width=0.5\textwidth]{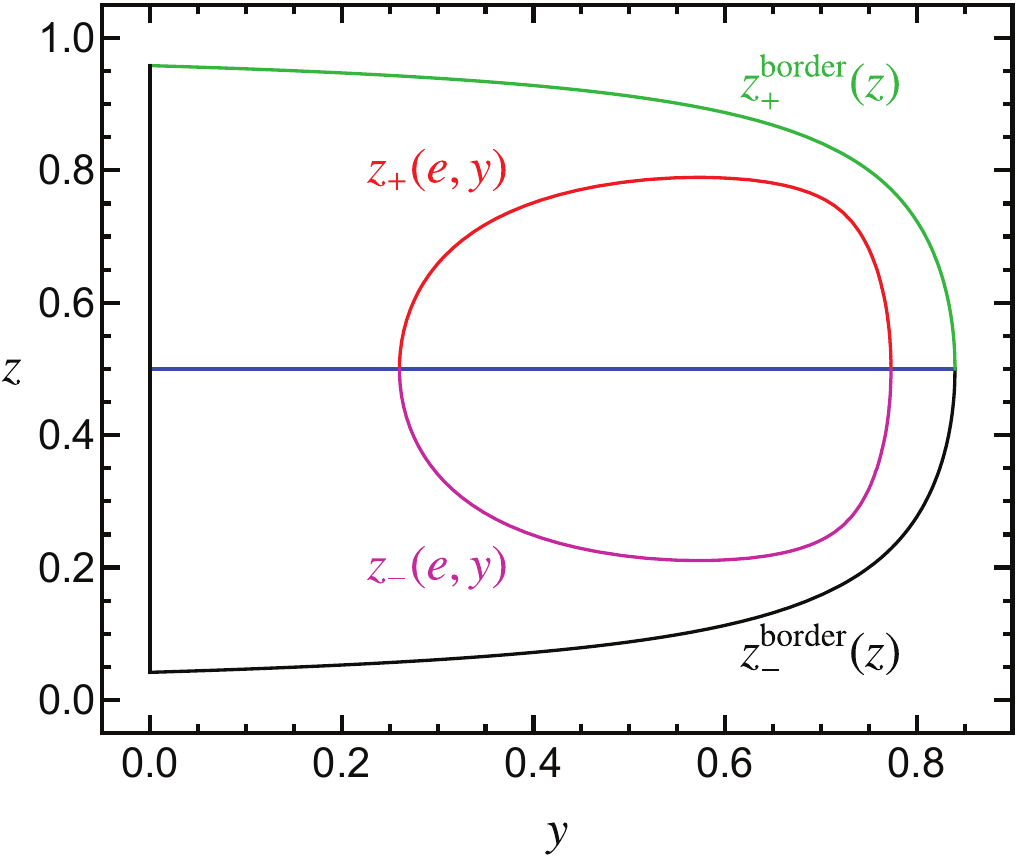}
\caption{Phase space diagram in $(y,z)$ coordinates depicted as green and black lines, corresponding to the functions $z_+^{\rm border}(z)$
and $z_-^{\rm border}(z)$, respectively. The red and magenta lines show a contour line with constant value of C-parameter,
corresponding to the functions $z_+(e,y)$ and $z_-(e,y)$, respectively. To generate this plot the numerical values $\hat m = 0.2$ and
$C=0.42$ were used.\label{fig:tilted}}
\end{figure}

To illustrate how the method works for ``real'' event shapes, let us assume for now that we are dealing with an observable such
that contour lines with constant event-shape value (that is, that the curve defined by the condition $\hat e(y,z)=e$ for some value of $e$
satisfying $e_{\rm min}<e<e_{\rm max}$) is continuous, convex and does not intersect with the phase-space boundary (the method can be
adapted for observables not satisfying this criteria, as will be explained later). Every event shape we considered, except for 2-jettiness,
\mbox{C-jettiness} and HJM (that intersect with the phase-space boundaries for some values of $e$), and \mbox{E-scheme} variables other than
C-parameter (that are not continuous) satisfy this condition. Curves for all event shapes which use the thrust axis have kinks, but this
does not pose a problem for the method. We proceed by integrating $y$ with the Dirac delta function first, while the $z$ integral is
performed numerically afterwards. The first step is finding the maximal and minimal value of $y$ in the contour line of constant
event-shape value, which we call $y_{\rm max}(e)$ and $y_{\rm min}(e)$, respectively. Since the curve is convex and symmetric
under $z\to 1-z$, these two values are attained for $z=1/2$. To find them, one has to solve the equation
\begin{equation}
\hat e\Bigl(y,\frac{1}{2}\Bigr) = e\,,
\end{equation}
which can be done e.g.\ with the Brent algorithm~\cite{brent2002algorithms}. In our numerical code we use the function \texttt{brentq} from the
\texttt{scipy.optimize} python module. The two roots are easily found since $y_{\rm min}$ is between $0$ and $y_{\rm middle}$, while $y_{\rm max}$
is between $y_{\rm middle}$ and $v = \sqrt{1 - 4 \hat m^2}$, with $(0.5,y_{\rm middle})$ being the point at which the lines that divide the phase space
into regions with the thrust axis pointing to the quark, anti-quark, or gluon $3$-momentum coincide:\,\footnote{This is the point
in phase space at which most event shapes obtain their maximal value $e_{\rm max}$, or a value very close to it. The
algorithm can be slightly refined figuring out ahead of time at which exact value of $y$ $e_{\rm max}$ is attained.}
\begin{align}
y_{\rm middle} = \frac{4}{3} \biggl( \!\sqrt{1 - 3 \hat m^2} - \frac{1}{2} \biggr)\,.
\end{align}
In the second step we obtain the contour lines of constant $e$ parametrized as two functions of $y$. This is obtained by solving the equation
$\hat e(y,z)=e$ for $z$ at a given value of $y$. There are two solutions to this equation, which we call $z_\pm(e,y)$, but we will
numerically obtain only $z_-(e,y)<1/2$, since the other solution can be obtained by symmetry.
Again, we employ the Brent algorithm, since we know the solution is always contained between $z=1/2$ and the phase-space boundary
$z_-^{\rm border}(y)$, which as a function of $y$ is written as
\begin{equation}
z_\pm^{\rm border}(y) = \frac{1}{2}\Biggl(1\pm\sqrt{\frac{1-y-4\hat m^2}{1-y}}\;\Biggr)\,.
\end{equation}
In our numerical code we again use the \texttt{brentq} function. Since we compute the distribution solving the Dirac delta function
in terms of the variable $y$, the next step is figuring out the lower integration limit in the $z$ variable,
dubbed $z_{\rm min}(e)$. Since the integrand is symmetric around $z=1/2$, we will integrate in the range $z=[\,z_{\rm min}(e),1/2]$
and double the result. The value of the point with the smallest $z$ value for a given event-shape value $e$,
$[\,z_{\rm min}(e),y_{\rm center}(e)\,]$, is obtained numerically as the minimum of the function
$z_-(e,y)$. The minimum lies between the values $y_{\rm max}(e)$ and $y_{\rm min}(e)$ previously determined,
and we use the Brent algorithm implemented in the \texttt{minimize\_scalar} function from the \texttt{scipy.optimize} python module, which
finds the minimum in a given interval. The maximum event shape value $e_{\rm max}$ satisfies
$z_{\rm min}(e_{\rm max})=z_{\rm max}(e_{\rm max}) = 1/2$. The last ingredient we need to determine before performing the numerical integral
in $z$ is the contour line of constant $e$ as a function of $z$. It is obtained by solving the equation $\hat e(y,z)=e$ for $y$ at a given
value of $z$, which has two solutions which we denote by $y_\pm(e,z)$, corresponding to the two zeroes of the Dirac delta function
argument when integrating the $y$ variable. The lower and upper solutions are contained in the intervals
$[\,y_{\rm min}(e),y_{\rm center}(e)\,]$ and $[\,y_{\rm center}(e),y_{\rm max}(e)\,]$, respectively, and are easily found numerically,
once again employing the Brent algorithm already described. In Fig.~\ref{fig:integration} we show graphically the position of $z_{\min,\max}(e)$,
$y_{\rm center}(e)$ and $y_{\min,\max}(e)$, as well as the curves $y_\pm(e,z)$, while Fig.~\ref{fig:tilted} shows the $z_\pm^{\rm border}(z)$ and
$z_\pm(e,z)$ lines. Putting everything together, the differential distribution can be written as\,\footnote{Note that
it is also possible to integrate the delta function in terms of $z$, leaving a numerical $y$ integral
\begin{equation}
F_e(e,\hat m) = 2\!\int_{y_{\rm min}(e)}^{y_{\rm max}(e)}\! \dd y\left.
\frac{M_C (y,z)}{y\,\big|\frac{\dd\hat e(y,z)}{\dd z}\big|}\right|_{z=z_-(e,y)}\,.
\end{equation}
We use this alternative expression to cross check our results. Both implementations agree within $15$ digits. We choose to integrate in $y$
first because then a)~event shapes with kinks in their curves with constant $e$ value can be treated with the algorithm just described,
b)~the algorithms for differential and cumulative distributions are very similar.}
\begin{equation}\label{eq:direct}
F_e(e,\hat m) =\! \int\! \dd z\, \dd y\,\frac{M_C (y,z)}{y}\,\delta[e - \hat e(y,z)]
= 2\!\int_{z_{\rm min}(e)}^{1/2}\! \dd z
\sum_{y=y_\pm(e,z)}\frac{M_C (y,z)}{y\,\Big|\frac{\dd\hat e(y,z)}{\dd y}\Big|}\,,
\end{equation}
where the sum in $y$ means that we evaluate the $y$-dependent expression for both $y=y_\pm(e,z)$ and add the results. The derivative
of the event-shape measurement function with respect to $y$ is performed analytically, while the value of $y_\pm(e,z)$ is obtained numerically
with the procedure outlined above. Our code computes the numerical integral using the python \texttt{quad} function, which
is the quadpack~\cite{RobertPiessensQasp} package implementation of the \texttt{scipy.integrate} module.

We close the section explaining how to modify the algorithm to compute E-scheme thrust, broadening and HJM. The contour lines for these event
shapes never intersect with the phase-space boundaries (at most they are tangent to it at a single point), but are not always continuous. It turns out
that if $\tau^E< \hat m$, $B^E < \hat m/2$ or $\rho^E < \hat m (1 - 2 \hat m)/(1 -\hat m)$, the event-shape contour lines are continuous and convex,
such that the algorithm described above can be used. For larger values, the lines of constant event-shape value show discontinuities
exactly along the lines that delimit the regions with the thrust axis pointing in the gluon or (anti-)quark direction. Therefore we find it convenient
to define event-shape measurement functions in each of the three regions: $\hat e_q(y,z)$, $\hat e_{\bar q}(y,z)$ and $\hat e_g(y,z)$. It is clear that
one should first determine the intersection points of the contour lines in the quark and gluon regions with $y_{\tau} (\hat{m}, z)$, defined in
Eq.~\eqref{eq:regions}, which we call $z_\pm^q(e)$ and $z_\pm^g(e)$, with $z_+^{q,g}(e)=1-z_-^{q,g}$. We find that in all cases
$z_-^g(e)>z_-^q(e)$. These points are computed solving the equations $\hat e_q[\,y_{\tau} (\hat{m}, z),z\,] = e$ and
$\hat e_g[\,y_{\tau} (\hat{m}, z),z\,] = e$ for $z$. In our code we employ the Brent algorithm again, and use
that the solutions are to be found in the interval $z\in[\hat m,1/2]$.

\begin{figure}[t]\centering
\includegraphics[width=0.5\textwidth]{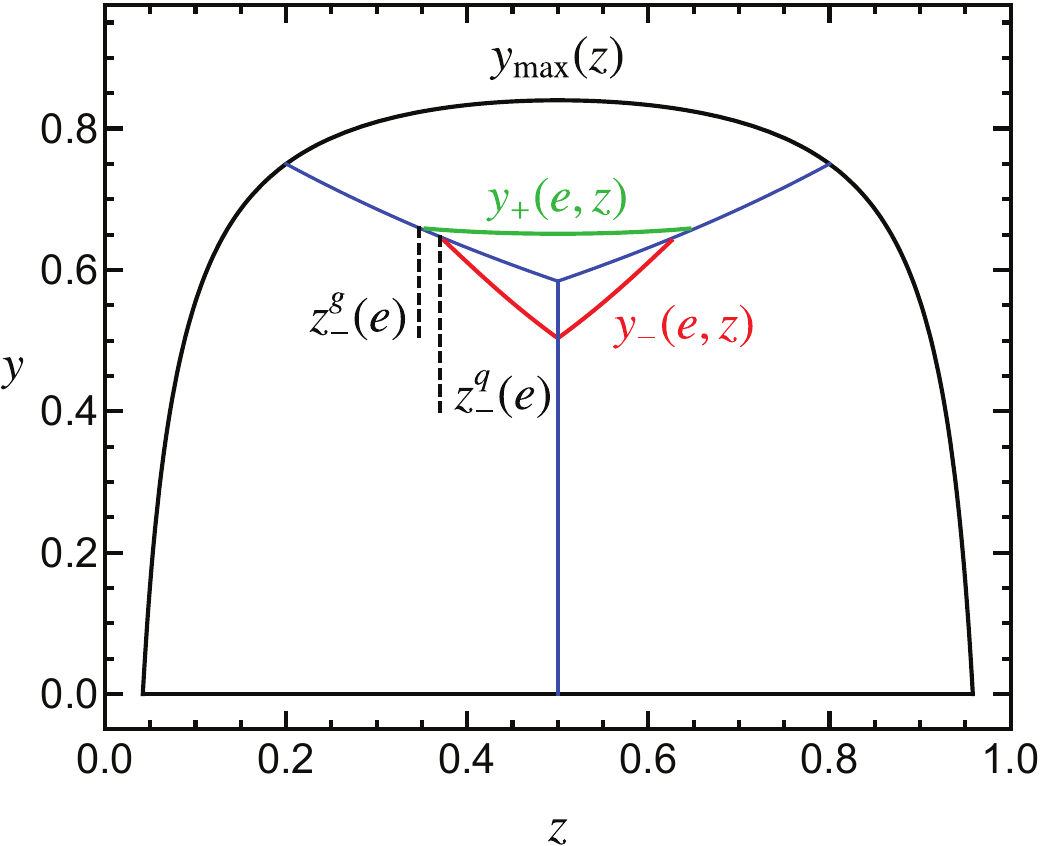}
\caption{Phase-space diagram in $(z,y)$ coordinates (black curves) split by blue lines into regions in which the thrust axis points in the
direction of the quark, anti-quark or gluon momenta. In green and red we show the contour lines for constant value of $\tau^E$
corresponding to the functions $y_+(e,z)$ and $y_-(e,z)$, respectively. The two curves do not meet at any point, but intersect with
the boundary separating the gluon and quark regions at the points $z_-^q(e)$ and $z_-^g(e)$, marked with dashed black lines.
To generate this plot the parameters $\hat m = 0.2$ and $\tau^E=0.27$ were used.\label{fig:E-thrust}}
\end{figure}

For E-scheme thrust and HJM one only needs to compute $y_-(e,z)$ and $y_+(e,z)$, which now coincide with the contours in the quark and
gluon regions, respectively. Therefore the equations to solve are $\hat e_q(y,z) = e$ and $\hat e_g(y,z) = e$, for which we use the Brent
algorithm, and use the fact that the solutions have to be contained in the ranges
$y\in[\,0,y_{\tau} (\hat{m}, z)\,]$ and $y\in[\,y_{\tau} (\hat{m}, z),y_{\rm max}(z)\,]$, respectively. In Fig.~\ref{fig:E-thrust}, for $\tau^E=0.27$ we
show the functions $y_\pm(e,z)$ in red and blue, as well as the points $z_\pm^q(e)$ and $z_\pm^g(e)$ with dashed black lines. The cross
section then is computed as
\begin{equation}\label{eq:dif-E}
F_e(e,\hat m) = 2\!\int_{z_-^q(e)}^{1/2}\! \dd z\, \Biggl|\frac{M_C (y,z)}{y\,\frac{\dd\hat e(y,z)}{\dd y}}\Biggr|_{y=y_-(e,z)} +
2\!\int_{z_-^g(e)}^{1/2}\! \dd z\, \Biggl|\frac{M_C (y,z)}{y\,\frac{\dd\hat e(y,z)}{\dd y}}\Biggr|_{y=y_+(e,z)}\,.
\end{equation}

\begin{figure}[t]\centering
\includegraphics[width=0.5\textwidth]{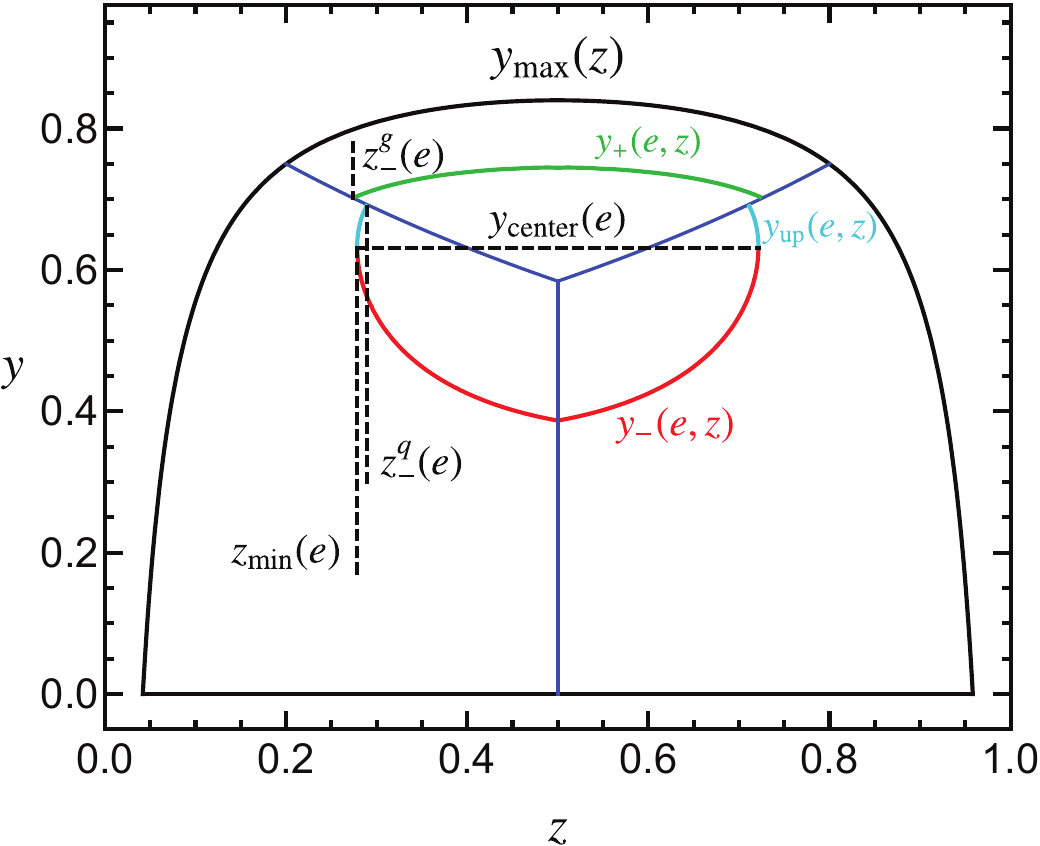}
\caption{Phase space diagram in $(z,y)$ coordinates (black curves) split by blue lines into regions in which the thrust axis points in the
direction of the quark, anti-quark or gluon momenta. In green, cyan and red we show the contour lines for constant value of $\tau^E$
corresponding to the functions $y_+(e,z)$, $y_{\rm up}(e,z)$ and $y_-(e,z)$, respectively. The cyan and green curves do not meet at any
point, but intersect with the boundary separating the gluon and quark regions at the points $z_-^q(e)$ and $z_-^g(e)$, marked with
dashed black lines. The cyan and red curves meet at the points $[\,z_{\min, \max}^q(e),y_{\rm center}(e) ]$.
To generate this plot the values $\hat m = 0.2$ and $B_T^E=0.2$ were used.\label{fig:E-broadening}}
\end{figure}

The most involved event shape is E-scheme broadening, which requires a specific algorithm. For $B_T^E>\hat m/2$ but smaller than a certain
critical value $B_T^{E,\rm crit}$, and for the small range $z\in[\,z_{\rm min}(e) , z_-^q(e)\,]$, there are two solutions to the equation
$\hat e_q(y,z) = e$, which we call $y_-(e,z)$ and $y_{\rm up}(e,z)$. The value of $z_{\rm min}(e)$ is computed using the algorithm already
explained for continuous event shapes, and the values of $z_-^{q,g}(e)$ are computed as described in the previous paragraph. In
Fig.~\ref{fig:E-broadening} we show the functions $y_\pm(e,z)$ and $y_{\rm up}(e,z)$ in green, red and cyan, respectively. The points
$z_{\rm min}(e)$ and $z_-^{q,g}(e)$ are marked with black, dashed lines. Therefore, while for $B_T>B_T^{E,\rm crit}$,\footnote{The
value of $B_T^{E,\rm crit}$ is obtained solving $z_{\rm min}(B_T^{E,\rm crit})=z_-^q(B_T^{E,\rm crit})$. In practice we do not compute it
explicitly, but simply use Eq.~\eqref{eq:dif-E} if $z_{\rm min}(e) > z_-^q(e)$.} one simply uses Eq.~\eqref{eq:dif-E}, if
$\hat m/2 < B_T^E < B_T^{E,\rm crit}$ the following expression has to be employed:
\begin{align}
F_e(e,\hat m)\, ={}& 2\!\int_{z_{\rm min}(e)}^{z_-^q(e)}\! \dd z
\sum_{y=y_-(e,z)}^{y_{\rm up}(e,z)}\frac{M_C (y,z)}{y\,\Big|\frac{\dd\hat e(y,z)}{\dd y}\Big|} +
2\!\int_{z_-^q(e)}^{1/2}\! \dd z\, \Biggl|\frac{M_C (y,z)}{y\,\frac{\dd\hat e(y,z)}{\dd y}}\Biggr|_{y=y_-(e,z)} \\
& + 2\!\int_{z_-^g(e)}^{1/2}\! \dd z\, \Biggl|\frac{M_C (y,z)}{y\,\frac{\dd\hat e(y,z)}{\dd y}}\Biggr|_{y=y_+(e,z)}\,.\nonumber
\end{align}

For $e$ larger than a certain value $e_{\rm th}$, the contour lines for thrust, HJM and broadening in the E-scheme exist only in the gluon
region, or in other words, $z_\pm^q(e_{\rm th})=1/2$, such that only the second term in Eq.~\eqref{eq:dif-E} contributes. $e_{\rm max}$
obviously satisfies $z_\pm^g(e_{\rm max})=1/2$. Some results for differential and cumulative distributions of 2-jettiness, C-jettiness and
HJM will be discussed in Sec.~\ref{sec:analytic}.
\subsection{Computation of the Cumulative Distribution}\label{sec:cumulative}
We define the cumulative distribution as
\begin{equation}
\Sigma(e_c) = \frac{1}{\sigma_0}\! \int_0^{e_c}\! \dd e \,\frac{\dd\sigma}{\dd e} =
R_0(\hat{m})\, \Theta(e_c-e_{\rm min}) + C_F \frac{\alpha_s}{\pi}\, \Sigma^1(e_c) + \Ord(\alpha_s^2)\,.
\end{equation}
Once again, since we can compute $\Sigma^1(e_c)$ with very high precision, the non-singular $\Sigma^{\rm NS}(e_c)$ function can be obtained
by removing the contributions from the delta and plus functions, which are known analytically. This is the last ingredient for the complete
description of the cumulative cross section at $\Ord(\alpha_s)$.

The cumulative distribution provides an alternative, numeric way of computing the delta coefficient, that will be
used as an additional cross check of our computation in Sec.~\ref{sec:numeric} (see Appendix~\ref{sec:indirect}) 
\begin{equation}\label{eq:A-num}
A_e(\hat m) = \lim_{e_c\to e_{\rm min}}\Sigma^1(e_c) - B_{\rm plus}(\hat m) \log(e_c-e_{\rm min})\,.
\end{equation}
In practice one can compute the one-loop contribution to the cumulative distribution by adding and subtracting
the total hadronic cross section $R_1^C(\hat{m})=\Sigma^1_C (e_{\max})$, and then using $1-\Theta(x) = \Theta(-x)$,
to obtain the relation
\begin{equation}\label{eq:cumu}
\Sigma^1_C (e_c) = R_1^C(\hat{m}) + \Sigma^1_C (e_c) - \Sigma^1_C (e_{\max})
=R_1^C(\hat{m}) -\! \int \!\dd y\, \dd z \, \Theta [\hat e(y, z) - e_c]\, \frac{M_C (y, z)}{y}\,. 
\end{equation}
The functions $R_1^C$ are shown graphically in Fig.~\ref{fig:R1}.
The cancellation of IR singularities is already realized in $R_1^C$ and the integral left over involves only real radiation.
Since the Heaviside function limits the $y$ integration on the lower side, the integral is convergent, and
consequently can be carried out using standard methods. For instance, the MC method described in Sec.~\ref{sec:MC}
can be easily adapted by simply summing up the events of all bins with lower endpoint larger larger than $e_c$.

In the rest of this section we describe a direct method along the lines of Sec.~\ref{sec:differential}. We again start with our
toy model, namely the event shape defined as the soft limit of another variable:
\begin{align}\label{eq:toy-int}
\Sigma^1_C (\bar{e}_c) & = R_1^C(\hat{m}) -\! \int_{z_-(\bar e_c)}^{z_+(\bar e_c)}\!\dd z
\int_{h(\bar e_c,z)}^{y_{\rm max}(z)}\dd y \, \frac{M_C (y, z)}{y} \\
& = R_1^C(\hat{m}) -\! \int_{z_-(\bar e_c)}^{z_+(\bar e_c)}\!\dd z\,
\mathcal{M}_C[\,y_{\rm max}(z),h(\bar e_c,z),z,\hat m]\,,\nonumber
\end{align}
where, given that the matrix element is a rank-two polynomial in $y$, the innermost integration can be performed analytically
\begin{align}
\!\int_{y_1}^{y_2}\!\dd y \, \frac{M_C (y, z)}{y} \equiv{} & \mathcal{M}_C(y_1,y_2,z,\hat m)
= M_C^0(z, \hat m) \log\biggl(\frac{y_1}{y_2}\biggr) \\
& + M_C^2(z,\hat m)\,(y_1-y_2) +\frac{1}{2}M_C^3(z,\hat m)\,(y_1^2-y_2^2)\,.\nonumber
\end{align}
The logarithm multiplying $M_C^0$ reflects the soft singularity and diverges if $\bar e_c=e_{\rm min}$.
The $z$ integration in Eq.~\eqref{eq:toy-int} can be easily performed numerically, and it corresponds to the
area marked with II in Fig.~\ref{fig:indirect}.\,\footnote{Note that a ``direct'' computation of the
cumulative distribution (not based on the difference with respect to the total hadronic cross
section), corresponds to the sum or areas marked as III and IV in Fig.~\ref{fig:indirect}, which
suffers from an IR singularity.}

For regular event shapes whose contour lines for constant $e$ are continuous, convex and do not intersect with the phase-space
boundaries, the cumulative distribution can be computed easily using the ingredients described in
Sec.~\ref{sec:direct}:
\begin{equation}\label{eq:cum-direct}
\Sigma^1_C (\bar{e}_c) = R_1^C(\hat{m}) - 2\!\int_{z_{\rm min}(e)}^{1/2}\!\dd z\,
\mathcal{M}_C[\,y_+(e,z),y_-(e,z),z,\hat m]\,.
\end{equation}
The $z$ integration is performed numerically as described in Sec.~\ref{sec:direct}. We have checked that taking a numerical
derivative of our cumulative distribution reproduces the differential cross section as computed in the previous section.

For E-scheme thrust and HJM the above formula has to be modified. From the analysis in the previous section and looking at
Fig.~\ref{fig:E-thrust}, one can readily conclude that between $z_-^g(e)$ and $z_-^q(e)$ the area to be integrated is limited by
$y_{\tau} (\hat{m}, z)$ and $y_+(e,z)$, while between between $z_-^q(e)$ and $1/2$ it is limited by $y_-(e,z)$ and $y_+(e,z)$:
\begin{align}
\Sigma^1_C (\bar{e}_c) ={} & R_1^C(\hat{m}) - 2\!\int_{z_-^g(e)}^{z_-^q(e)}\!\dd z\,\mathcal{M}_C[\,y_+(e,z),y_{\tau} (\hat{m}, z),z,\hat m] \\
&- 2\!\int_{z_-^q(e)}^{1/2}\!\dd z\,\mathcal{M}_C[\,y_+(e,z),y_-(e,z),z,\hat m] \,.\nonumber
\end{align}
Finally, for E-scheme broadening, if $\hat m/2 < B_T^E < B_T^{E,\rm crit}$ one has to use the following expression:
\begin{align}
\Sigma^1_C (\bar{e}_c) ={} & R_1^C(\hat{m}) - 2\!\int_{z_{\rm min}(e)}^{z_-^q(e)}\!\dd z\,
\mathcal{M}_C[\,y_{\rm up}(e,z),y_-(e,z),z,\hat m]\nonumber\\
& -2\!\int_{z_-^g(e)}^{z_-^q(e)}\!\dd z\,\mathcal{M}_C[\,y_+(e,z),y_{\tau} (\hat{m}, z),z,\hat m] \\
&- 2\!\int_{z_-^q(e)}^{1/2}\!\dd z\,\mathcal{M}_C[\,y_+(e,z),y_-(e,z),z,\hat m] \,.\nonumber
\end{align}
All the $z$ numerical integrals in this section are performed using the python quadpack implementation in the \texttt{scipy} module.

The cumulative distribution for event shapes whose contour lines intersect with the phase space will be discussed in the next
section taking 2-jettiness as an example.

\section{Cross Sections for Mass-Sensitive Event Shapes}\label{sec:analytic}
As an application of the approach presented in this work, we now work out differential and cumulative cross sections for those event
shapes whose contour lines intersect with the phase-space boundaries: HJM, 2-jettiness and C-jettiness. These happen to be the most
sensitive to the heavy quark mass, since their threshold gets displaced from zero to a mass-dependent position. Analytic results for the
differential 2-jettiness~\cite{Dehnadi:2016snl} and \mbox{C-jettiness} distributions~\cite{Preisser:mthesis}\,\footnote{In this reference the coefficient of the
delta function was determined numerically.} are already known, but in this section we will carry out the manipulations necessary to bring their
computation into 1D integrals over $z$. We shall show how to obtain analytical results for the differential heavy jet mass and cumulative 2-jettiness
distributions. Since the coefficients of the delta functions are already collected in Appendix~\ref{app:anadel} and the plus-distribution coefficient is
universal, the only missing piece is the non-singular contribution, which can be computed in $d=4$ dimensions.

\subsection{Differential HJM}\label{sec:HJM}
\begin{figure}[t]\centering
\includegraphics[width=0.5\textwidth]{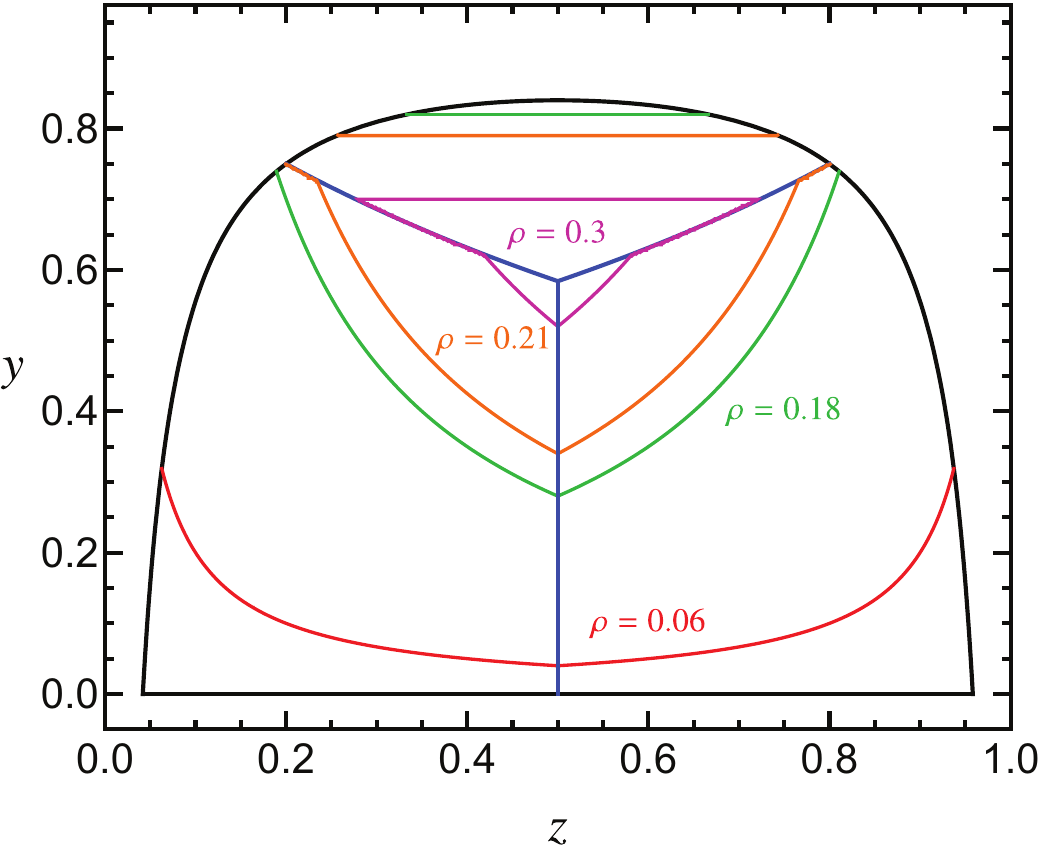}
\caption{Three-particle phase space (black lines), thrust boundaries (blue lines) and constant $\rho$ value contour lines, with $\hat m = 0.2$.\label{fig:HJM}}
\end{figure}

In order to derive the HJM differential distribution it is useful to split the phase space into the usual
three regions, as described by Eq.~\eqref{eq:regions} and shown in Fig.~\ref{fig:Phase-space}. For three particles, one of them massless and the
other two with same mass $m$, $\rho(y, z)$ takes the following form in these regions:\,\footnote{The sum of hemisphere masses takes the same
form in the last region, while in the first two regions the squared mass gets a factor of $2$. Therefore, the calculation presented in this section can
be easily modified to give results for this event shape.}
\begin{equation}
\rho_q = \hat m^2 + y z\,, \qquad
\rho_{\bar q} = \hat m^2 + y (1-z)\,,\qquad
\rho_g = 1 - y \,.
\end{equation}
Consequently, the measurement delta functions in the respective regions are very simple and given by
\begin{align}
\delta (\rho - y z - \hat{m}^2) = \frac{1}{z} \delta \biggl( y - \frac{\rho -
\hat{m}^2}{z} \biggr)\,,& \qquad \delta [\rho - y (1 - z) - \hat{m}^2] =
\frac{1}{1 - z} \delta \biggl( y - \frac{\rho - \hat{m}^2}{1 - z} \biggr)\,,\nonumber\\
&\delta(\rho - 1 + y)\,.
\end{align}
To figure out the correct integration boundaries for $y$ and $z$ we consider the quark and anti-quark region first. In this region, the lines of
constant $\rho$ meet either the phase-space boundary [\,for $\rho < \hat{m} (1 - \hat{m} - \hat{m}^2) / (1 - \hat{m})$\,], or the boundary to the gluon
region (for larger values). Defining $\xi_\rho \equiv \sqrt{t_\rho^2 - 4 \rho}$ and $t_\rho = 1 + \rho - \hat{m}^2$, the intersections with the phase-space
boundary are located at $z = z_1 \equiv (t - \xi_\rho)/ 2$ and $z = 1 - z_1$, while the intersections with the gluon-region boundary are located at
$z = z_2 \equiv (t_\rho - 1) / \sqrt{(1 - \rho)^2 - 2 \hat{m}^2 (1 + \rho) + \hat{m}^4}$ and $z = 1 - z_2$. The gluon region, on the other hand, contributes
only for $\rho > 4 \hat{m}^2$, intersecting the phase-space boundary [\,for $\rho < \hat{m} / (1 - \hat{m})$\,] or the region boundary (for larger values). The
phase-space boundary intersections are located at $z = z^2_- \equiv \bigl[\, 1 - \sqrt{1 - 4 \hat{m}^2 / \rho}\;\bigr] / 2$ and $z = z^2_+ \equiv 1 - z^2_-$
for $\rho < \hat{m} / (1 - \hat{m})$, while the region boundary intersections are located at $z = z^3_- \equiv \bigl[\,1 - \sqrt{(1-\rho^2)+4\hat m^2}\;\bigr] / (1-\rho)$
and $z = z^3_+ \equiv 1 - z^3_-$. All those cases are shown in Fig.~\ref{fig:HJM}.
Note that that the constant event-shape lines of the (anti-)quark and the gluon region do not meet at the region boundaries.

Taking all this into account, the differential cross section in $d = 4$ is given by
\begin{align}
F_\rho(\rho,\hat m) ={}&
2\! \int^{\frac{1}{2}}_{\max (z_1,
z_2)} \frac{\text{d} z}{\rho - \hat{m}^2} M_C \biggl( \frac{\rho -
\hat{m}^2}{z}, z \biggr)
+ \frac{2\, \Theta (\rho - 4
\hat{m}^2)}{1 - \rho} \!\int^{\frac{1}{2}}_{\max (z^2_-, z^3_-)} \!\!\text{d} z
\, M_C (1 - \rho, z) \nonumber\\
={}& f_1[\max (z_1, z_2)] + f_2 [\max (z^2_-,z^3_-)]\,,
\end{align}
with the integral solutions
\begin{align}
f^V_1 (z) ={} & \frac{(1 - 2 z) \bigl[ \, \rho (1 - z) (\rho - 4 z) - 2
\hat{m}^2 (\rho + 2 (1 - 2 \rho) z^2 + 3 \rho z) + \hat{m}^4 (1 - z - 8
z^2) \, \bigr]}{2 (1 - z) z^2 (\rho - \hat{m}^2)}\nonumber\\
& + \biggl( \rho - 2 - 5 \hat{m}^2 + 2 \frac{1 - 4 \hat{m}^4}{\rho -
\hat{m}^2} \biggr) \log \biggl( \frac{1 - z}{z} \biggr)\,,\nonumber\\
f^A_1 (z) ={}& 4 - 8 \hat{m}^2 (2 + \rho) + 8 \hat{m}^4 + \frac{1}{2 (\rho
- \hat{m}^2)} \biggl\{ \frac{(1 + 2 \hat{m}^2) (\rho - \hat{m}^2)^2}{z^2} +
\frac{4 (1 - 4 \hat{m}^2) \hat{m}^2}{1 - z} \nonumber\\
& - 2\, \frac{\rho (2 + \rho) - 2 \hat{m}^2 \rho (5 + \rho) + \hat{m}^4
(1 + 4 \rho) - 2 \hat{m}^6}{z}\nonumber\\
& + 2 \bigl[ \, 2 - (2 - \rho) \rho + 2 \hat{m}^2 (\rho (\rho +
3) - 5) + \hat{m}^4 (9 - 4 \rho) + 2 \hat{m}^6 \bigr] \log\biggl( \frac{1
- z}{z} \biggr) \biggl\}\,,\nonumber\\
f^V_2 (z) ={}& \frac{1}{t} \biggl\{ [2 - (2 - t) t - 4 \hat{m}^2 t - 8
\hat{m}^4] \log\biggl( \frac{1}{z} - 1 \biggr) - \frac{(1 - 2 z) \bigl[ (1 -
z) z \, t^2 + 2 \hat{m}^2 + 4 \hat{m}^4 \bigr]}{(1 - z) z} \biggl\}\,,
\nonumber\\
f^A_2 (z) ={}& \frac{\bigl[2 - 2 t + t^2 + 2 \hat{m}^2 (t^2 + 4 t - 6) + 16
\hat{m}^4\bigr] \log \biggl( \frac{1}{z} - 1 \biggr) - \frac{(1 - 2 z) [t^2 (1 -
z) z + 2 \hat{m}^2 - 8 \hat{m}^4]}{(1 - z) z}}{t}\, . \nonumber
\end{align}
The integral boundaries arise as follows: while making use of the symmetry axis $z=1/2$ by integrating $z$ only up to $1/2$ and
multiplying by $2$, the boundaries automatically choose the appropriate intersection of the constant event-shape line with the
phase-space or region boundary.

\subsection{Cumulative 2-Jettiness}\label{sec:2J}

2-jettiness is defined as
\begin{equation}
\tau_J = \frac{1}{Q} \min_{\hat{n}} \sum_i (E_i - | \hat{n} \cdot
\vec{p}_i |)\,,
\end{equation}
which, for the same setup of three partons, two of them massive, takes the simple form
\begin{equation}
\tau_J = \min \bigl[\, 1 - y, 1 - \, \text{mod} (y, z), 1 - \, \text{mod} (y, 1 - z)\, \bigr]\,,
\end{equation}
with $\text{mod} (y, z) \equiv \sqrt{(1 - y z)^2 - 4 \hat{m}^2}$. The three values in the list correspond to the thrust axis pointing
into the direction of the gluon, quark and anti-quark momenta, respectively, see Eq.~\eqref{eq:regions}.

Therefore, in these three regions the measurement $\delta$ functions read
\begin{align}
&\delta (\tau_J + y - 1)\,,\nonumber\\
&\delta [\tau_J + \text{mod} (y, z) - 1] = \frac{t_\tau}{z \xi_\tau}\, \delta \biggl( y - \frac{1 - \xi_\tau}{z} \biggr)\,,\\
&\delta [\tau_J + \text{mod} (y, z) - 1] = \frac{t_\tau}{(1 - z) \xi_\tau} \,\delta \biggl( y - \frac{1 - \xi_\tau}{1 - z} \biggr)\,,\nonumber
\end{align}
with $t_\tau = 1 - \tau_J$, $\xi_\tau = \sqrt{t_\tau^2 + 4 \hat{m}^2}$.
In analogy to the previous section, we analyze the various intersections of the constant event-shape line with the phase-space and region
boundaries to figure out the correct integration limits: the constant event-shape line in the gluon region is equivalent to the one for heavy
jet mass in the previous section and can be adopted. In the (anti-)quark region, the constant event-shape line intersects the phase-space
boundary for $\tau_J < \hat{m} / (1 - \hat{m})$, or the boundary to the gluon region for larger values. These intersections take place at
$z = z^1_- \equiv (1 + \tau_J - \xi_\tau) / 2$, $z = z^1_+ \equiv 1 - z^1_-$, and $z = z^3_- \equiv (1 - \xi_\tau) / t$, $z = z^3_+ \equiv 1 - z^3_-$,
respectively. In contrast to heavy jet mass, the constant $\tau_J$-lines meet at the region boundaries, making the structure of the integration
boundaries simpler.

\begin{figure}[t]\centering
\includegraphics[width=0.5\textwidth]{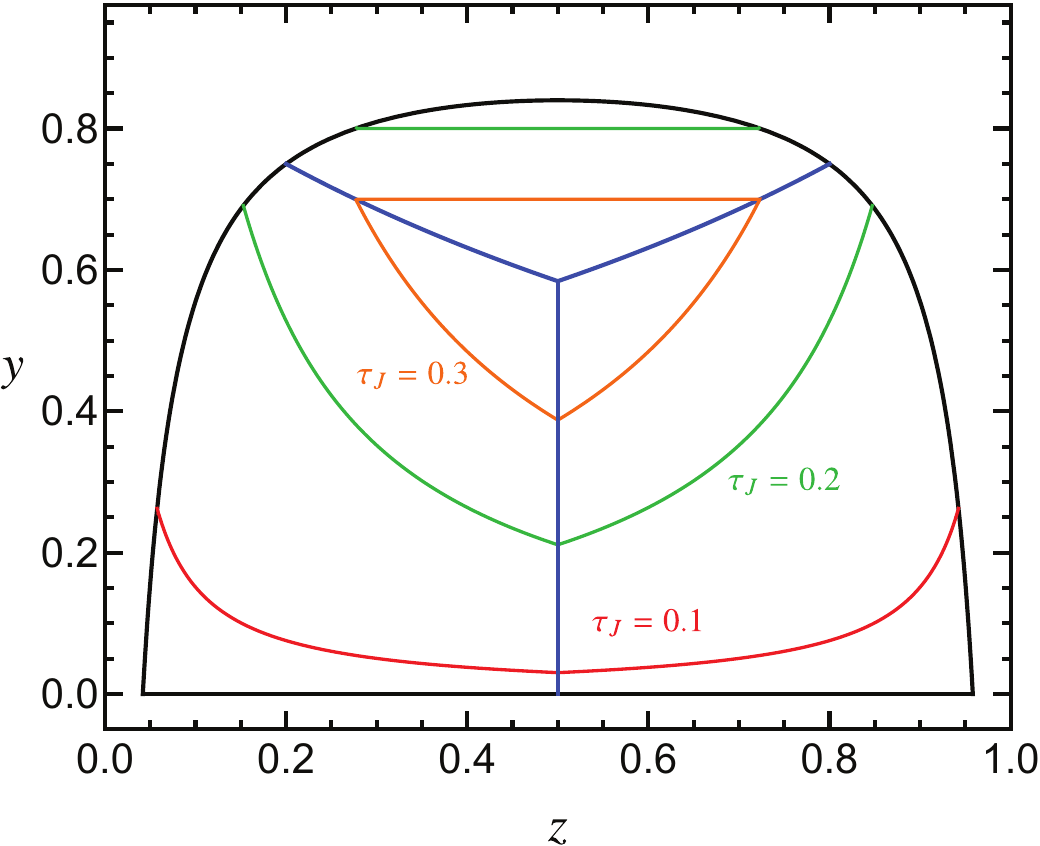}
\caption{Three-particle phase space (black lines), thrust boundaries (blue lines) and constant $\tau_J$ value contour lines, with $\hat m = 0.2$.\label{fig:TauJ}}
\end{figure}

Using the expression in Eq.~\eqref{eq:cumu} the cumulative distribution is given by
\begin{align}
\Sigma^1_C (\tau^c_J < 4 \hat{m}^2) ={}& R_1^C (\hat{m}) - 2
\int^{\frac{1}{2}}_{z^1_-} \text{d} z \int^{y_{\max} (z)}_{\frac{1 -
\xi}{z}} \text{d} y\, \frac{M_C(y, z)}{y}, \\
\Sigma^1_C (\tau^c_J > 4 \hat{m}^2) ={}& R_1^C (\hat{m}) - 2
\int^{\frac{1}{2}}_{\max (z^1_-, z^3_-)} \text{d} z \int^{\min [y_{\max}
(z), t]}_{\frac{1 - \xi}{z}} \,_{} \text{d} y\, \frac{M_C(y, z)}{y}, \nonumber
\end{align}
where the integral corresponds to the region in between the limiting value
$\tau^c_J$ and (potentially) the phase-space boundary, see Fig.~\ref{fig:TauJ}.
All integrals can be computed analytically, resulting in a long expression containing logs and dilogs. The expressions are available from the authors on request.

\subsection{C-Jettiness}
\begin{figure}[t]\centering
\includegraphics[width=0.5\textwidth]{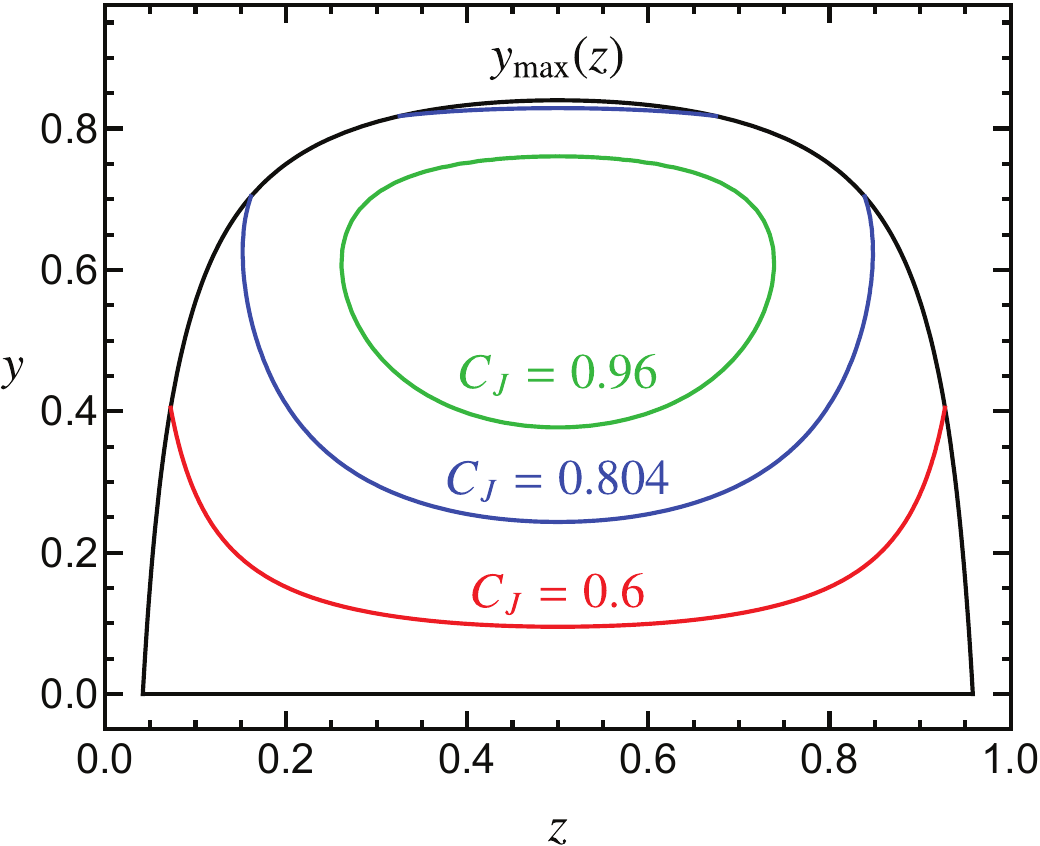}
\caption{Three-particle phase space (black lines) showing contour lines for the C-jettiness event shape in red ($C_J =0.6$), blue
($C_J= 0.804$) and green ($C_J=0.96$). The plot uses the numerical value $\hat m = 0.3$.\label{fig:CJ}}
\end{figure}

We finish the discussion on the computation of event-shape differential and cumulative distributions with the special case of
$C_J$, whose contour lines are continuous and smooth, but can intersect with the phase-space boundaries in various ways,
some of which are shown in Fig.~\ref{fig:CJ}. 
For the cumulative distribution one can find a simple expression that covers all possible scenarios
\begin{equation}
\Sigma^1_C (\bar{e}_c) = R_1 (\hat{m}) - 2\!\int_{z_m(e)}^{1/2}\! \dd z\,
\mathcal{M}_C[\,\min\{y_{\rm max}(z),y_+(e,z)\},y_-(e,z),z,\hat m]\,,
\end{equation}
with $z_m$ the minimal value that $z$ can attain in the contour line within the phase-space boundaries. Therefore $z_m$ can be either
$z_{\rm min}(e)$, the point at which the contour line has infinite slope, or $z_-^{\rm cut}$, the point at which it intersects
with the phase-space boundary on the left side (if it intersects more than once, then the lower intersection has always smaller $z$ value).
Analytic expressions for $y_{\pm}(z)$ can also be found. A careful examination of those and their interplay with $y_{\rm max}$ allows to find
a general analytic expression for $z_m$. These results are given in Ref.~\cite{Preisser:mthesis}. 
The $z$ integration has to be performed numerically, and we use the quadpack package for that.

For the differential cross section one can also write down a unique expression by carefully defining the ``upper'' and lower contour lines
\begin{equation}\label{eq:CJ-dif}
F_e(e,\hat m) \,= \, 2\!\int_{z_m(e)}^{1/2}\! \dd z
\sum_{y=y_\pm(e,z)}\Theta[y_{\rm max}(z) - y]\,\frac{M_C (y,z)}{y\,\Big|\frac{\dd\hat e(y,z)}{\dd y}\Big|}\,.
\end{equation}
Here the Heaviside function splits the two $z$ integrals in various sub-integrals corresponding to different segments, which are sometimes
disconnected. After a careful analysis one can disentangle all possible scenarios that Eq.~\eqref{eq:CJ-dif} encompasses. These depend of
course on the value of $C_J$, but also on $\hat m$. After working those out, the resulting $z$ integrals can be performed analytically in
terms of incomplete Elliptic functions. A detailed computation, together with the final analytic expressions, is given in Ref.~\cite{Preisser:mthesis}.

\section{Numerical Analysis}\label{sec:checks}
\begin{figure*}[t!]
\subfigure[]
{
\includegraphics[width=0.48\textwidth]{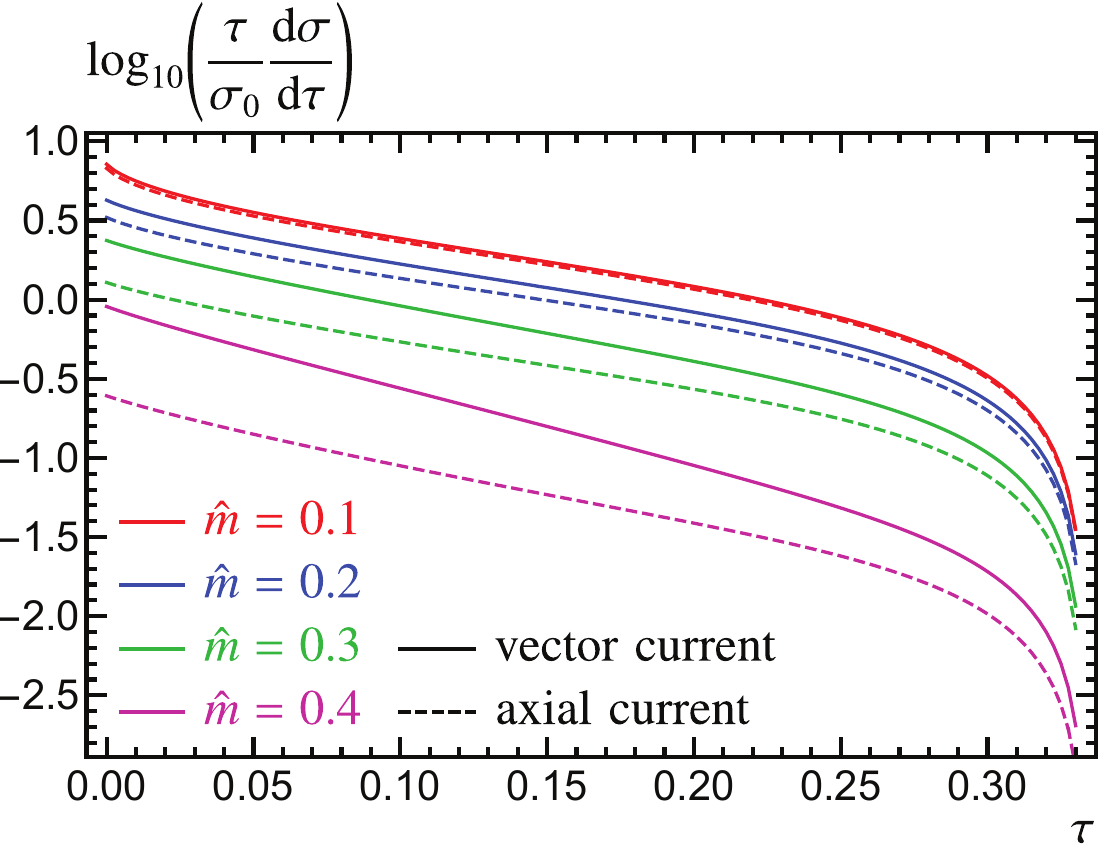}
\label{fig:DistTau}
}
\subfigure[]{
\includegraphics[width=0.48\textwidth]{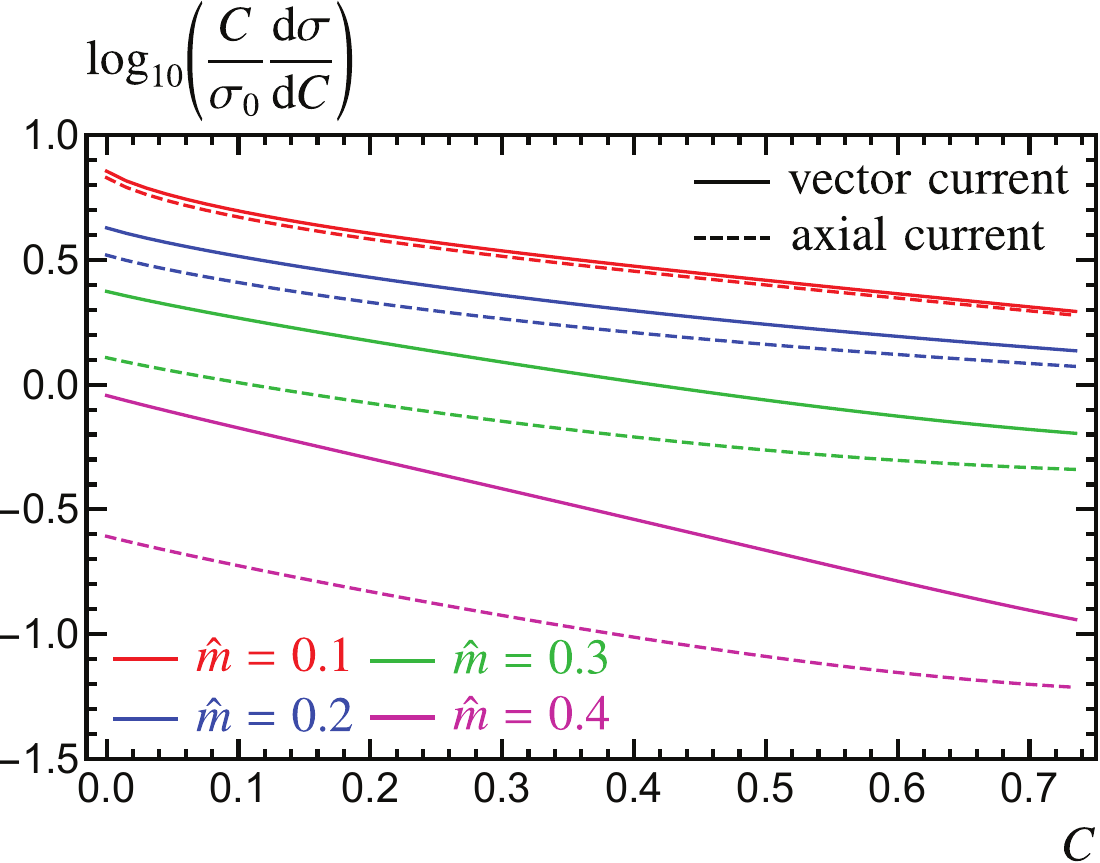}
\label{fig:DistC}
}
\subfigure[]{
\includegraphics[width=0.48\textwidth]{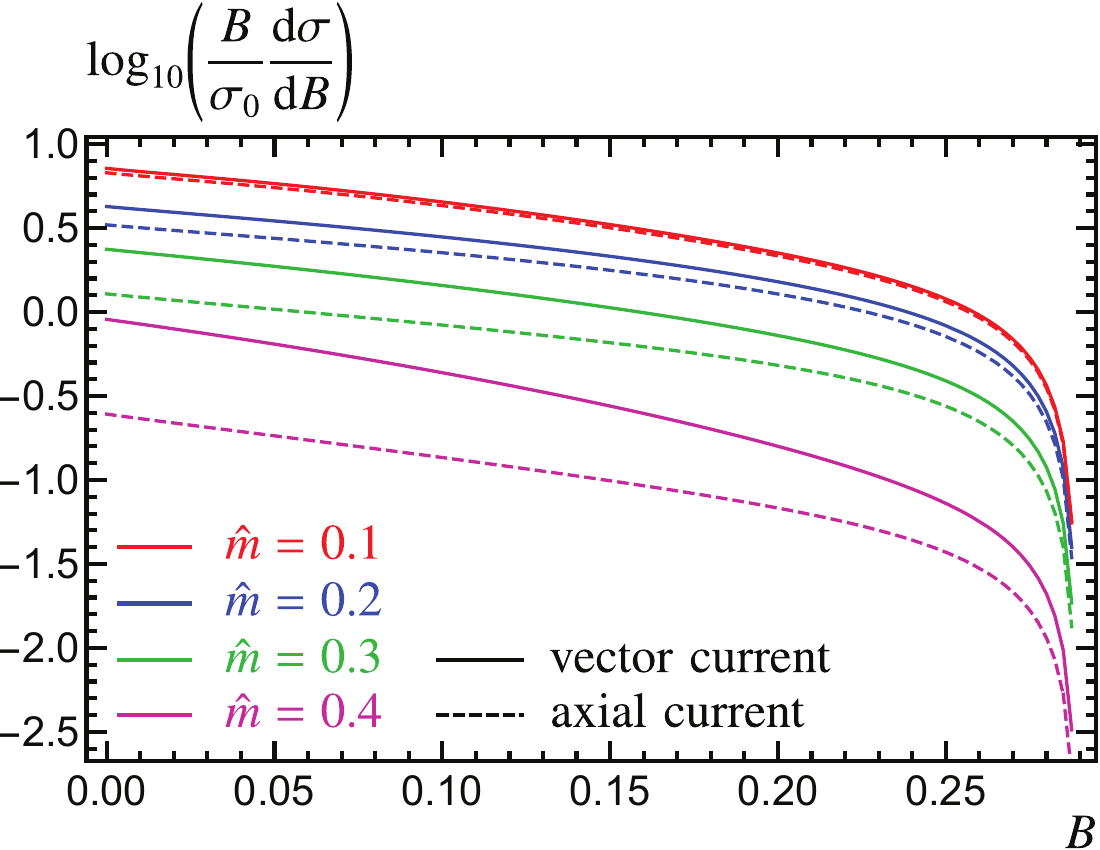}
\label{fig:DistB}
}
\subfigure[]{
\includegraphics[width=0.48\textwidth]{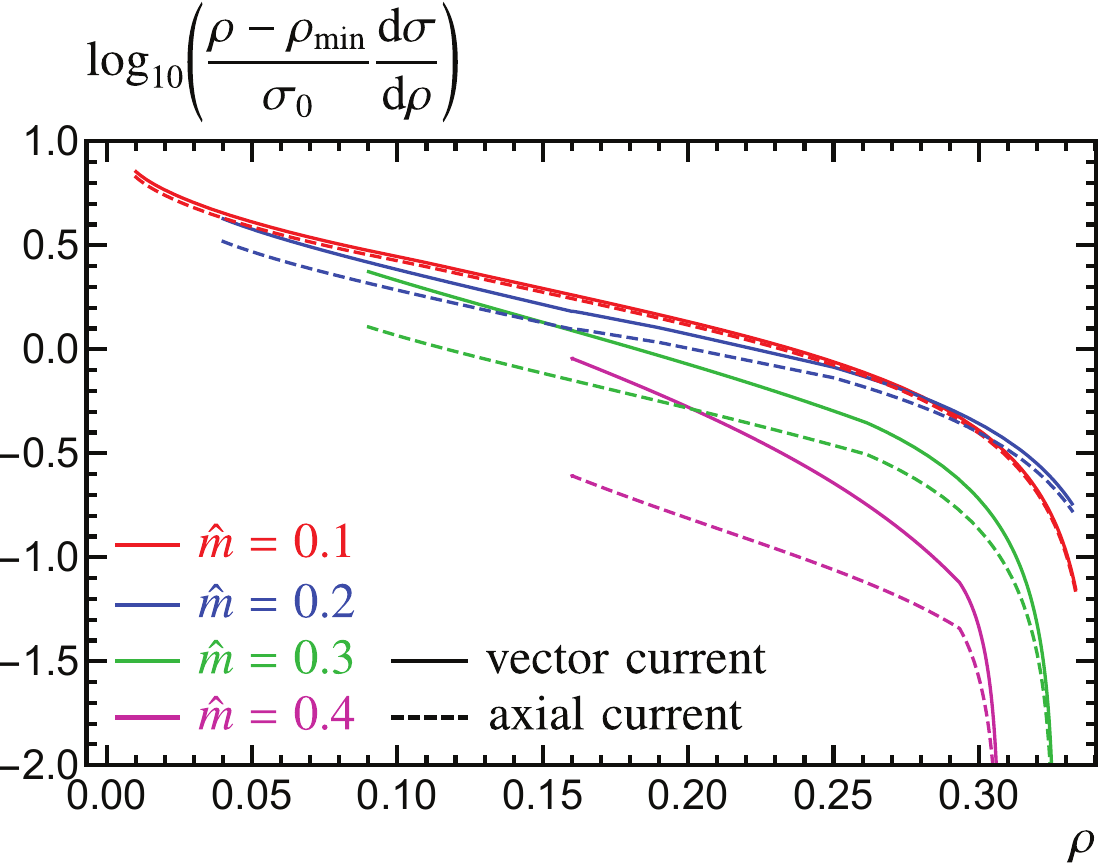}
\label{fig:DistRho}
}
\caption{event-shape distributions for vector (solid curves) and axial-vector (dashed curves) currents. Panels (a), (b), (c), and (d) show
results for thrust, C-parameter, jet broadening, and heavy jet mass, respectively. All curves are multiplied by $e-e_{\rm min}$, with
$e$ the event-shape value and $e_{\rm min}$ its minimal value, such that the cross section is finite for $e=e_{\rm min}$. Red, blue,
green and magenta lines show the results for $\hat m=0.1,\,0.2,\,0.3$ and $0.4$, respectively.}
\label{fig:distributions}
\end{figure*}
We start this section by showing differential cross section results for different values of $m/Q$ in Fig.~\ref{fig:distributions}. Since the only
singular term at threshold diverges like $\sim 1/(e-e_{\rm min})$, we plot $(e-e_{\rm min})$ times the distribution, such that the curves are finite
in the whole range. We also use a logarithmic scale on the $y$ axis to make the curves with small reduced mass value more visible. We show
results for the most common event shapes, namely thrust, C-parameter, jet broadening and heavy jet mass in their original definition, although our
code can yield results for those in any other scheme\,\footnote{While the aim of Fig.~\ref{fig:EvsP} was to highlight the difference between schemes, here
we want to show the difference between vector and axial-vector currents, together with the mass dependence.}. 
With the exception of HJM, $e_{\rm min}$ and $e_{\rm max}$
are mass-independent, and therefore the main sensitivity of the cross section is through the magnitude of the curves:
smaller masses result in larger cross sections in the tail. Vector and axial-vector cross sections are similar for small values of the reduced
mass, but clearly different for e.g.\ $\hat m=0.4$. We observe that the axial-vector distributions are always lower than their
vector counterparts. That hierarchy is also true for the plus and delta-function coefficients, as well as for the cumulative distribution.
The HJM endpoints are mass-dependent, such that $\rho_{\min}$ [$\rho_{\max}$] increases [decreases] with $\hat m$. Therefore for
this event shape (and all other mass-dependent ones) the mass sensitivity comes partly from the cross-section magnitude, but mainly
from the peak position [the peak is not visible in the plots because of the $(\rho-\rho_{\rm min})$ factor, and the missing resummation
and convolution with a non-perturbative shape function].
\begin{figure}[t]\centering
\includegraphics[width=0.5\textwidth]{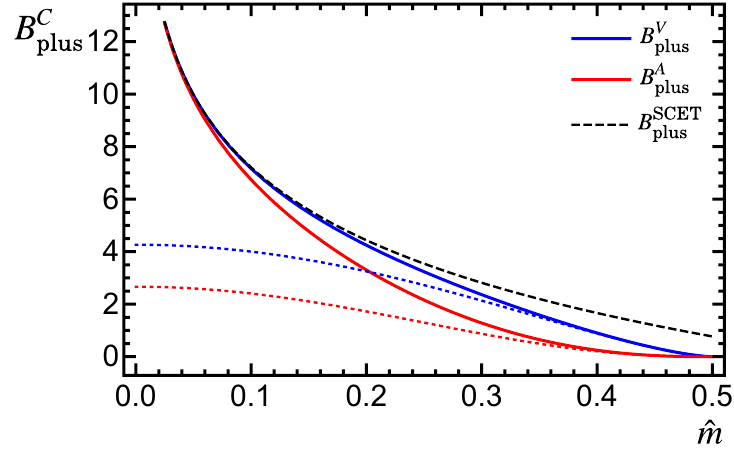}
\caption{The universal coefficient of the plus distribution $B^C_{\rm plus}$ as a function of the reduced mass $\hat m$. We show the
coefficient in the case of an axial-vector current (red solid line), a vector current (blue solid line) and the SCET limit (black dashed line). The
SCET limit agrees with both the full QCD vector and axial cases in the limit $\hat m \to 0$. We also show the respective threshold limits
as dotted lines, expanded up to $\Ord(v^5)$. Both full QCD expressions approach zero in this limit, as expected. \label{fig:plusCoef}}
\end{figure}

\begin{figure*}[t!]
\subfigure[]
{
\includegraphics[width=0.465\textwidth]{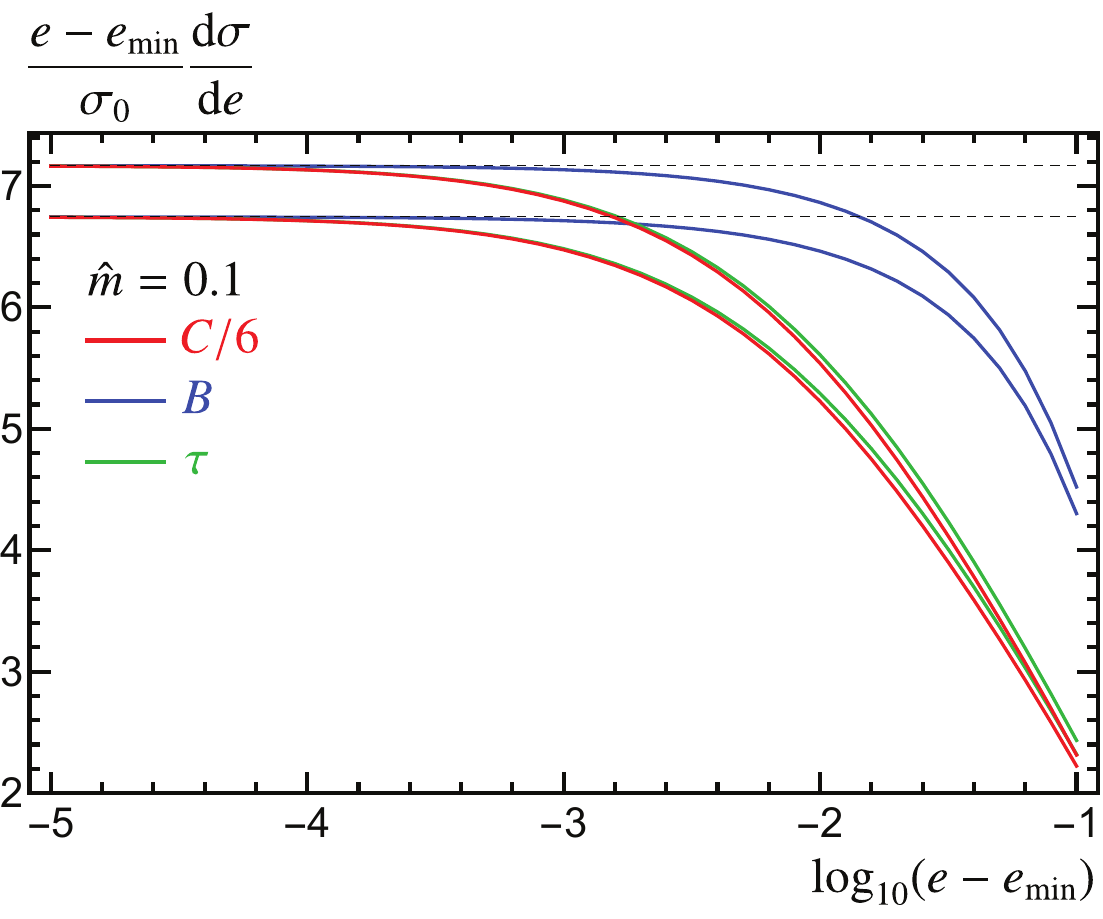}
\label{fig:Residue01}
}
\subfigure[]{
\includegraphics[width=0.48\textwidth]{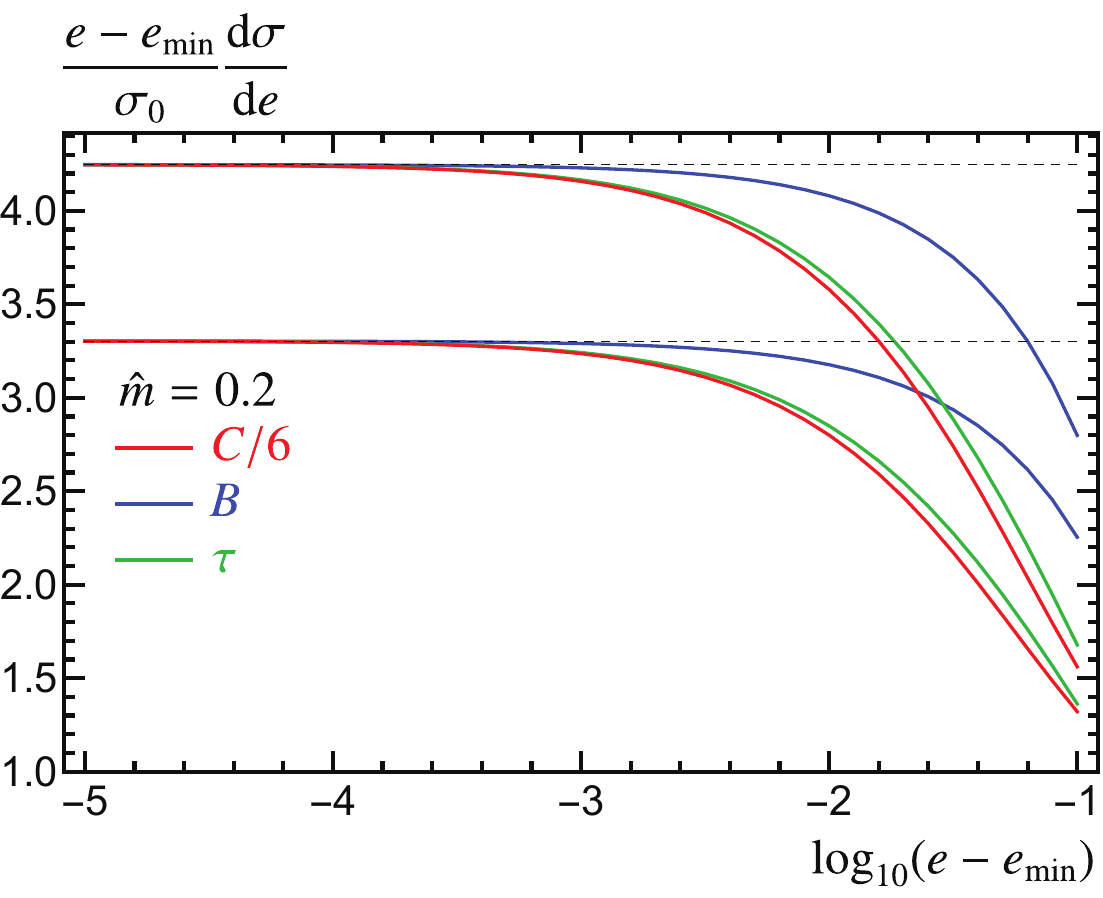}
\label{fig:Residue02}
}
\subfigure[]{
\includegraphics[width=0.48\textwidth]{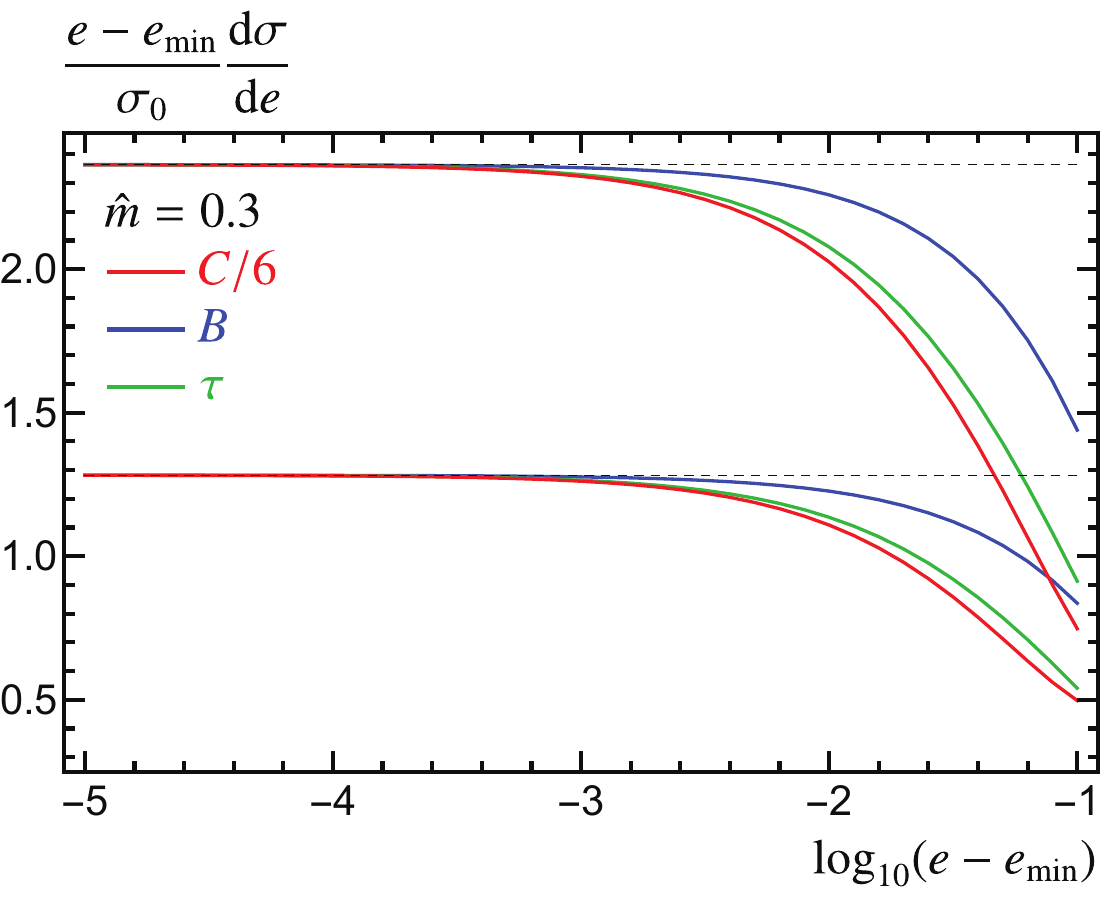}
\label{fig:Residue03}
}
\subfigure[]{
\includegraphics[width=0.48\textwidth]{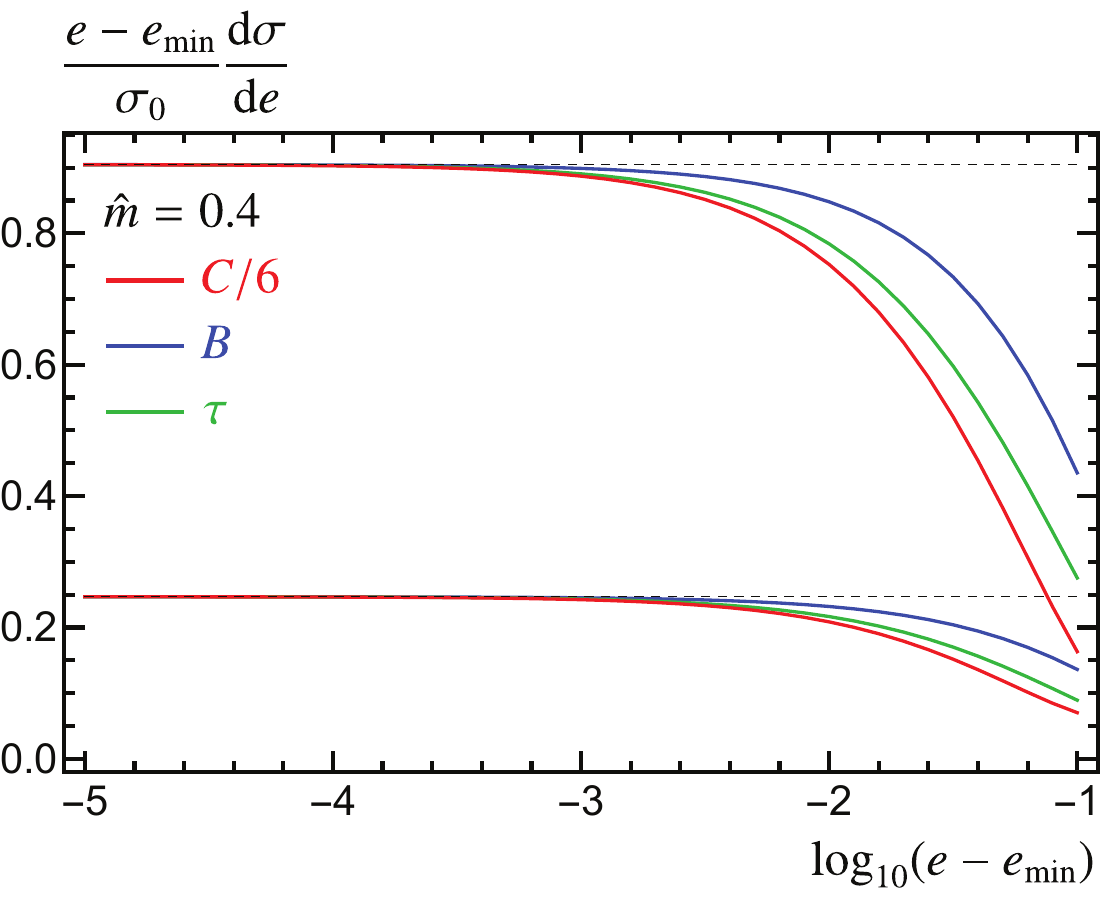}
\label{fig:Residue04}
}
\caption{Numerical determination of the plus-function coefficient. Panels (a), (b), (c) and (d) correspond to $\hat m=0.1,\,0.2,\,0.3$ and $0.4$, respectively.
In each panel curves converging to the upper (lower) horizontal dashed line show vector (axial-vector) current results. The dashed lines show the analytical
result of $B_{\rm plus}$ for both currents. Thrust is shown in green, jet broadening in blue, and reduced C-parameter ($\widetilde C=C/6$) in red.}
\label{fig:Residue}
\end{figure*}

In Fig.~\ref{fig:plusCoef} we show the dependence of the plus-function coefficient on $\hat m$. For very small reduced masses one recovers
the result predicted by SCET or bHQET,
\begin{equation}\label{eq:BSCET}
B_\text{plus}^\text{SCET} = -2\, [\,1 + 2\log(\hat m)\,]\,,
\end{equation}
in which powers of $\hat m$ are suppressed and the mass-dependence is purely logarithmic. Since the squared matrix elements in QCD do not
depend on $\log(\hat m)$, the mass dependence of $B_\text{plus}^\text{SCET}$ must come from phase-space restrictions and the event shape
definition. 
Hence, the massless limit has to be the same for both vector and axial-vector currents, as can be seen in Fig.~\ref{fig:plusCoef} or by Taylor
expanding the corresponding analytic formulas [\,as in Eq.~\eqref{eq:BSCET}\,]. For $v\to 0$ both vector and axial-vector
versions of $B_{\rm plus}$ tend to zero (faster for the axial current), which can be explained by physical arguments. In the threshold limit, all the
energy coming from the $e^+e^-$ collision is invested in creating a heavy $q\bar q$ pair at rest. Therefore either no gluon (or massless quark) is
radiated, or they have zero energy and momentum. No extra massive particles can be created. In this situation it is clear that
$e=e_{\rm min}=e_{\rm max}$, and therefore there is no radiative tail. Hence both the plus distribution and non-singular cross section identically
vanish, and only the delta function can remain.
As a numerical check of the universality of $B^C_{\rm plus}$, shown in Fig.~\ref{fig:plusCoef},
we use the fact that
\begin{equation}
B_{\rm plus} = \lim_{e\to e_{\rm min}} (e - e_{\rm min}) F_e(e, {\hat m})\,,
\end{equation}
to determine graphically the plus-function coefficient for three distinct event shapes. For the vector and axial-vector
currents separately, the cross sections approach the same horizontal line, matching our theoretical prediction. We have shown this behavior for
$\hat m=0.1,\,0.2,\,0.3$ and $0.4$ in the four panels of Fig.~\ref{fig:Residue}. We have checked that universality holds for all $16$ event shapes
considered in this article, and for $50$ values of the reduced mass.\,\footnote{We have also checked analytically that $B_{\rm plus}$ coincides
with the plus-distribution coefficient implied by the high-energy limit of the bare one-loop massive hemisphere soft function, given in Eq.~(59) of
Ref.~\cite{vonManteuffel:2014mva}.}

\begin{figure*}[t!]
\subfigure[]
{
\includegraphics[width=0.48\textwidth]{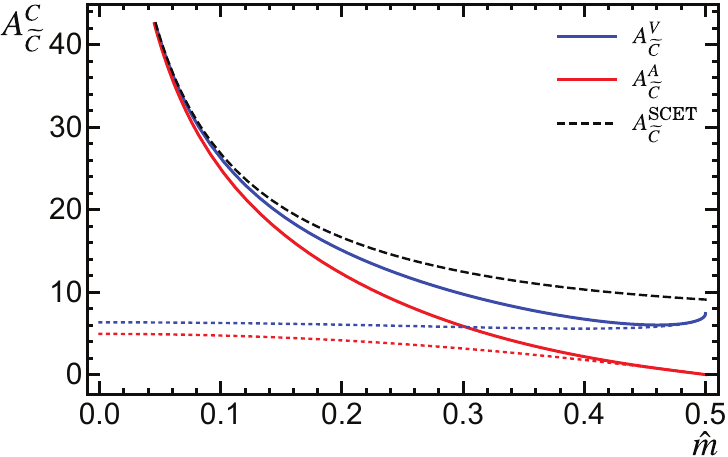}
\label{fig:CDelta}
}
\subfigure[]{
\includegraphics[width=0.475\textwidth]{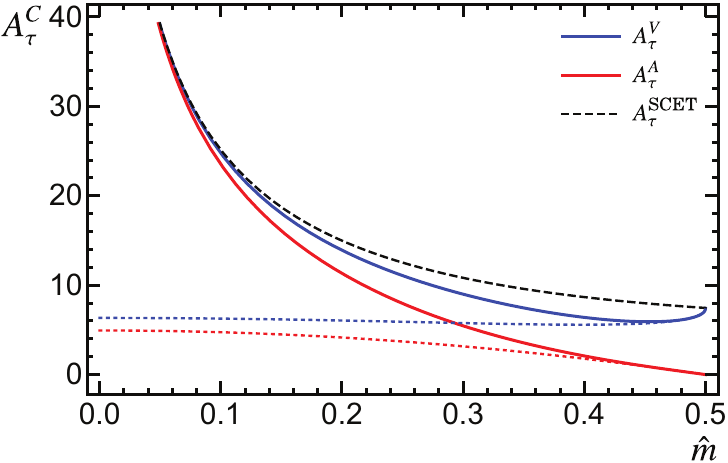}
\label{fig:ThrustDelta}
}
\subfigure[]{
\includegraphics[width=0.48\textwidth]{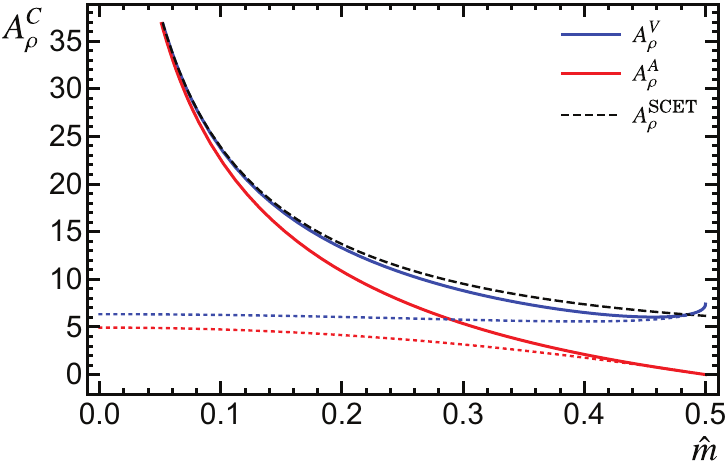}
\label{fig:HJMDelta}
}
\subfigure[]{
\includegraphics[width=0.48\textwidth]{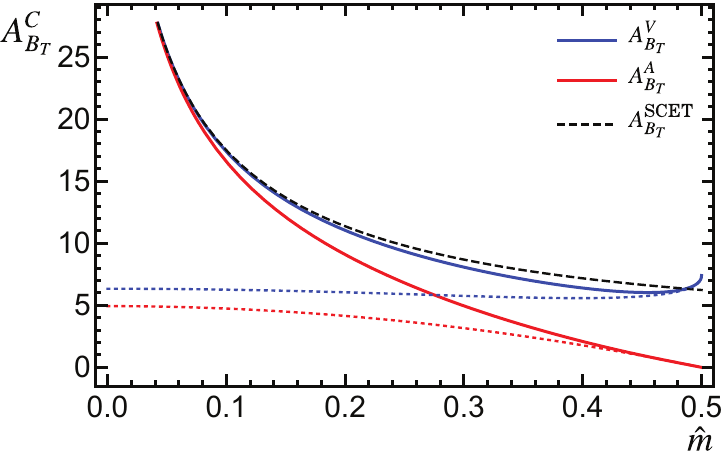}
\label{fig:BTDelta}
}
\caption{The delta-function coefficients for reduced C-parameter, thrust, heavy jet mass and jet broadening (all in their original definitions) as a
function of the reduced mass $\hat m$. The red solid lines show the coefficients for an axial-vector current, the blue solid lines for a vector current.
We also show the respective threshold limits as dotted lines, including terms up to $\Ord(v^2)$. The SCET limit, shown as black dashed
lines, coincides with the vector as well as axial currents for $\hat m \to 0$. The values in the threshold limit are universal, approaching
$3\pi^2/4$ and zero in the vector and axial-vector cases, respectively.}
\label{fig:delCoef}
\end{figure*}

\begin{figure*}[t!]
\subfigure[]
{
\includegraphics[width=0.48\textwidth]{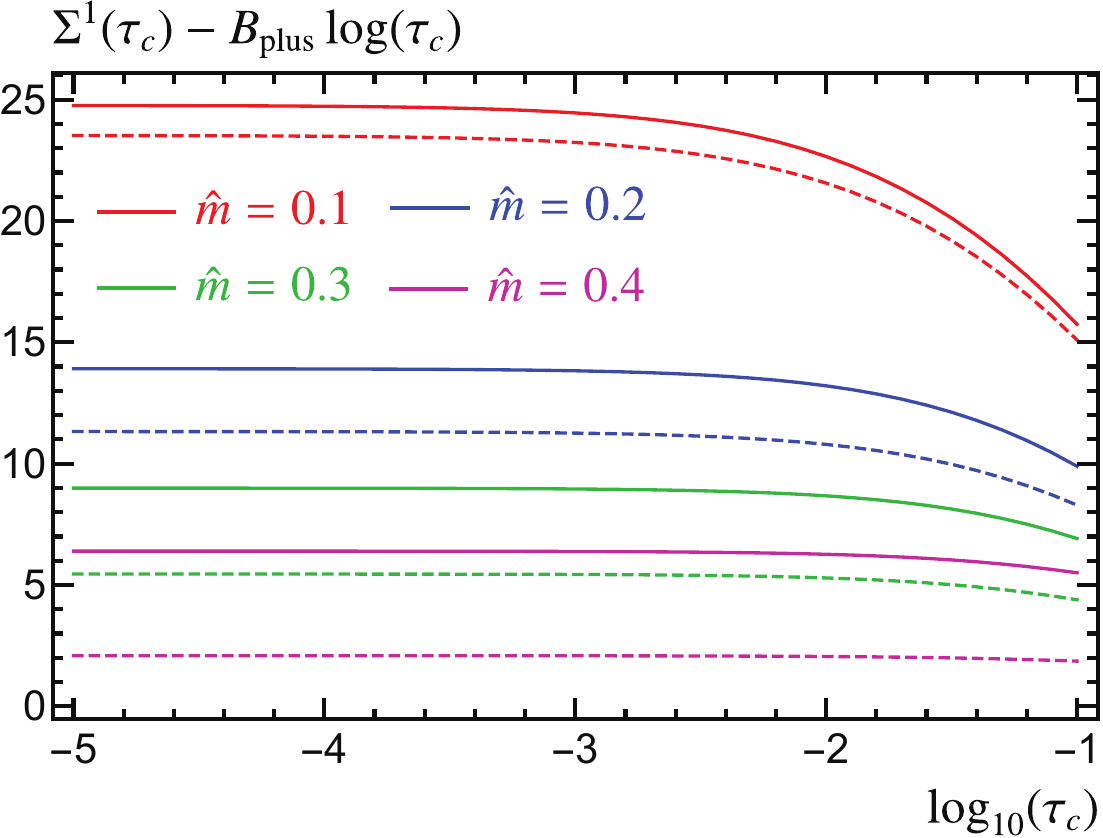}
\label{fig:NumDeltaTau}
}
\subfigure[]{
\includegraphics[width=0.475\textwidth]{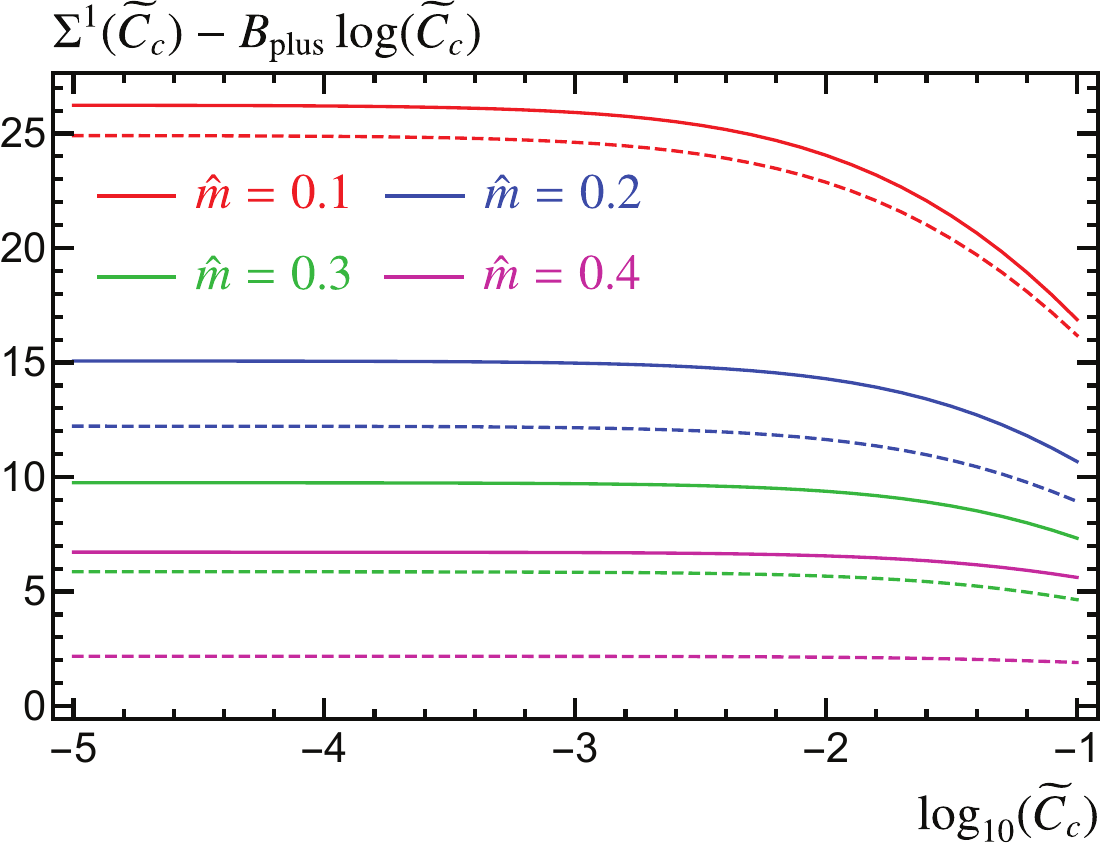}
\label{fig:NumDeltaC}
}
\subfigure[]{
\includegraphics[width=0.48\textwidth]{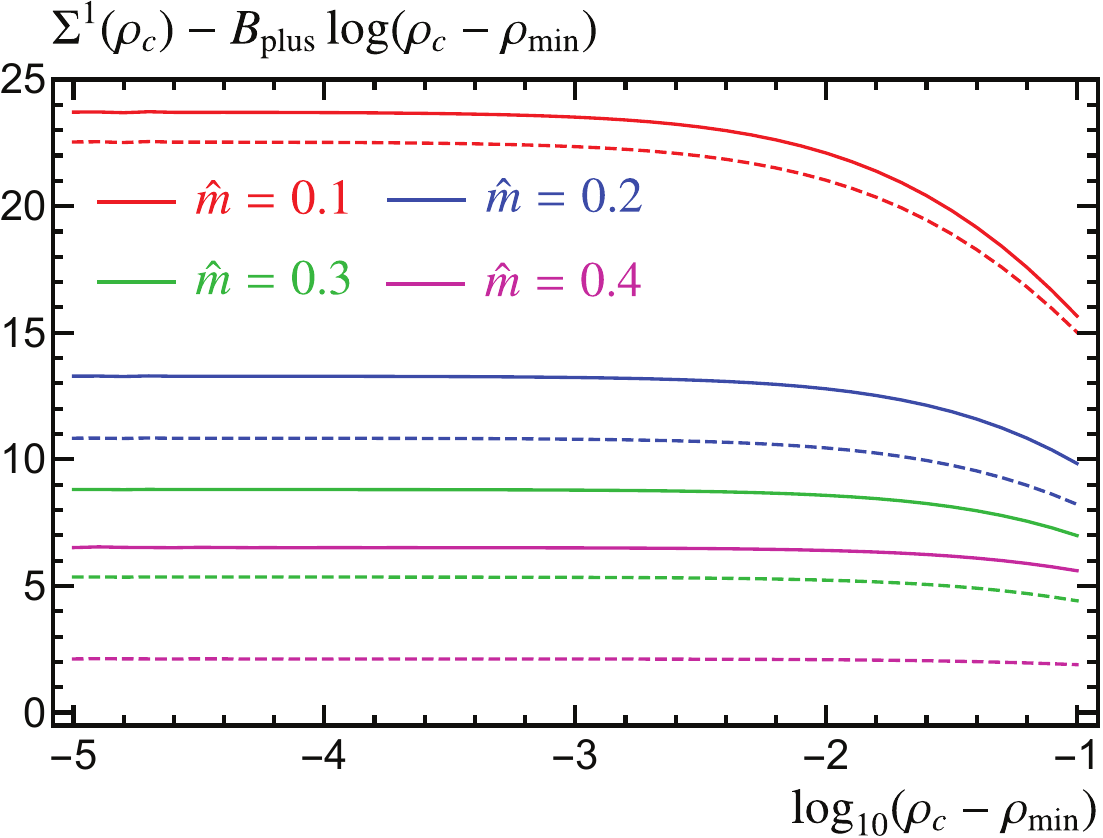}
\label{fig:NumDeltaRho}
}
\subfigure[]{
\includegraphics[width=0.49\textwidth]{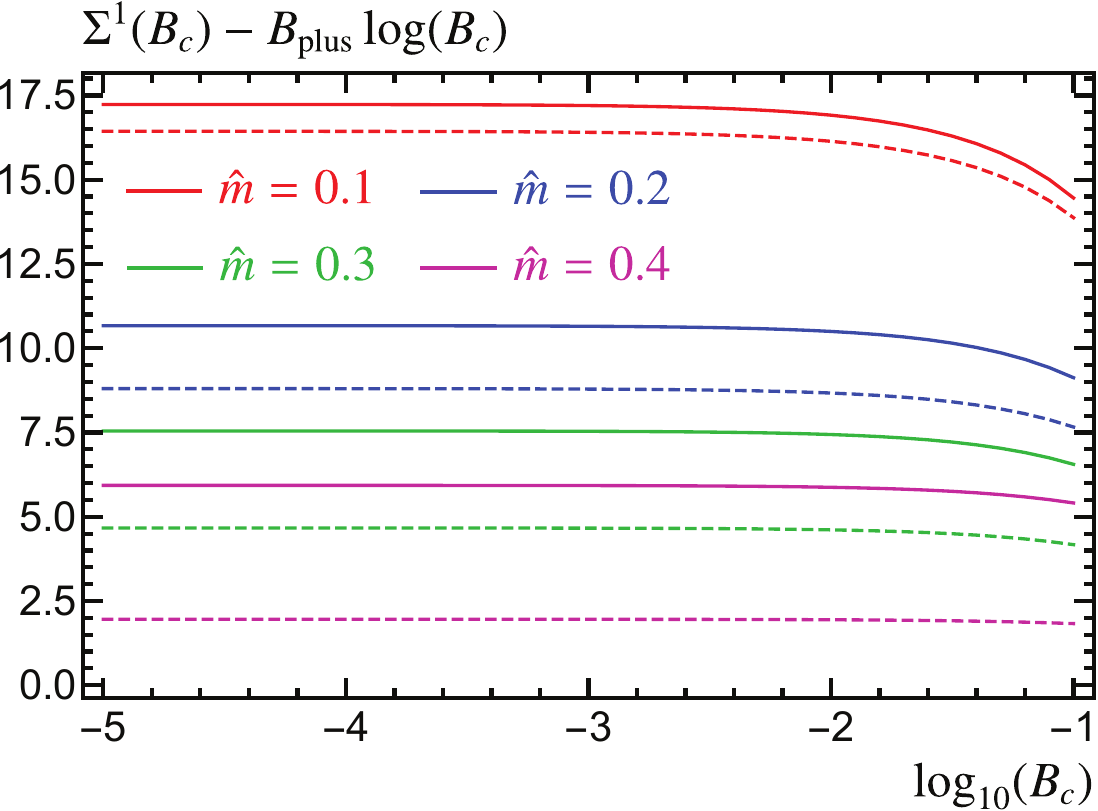}
\label{fig:NumDeltaB}
}
\caption{Numerical determination of the delta-function coefficient. Panel (a) is for thrust, (b) for reduced C-parameter, (c) for heavy
jet mass and (d) for jet broadening, respectively. Solid (dashed) lines show the vector (axial-vector) current. Red, blue, green and
magenta have reduced masses of $\hat m=0.1,\,0.2\,0.3$ and $0.4$, respectively. Each curve becomes a horizontal line as the
event shape approaches its minimal value, reproducing exactly our analytic computations for $A_e$.}
\label{fig:NumDelta}
\end{figure*}

In Fig.~\ref{fig:delCoef} we show the $A_{e}^C$ coefficients as a function of the reduced mass for thrust, C-parameter, jet broadening
and heavy jet mass. We again see that in the SCET/bHQET limit ($\hat m\to 0$), the result for both currents is the same. This follows from the
same reasoning as in the previous paragraph. Moreover, since the axial total hadronic cross section vanishes for $v\to0$, $A^A_{e}$
also vanishes in this limit. Furthermore, $A_{e}^V(\hat m = 1/2)=R_1^V(v=0)=3\pi^2/4$ takes the same value for all event shapes in the limit $v\to0$. This can be
easily understood since the event-shape dependent integral in Eq.~\eqref{eq:ES-dependent} appears multiplied in Eq.~\eqref{eq:main-result}
precisely by the plus-function coefficient, and we already argued that in the threshold limit $B_{\rm plus}\to 0$. Another interesting property of
the Dirac delta-function coefficient is that
\begin{eqnarray}
\lim_{\hat m\to 0} [A_{e^M}(\hat m) -A_{e^P}(\hat m)] = \lim_{\hat m\to 0} [A_{e^M}(\hat m) -A_{e^E}(\hat m)] = {\rm constant}\,,
\end{eqnarray}
where the constant is the difference of the Fourier-space jet functions in the massive and E/P schemes that appear in the bHQET factorization
theorem for massive event shapes. This shall be shown in more detail in Ref.~\cite{Alejandro-Vicent}, but it is based on the fact that the hard
and $H_m$ functions are the same for any event shape, and the soft function is the same in any scheme since it is mass independent at one
loop order. Therefore the consistency condition requires that the jet anomalous dimensions and hence also the logs of $\hat m$ are the
same in any scheme. Specifically, the logarithmic terms of the delta-function coefficients in the SCET limit $\hat m\to 0$ take the form
\begin{equation}
A_{e_\mathrm{I}}^\text{SCET} \Big|_{\log(\hat m)} = 4\log^2(\hat m) + \log(\hat m)\,,
\end{equation}
for pure $\mathrm{SCET_I}$ type event shapes like thrust, heavy jet mass and C-parameter, and
\begin{equation}
A_{e_\mathrm{II}}^\text{SCET} \Big|_{\log(\hat m)} = 2\log^2(\hat m) -\log(\hat m)\,,
\end{equation}
for $\mathrm{SCET_{II}}$ type ones like jet broadening. For similar
reasons, the SCET/bHQET limit is equal for the same event shape in the E or P scheme, as well as with normalization $Q$ or $Q_P$.
In Fig.~\ref{fig:NumDelta} we use Eq.~\eqref{eq:A-num} to graphically determine the delta-function coefficient, checking that it agrees with our analytic
computation. To make it visually clearer, we use logarithmic scaling in the horizontal axis, such that the log-subtracted cumulative distribution becomes a
horizontal line as the event-shape value approaches the threshold. This plot also shows how reliable our numerical code is, even for very small values of
the event shape. We choose the same event-shape measurement functions and reduced-mass values as for the rest of analyses in this section.

\section{Conclusions}\label{sec:conclusions}
In this article we have shown how to accurately compute differential and cumulative massive event-shape distributions at $\Ord(\alpha_s)$.
We have analytically calculated all singular terms: the plus-distribution coefficient, which we found to be the same for all event shapes linearly
sensitive to soft momenta, and the delta-function coefficient. 
In our computation of the latter we add the contributions of virtual- and real-radiation diagrams, which are individually IR-divergent, although the sum
is finite. 
We find that the delta-function coefficient only depends on the soft limit of the event-shape measurement function,
and can be expressed as the sum of a universal term plus an observable-dependent integral times the plus-function coefficient.

We have developed a numerical algorithm to compute the non-singular distribution efficiently and with high accuracy which does not require binning
the distributions. The method directly solves the measurement delta function and figures out the integration limits using standard numerical methods to find the
roots of an equation or to minimize functions. The remaining integration is performed numerically. Our method serves for both differential and
cumulative cross sections, can be used for massless quarks, and provides very accurate results even for extreme dijet configurations. Possible
additional applications of the algorithm are a)~event-shape distributions depending on a continuous parameter, such as angularities; b)~non-global
observables implying a jet algorithm and grooming, such as Soft Drop~\cite{Larkoski:2014wba}. Our approach is very different from conventional
MC methods, which are based on binned distributions. Although at $\Ord(\alpha_s)$ one does not yet need to implement subtractions, our
strategy to compute the delta-function coefficient could be implemented in NLO parton-level MCs to achieve a much more effective cancellation of IR
divergences, that could happen at an early stage.

Extending our calculation of the singular terms to $\Ord(\alpha_s^2)$ could be possible, provided one can find an analytic form for the
event-shape measurement function in the soft limit. 
Computing the next order could provide hints to figure out if the universality of the plus distribution holds to all orders.
Another approach to compute the delta and plus distributions at two loops is using EFT
language. Since these parameters are defined in the limit of very soft gluon momenta, one can imagine building an EFT of heavy quarks
(not necessarily boosted) interacting with soft particles (massless quarks or gluons), such that $E_{\rm soft}\ll m_q$. In this way one
could derive a factorized form for the cross section in terms of a universal matching coefficient and an event-shape dependent
soft function. This could simplify the computation, since the cross section provided by the EFT would be already purely singular, and
the problem would be naturally split into simpler pieces, treating one scale at a time. Moreover, such a theory would allow to sum
large logarithms of ratios of scales to all orders in perturbation theory. A step in this direction has been taken already in
Ref.~\cite{vonManteuffel:2014mva}.

Our results will be a reference for ongoing and future research carried out in the context of event shapes with massive particles. In particular
they will play an important role in the calibration of the MC top quark mass parameter, and will be even more relevant for the top quark mass
measurement program at future linear colliders. In this direction, the results presented in this article will help computing efficiently $e^+e^-$
event-shape distributions at $\Ord(\alpha_s)$ for unstable top quarks, since in that case, the delta function that sits at threshold for
stable quarks radiates into the tail through the top decay products.
Therefore, already at $\Ord(\alpha_s)$ one has to deal with the cancellation of IR divergences away from threshold, and our strategy
for computing the delta-function coefficient could be adapted for this situation.

\acknowledgments
This work was supported in part by FWF Austrian Science Fund under the Project No. P28535- N27, the Spanish MINECO Ram\'on y Cajal program
(RYC-2014-16022), the MECD grant FPA2016-78645-P, the IFT Centro de Excelencia Severo Ochoa Program under Grant SEV-2012-0249, the EU
STRONG-2020 project under the program H2020-INFRAIA-2018-1, grant agreement no. 824093 and the COST Action CA16201 PARTICLEFACE.
CL is supported by the FWF Doctoral Program ``Particles and Interactions'' No. W1252-N2.
CL thanks the University of Salamanca for hospitality while parts of this work were completed.
We thank A.\,H.\ Hoang for many valuable comments.

\appendix

\section{Indirect Computation}\label{sec:indirect}
\begin{figure}[t]\centering
\includegraphics[width=0.5\textwidth]{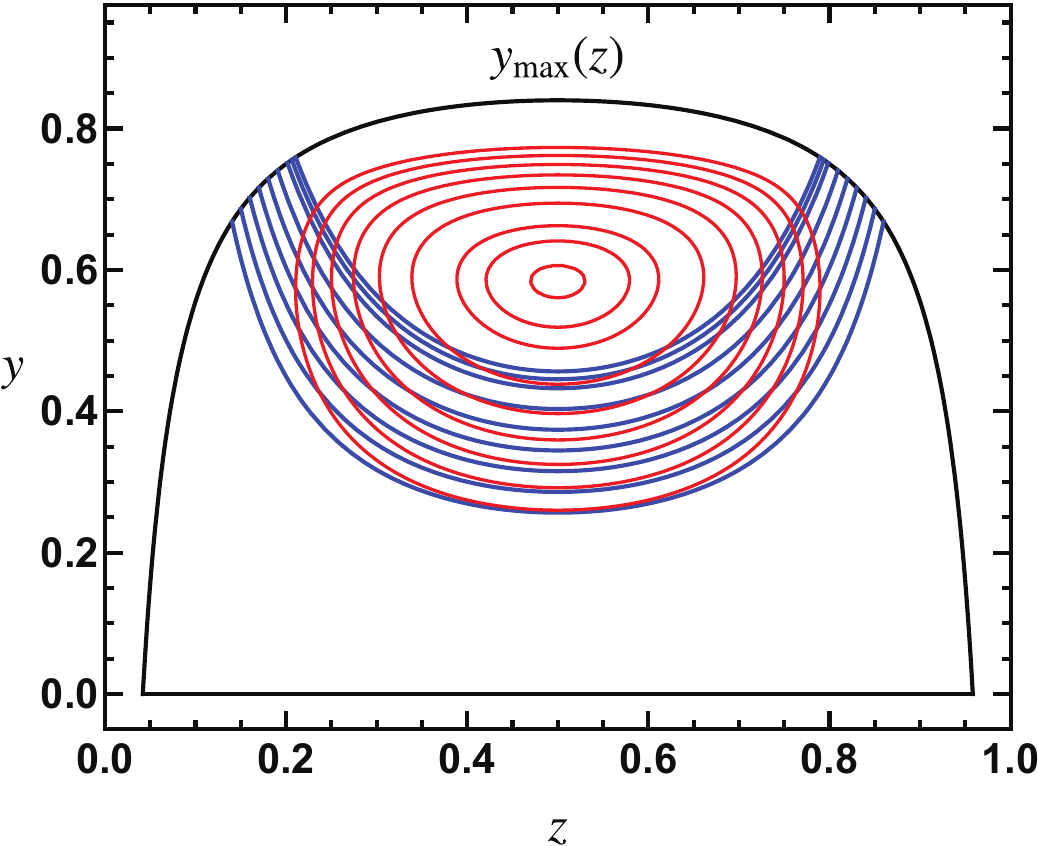}
\caption{Phase space diagram in $(z,y)$ coordinates (black lines) showing curves with constant value of the C-parameter event-shape
measurement function (in red) and for its soft limit (in blue), for the same set of $C$ values $\{0.42, 0.468, 0.516, 0.564, 0.612, 0.66, 0.708,
0.729, 0.747\}$, while $\hat m = 0.2$. Red curves are always contained withing the phase-space boundaries, and the enclosed area
becomes smaller as the value of $C$
approaches its maximum value, when it collapses to a single point. Blue curves always intersect with the phase-space boundary, and have a finite extension
even for $C=C_{\rm max}$. Blue and red curves are similar for small values of $y$. \label{fig:multiline}}
\end{figure}
In this appendix we recover the main result of Eq.~\eqref{eq:main-result} imposing that the integrated event-shape distribution over its
entire domain reproduces the known total hadronic cross section. This method was discussed in the introduction as a way of
obtaining, in a numeric form, the delta-function coefficient if an analytic form for the differential cross section is available in $d=4$ dimensions. Here we
show that if the event shape integration is performed before the phase-space integrals, analytic results can be obtained for any
event shape considered in this article. Moreover, the result is equivalent to what we have already found by the direct computation.

\begin{figure}[t]\centering
\includegraphics[width=0.5\textwidth]{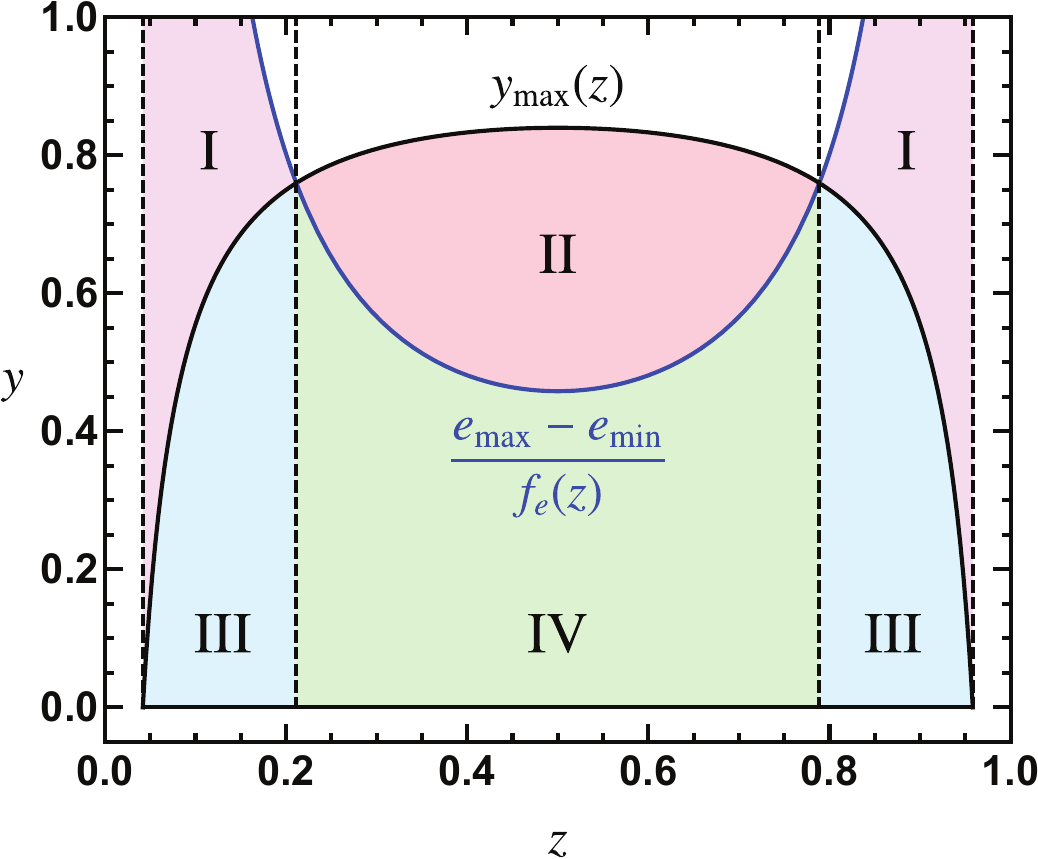}
\caption{Phase space diagram in $(z,y)$ coordinates (black lines) showing the curve for which the C-parameter soft measurement function
equals the maximum allowed value $C_{\rm max} = 0.75$. Different colors correspond to the different integration regions of the
soft non-singular distribution. To make this plot we use the numerical value $\hat m = 0.2$. \label{fig:indirect}}
\end{figure}

Writing the massive total hadronic cross section as
\begin{equation}
R (\hat{m}) = R_0 (\hat{m}) + C_F \frac{\alpha_s}{\pi} R_1 (\hat{m}) + \Ord (\alpha^2_s)\,,
\end{equation}
one observes the non-trivial constraint
\begin{equation}
R_1 (\hat{m}) = A_e (\hat{m}) + B_{\rm plus} (\hat{m}) \log (e_{\max} - e_{\min}) +
\int^{e_{\max}}_{e_{\min}} \dd e\, F^{\rm NS} (e, \hat{m})\,,
\end{equation}
where $F^{\rm NS}_e $, defined in Eq.~\eqref{eq:real}, has to be integrated in the whole $e$ domain $[e_{\min}, e_{\max}]$. This is done before
integrating in $y$ and $z$. The $e$ integration in $ F^{\rm NS}_{\rm hard}$ effectively only replaces the Dirac delta function by $1$, since the
entire phase space is exactly covered by the full event shape range,
\begin{equation}
\int^{e_{\max}}_{e_{\min}} \!\dd e\, F^{\rm NS}_{\rm hard} (e, \hat{m})=\!
\int\! \dd z\,\dd y\, M^{\rm hard}_C (y, z, \hat{m})\,.
\end{equation}
It is convenient to write the soft non-singular distribution with explicit Heaviside functions in $y$, such that the two terms
with the soft measurement function in Eq.~\eqref{eq:real} can be added together
\begin{equation}
F^{\rm NS}_{\rm soft} = \int\! \dd z\, \dd y \,\frac{ M_C^0(z, \hat{m})}{y}\,
\Theta(y) \biggl\{\Theta [y_{\max} (z) - y]\,\delta [e - \hat e(y,z)]
- \delta [e - \bar e(y,z)]\biggl\}\,,
\end{equation}
understanding that the $z$ integration is still between $z_-$ and $z_+$. To evaluate the $ F^{\rm NS}_{\rm soft}$ contribution it is convenient to
insert \mbox{$\Theta (e_{\max} - e)$} before performing the $e$ integral over the Dirac delta functions. In this way we obtain
\begin{equation}
\int^{e_{\max}}_{e_{\min}} \dd e\, F^{\rm NS}_{\rm soft} (e, \hat{m}) =
\int\! \dd z\,\dd y \, \frac{M^0_C (z, \hat{m})}{y}\, \Theta(y)\Bigl\{\Theta [y_{\max} (z) - y] - \Theta [h(e_{\rm max},z) - y] \Bigl\} \,.
\end{equation}
Graphically, the integral of the first term corresponds to the areas marked as II, III and IV in Fig.~\ref{fig:indirect}, while the second
integral is the sum of regions I, III and IV. The sum of III and IV corresponds to integrating the last term in Eq.~\eqref{eq:real},
while the pieces marked with I are the integration of the one-to-last term in Eq.~\eqref{eq:real}.
Adding the two terms, the resulting $y$ integration has the boundaries $[h (e_{\max}, z), y_{\max} (z)]$ in the whole $z$ range, which can be expressed as
\begin{equation}
\int^{e_{\max}}_{e_{\min}} \dd e\, F^{\rm NS}_{\rm soft} (e, \hat{m}) =
\int_{z_-}^{z_+}\! \dd z\, M^0_C (z, \hat{m}) \log\Biggl[
\frac{1 - \frac{\hat{m}^2}{z (1 - z)}}{\frac{e_{\max} - e_{\min}}{f_e (z,\hat{m})}} \Biggr]\,,
\end{equation}
and graphically corresponds to subtracting the portions marked with I from the area labeled II in Fig.~\ref{fig:indirect}.
Combining this result with the plus function term and using Eq.~\eqref{eq:real} the dependence on $e_{\rm max}$ cancels. The $y$ integration
can be performed analytically, giving
\begin{align}
&B_{\rm plus}(\hat{m}) \log (e_{\max} - e_{\min}) +\! \int^{e_{\max}}_{e_{\min}}\! \dd e\, F^{\rm NS}_{\rm soft} (e, \hat{m})\nonumber \\
&=\int\! \dd z\, M^0_C (z, \hat{m}) \biggl\{\log \biggl[ 1 - \frac{\hat{m}^2}{z (1 - z)} \biggr]+ \log [f_e (z)]\biggl\}\,,\\
&\!\!\!\int\! \dd y\, \dd z\, M^{{\rm hard}}_C (y, z) =\! \int\!
\dd z\, \biggl( 1 - \frac{\hat{m}^2}{z (1 - z)} \biggr) \biggl[ M^2_C (z,
\hat{m}) + \frac{1}{2} M^3_C (z, \hat{m}) \biggl( 1 - \frac{\hat{m}^2}{z (1 - z)} \biggr) \biggr].\nonumber
\end{align}
With this we conclude with the alternative, but analytically equivalent, expression for the $\delta$ coefficient
\begin{align}
A_e (\hat{m}) ={} & R_1 (\hat{m}) - \!\int_{z_-}^{z_+} \!\dd z\, \biggl\{ M^0_C (z, \hat{m}) \bigl[ \log \bigl[z (1 - z) - m^2\bigr] -
\log[z (1 - z)] + \log [f_e (z)] \, \bigr] \nonumber \\
& + \biggl( 1 - \frac{\hat{m}^2}{z (1 - z)} \biggr) \biggl[ M^2_C
(z, \hat{m}) + \frac{1}{2} M^3_C (z, \hat{m}) \biggl( 1 - \frac{\hat{m}^2}{z(1 - z)} \biggr)\biggr] \biggl\} .
\end{align}
The integral over hard matrix elements can be performed using Eqs.~\eqref{eq:noES}, while the other integrals are
explicitly given in Eqs.~\eqref{eq:real-integrals}. Using these analytic results we arrive at
\begin{eqnarray}\label{eq:alternative}
A^V_e (\hat{m}) & = & R^V_1 (\hat{m}) - (3 - v^2) I_e (\hat{m}) -
(1 + 2 \hat{m}^4)L_v - \frac{v}{4}
\biggl[ 11 + 34 \hat{m}^2 - 8 (3 - v^2) \log \Bigl( \frac{v}{\hat{m}} \Bigr)\biggr]\nonumber\\
& & \!\!\!\!\! + 2 \,v^2 \biggl[ {\rm Li}_2 \biggl( \frac{2 v}{1 + v}
\biggr) - 2 \log \Bigl( \frac{v^2}{\hat{m}} \Bigr)L_v + \, {\rm Li}_2 \biggl( \frac{1 - v}{2} \biggr) - {\rm Li}_2
\biggl( \frac{2 v}{v - 1} \biggr) - {\rm Li}_2 \biggl( \frac{1 + v}{2}
\biggr) \biggr], \nonumber\\
A^A_e (\hat{m}) & = & R^A_1 (\hat{m}) - 2 v^2 I_e (\hat{m}) - (1 + 2
\hat{m}^2 - 6 \hat{m}^4 + 12 \hat{m}^6) L_v - \frac{v}{4} \biggl[ 11 - 68 \hat{m}^2+ 12 \hat{m}^4
\nonumber\\
& &\!\!\!\!\! - 16 v^2 \log\Bigl( \frac{v}{\hat{m}} \Bigr)
\biggr] + 2 (1 - 6 \hat m^2 + 8 \hat{m}^4) \biggl[ {\rm Li}_2 \biggl( \frac{2 v}{1
+ v} \biggr) + {\rm Li}_2 \biggl( \frac{1 - v}{2} \biggr)
\nonumber\\
& &\!\!\!\!\! - \, 2 \log\Bigl( \frac{v^2}{\hat{m}} \Bigr) L_v - {\rm Li}_2 \biggl( \frac{2 v}{v - 1}
\biggr) - {\rm Li}_2 \biggl( \frac{1 + v}{2} \biggr) \biggr].
\end{eqnarray}
Using the known results for the total hadronic cross section, collected for convenience in Eq.~\eqref{eq:Rhad} of Appendix~\ref{sec:total},
it can be checked that the above result is analytically equivalent to Eqs.~\eqref{eq:main-result}.

\section{Phase-Space Integrals}\label{app:integrals}
Almost every integral in the phase-space variable $z$ we have computed for various event-shape measurement functions can, due to the
symmetry under $z\to 1-z$, be cast in the one these forms:
\begin{align}
\int^{z_+}_{\frac{1}{2}} \dd z \frac{\log (z - a)}{z} ={}& {\rm Li}_2\biggl(1 - \frac{1 + v}{2 a} \biggr) +
\log\Bigl(\frac{1 + v}{2 a} \Bigr) \log \biggl( \frac{1 - 2 a + v}{2} \biggr)\\
& - {\rm Li}_2 \biggl(1 - \frac{1}{2 a} \biggr) + \log \biggl(\frac{1}{2} - a \biggr)
\log (2 a)\,, \quad a \leq \frac{1}{2}\,,\nonumber\\
\int^{z_+}_{\frac{1}{2}} \dd z \frac{\log (a - z)}{z} ={}& {\rm Li}_2
\biggl(1 - \frac{1 + v}{2 a} \biggr) + \log\Bigl(
\frac{1 + v}{2 a} \Bigr) \log\Bigl( \frac{2 a - 1 -v}{2} \Bigr)\nonumber\\
& - {\rm Li}_2 \biggl(1 - \frac{1}{2 a} \biggr) + \log \Bigl( a -
\frac{1}{2} \Bigr) \log (2 a)\,, \quad a \geq z_+\,,\nonumber\\
\int^{z_+}_{\frac{1}{2}} \dd z \frac{\log (z - a)}{z^2} ={}&
\frac{1}{a} \Biggl\{\log \Bigl( \frac{1 - 2 a + v}{1 + v} \Bigr) - 2 a \log (2) - (1 - 2 a) \log (1 - 2 a)
\nonumber\\
& - \, \frac{2 a \log \Bigl( \frac{1 - 2 a +v}{2} \Bigr)}{1 + v} \Biggl\}, \quad a
\leq \frac{1}{2}\,,\nonumber\\
\int^{z_+}_{\frac{1}{2}} \dd z \frac{\log (a - z)}{z^2} ={}&
\frac{1}{a} \Biggl\{\log\Bigl( -\frac{1 - 2 a + v}{1+ v} \Bigr) - 2 \,a \log (2) - (1 - 2 a) \log (2 a - 1)
\nonumber\\
& - \, \frac{2 a \log\Bigl( -\frac{1 - 2 a + v}{2} \Bigr)}{1 + v} \Biggl\}, \quad a
\geq z_+ \,.\nonumber\end{align}
For integrals in which no event-shape measurement function is involved, the results are even simpler
\begin{align}\label{eq:noES}
\int^{z_+}_{z_{_-}}\!\! \dd z &=v\,, \qquad\qquad\qquad \int^{z_+}_{z_{_-}} \!\!\frac{\dd z}{z} = \!\int^{z_+}_{z_{_-}}\! \frac{\dd z}{1 - z} = 2\, L_v\,,\\
\int^{z_+}_{z_{_-}}\! \frac{\dd z}{z^2} &= \int^{z_+}_{z_{_-}} \!\frac{\dd z}{(1 - z)^2} = \frac{v}{\hat{m}^2}, \qquad
\int^{z_+}_{z_{_-}}\!\frac{ \dd z }{z^3} = \int^{z_+}_{z_{_-}}\!\frac{ \dd z}{(1 - z)^3} = \frac{v}{2 \hat{m}^4}\,,\nonumber\\
\int^{z_+}_{z_{_-}}\!\! \dd z \frac{\log [z (1 - z) - \hat{m}^2]}{z^2} & =\!
\int^{z_+}_{z_{_-}} \!\!\dd z \frac{\log [z (1 - z) - \hat{m}^2]}{(1 - z)^2}= \frac{2 v \log (v)-2L_v}{\hat{m}^2}\,, \nonumber\\
\int^{z_+}_{z_{_-}}\!\! \dd z \frac{\log [z (1 - z) - \hat{m}^2]}{z} & =\!
\int^{z_+}_{z_{_-}}\! \!\dd z \frac{\log [z (1 - z) - \hat{m}^2]}{1 - z} =
{\rm Li}_2 \Bigl( \frac{2 v}{v - 1} \Bigr) - {\rm Li}_2 \Bigl( \frac{2v}{1+ v} \Bigr) + 4 \log (v) L_v\,.\nonumber
\end{align}
With these one can easily compute the integrals for the real radiation contribution
\begin{align}\label{eq:real-integrals}
&\qquad\qquad\qquad\qquad\int^{z_+}_{z_{_-}}\!\! \dd z\,\frac{(1 - z) z - \hat{m}^2 }{(1 - z)^2 z^2 } =\, 2\bigl[ (1 + v^2) L_v - v\bigr]\nonumber\\
&\int^{z_+}_{z_{_-}}\!\! \dd z\,\frac{(1 - z) z - \hat{m}^2 }{(1 - z)^2 z^2 } \log [z (1 - z) - \hat{m}^2] =\,
4\,L_v -4\,v\log(v) +(1+v^2)\biggl[ {\rm Li}_2 \Bigl( \frac{2 v}{v - 1} \Bigr) \nonumber\\
&\qquad\qquad\qquad\qquad\qquad-{\rm Li}_2 \Bigl( \frac{2 v}{1 + v} \Bigr) + 4\log(v)\,L_v\biggr]\,,\\
&\int^{z_+}_{z_{_-}}\!\! \dd z\,\frac{(1 - z) z - \hat{m}^2 }{(1 - z)^2 z^2 } \log [z (1 - z)] =\,
(1 +v^2) \biggl[ {\rm Li}_2 \Bigl( \frac{1 -v}{2} \Bigr) - {\rm Li}_2 \Bigl( \frac{1 +v}{2} \Bigr) \biggr] \nonumber\\
& \qquad\qquad\qquad\qquad\qquad+2\bigl[ (1 +v^2) \log (\hat{m}) + 4 \hat{m}^2 \bigr]L_v - 2v \bigl[ 1 + 2 \log (\hat{m}) \bigr]\,.\nonumber
\end{align}
Note that the integrands given in Eq.~\eqref{eq:real-integrals} are invariant under the substitution $z\to (1-z)$ and
therefore $\int^{z_+}_{z_{_-}}\dd z = 2\int^{1/2}_{z_-}\dd z = 2\int^{z_+}_{1/2}\dd z$.

\section{Total Hadronic Cross Section}\label{sec:total}
It is convenient to write the total hadronic cross section in the following way, such that it can be implemented in Eq.~\eqref{eq:alternative}
\begin{eqnarray}\label{eq:Rhad}
R^V_1 (\hat{m}) & = & 4 (1 - 4 \hat{m}^4) \biggl[ 3 L_v^2 + 2 \,{\rm Li}_2 \biggl( \frac{1 - v}{v + 1} \biggr) +
{\rm Li}_2 \biggl( \frac{v - 1}{v + 1} \biggr) - \log \Bigl(
\frac{v^2}{\hat{m}^3} \Bigr) L_v
\biggr]\\
& & + \, 2 (3 - 2 \hat{m}^2 - 7 \hat{m}^4) L_v + v \biggl[ \frac{3}{4} (1 + 6 \hat{m}^2) -
(3 - v^2) \log\Bigl( \frac{v^2}{\hat{m}^3} \Bigr) \biggr], \nonumber\\
R^A_1 (\hat{m}) & = & 2 \, v^2 (1 +v^2) \biggl[ 3\, L_v^2 + 2 \,{\rm Li}_2 \biggl( \frac{1 - v}{v + 1}
\biggr) + {\rm Li}_2 \biggl( \frac{v - 1}{v + 1} \biggr) - \log\Bigl(\frac{v^2}{\hat{m}^3} \Bigr) L_v
\biggr] \nonumber\\
& & +\, 2 (3 - 11 \hat{m}^2 + 5 \hat{m}^4 + 6 \hat{m}^6) L_v- \frac{21}{16} v - v^3 \biggl[ 2 \log\Bigl(
\frac{v^2}{\hat{m}^3} \Bigr) - \frac{15}{8} - \frac{3}{16} v^2 \biggr] ,
\nonumber
\end{eqnarray}
coinciding with Ref.~\cite{Chetyrkin:1994js}.

\section{Analytic Delta-Function Coefficients for some Event Shapes}\label{app:anadel}
In this appendix we present some analytic results for the delta-function coefficient of massive event-shape
differential cross sections. Specifically, we list the following information:
\begin{itemize}
\item The general definition of the event shape $e$.
\item The expression $e_q(y,z)$ describing the event shape in the phase-space region in which the thrust axis $\hat t$
is aligned with the quark momentum in terms of $y, z$. When evaluating the integral $I_e$, we use the $z\leftrightarrow (1-z)$
symmetry and integrate only in this region.
\item The minimal value of the event shape $e_\text{min}$, valid to all orders in perturbation theory.
\item The maximal value of the event shape $e_\text{max}$, valid at one-loop level.
\item The soft term $f_e(z)$ of the event shape required for the integral $I_e$, derived from $e_q(y,z)$ and therefore valid in the
same phase-space region only. It is straight forward to obtain the expression for the region where $\hat t$ is proportional to the
anti-quark momentum by applying the substitution $z\to (1-z)$.
\item The solution of the integrals $I_e(\hat m)$, which can be inserted into Eq.~\eqref{eq:main-result} or \eqref{eq:alternative} to
obtain the delta-function coefficient $A_e^C$.
\end{itemize}

For simplicity, we define
\begin{align}
r ={}& \sqrt{1 - 3 \hat m^2}\,,\qquad \text{mod}(y, z) = \sqrt{(1 - yz)^2 - 4 \hat m^2}\,,\nonumber\\
\mathcal I(\hat m)\equiv{}&\int_{z_-}^{\frac{1}{2}}\dd z\,\frac{z(1-z) - \hat m^2}{z^2(1-z)^2} = \bigl(1 + v^2\Bigr) L_v - v\,,
\end{align}
frequently appearing in expressions for $e_\text{max}$, $e_q(y,z)$ and relations between event shapes, respectively. Moreover, we use the
pseudo-rapidity $\eta$, transverse momentum $p_\bot \equiv |\vec p_\bot|$ and transverse mass $m_\bot \equiv \sqrt{p_\bot^2 + m^2}$ in some
event-shape definitions, where the transverse momentum is measured with respect to the thrust axis. The solutions of
the integrals needed to compute the various expressions for $I_e(\hat m)$, are provided in Appendix~\ref{app:integrals}.

Sometimes it can be useful to define a P-scheme event shape $e^{P,Q}$ with $1/Q^n$ normalization instead of $1/Q_P^n$. The relation between
the related event shape dependent integrals $I_e(\hat m)$ is trivial, as long as $e_\text{min}=0$:
\begin{equation}
I_{e^{P,Q}}(\hat m) = I_{e^{P}}(\hat m) + n \log(v)\, \mathcal I(\hat m)\,.
\end{equation}

\subsection*{Heavy Jet Mass}

\noindent\underline{Original definition}
\begin{itemize}
\item $\rho = \frac{1}{Q^2}\Bigl(\sum_{i \in \text{heavy}}p_i\Bigr)^{\!2}$
\item $\rho_q(y, z) = \hat m^2 + y z$
\item $\rho_\text{min} = \hat m^2$
\item $\rho_\text{max} = ( 5 - 4 r )/3$
\item $f_\rho(z) = z$
\item $\begin{aligned}[t]
I_\rho(\hat m) \,={} & \frac{1}{4}\biggl[(1+v^2) \text{Li}_2\biggl(\frac{v+1}{v-1}\biggr)+\frac{\pi ^2}{6} (1+v^2)-2 v (v+1)\\
& - [1 + (4-v) v] \log\biggl(\frac{1-v}{1+v}\biggr)-4 v \log \biggl(\frac{v+1}{2}\biggr)\biggr]
\end{aligned} $
\end{itemize}

\noindent\underline{P-scheme}
\begin{itemize}
\item $\rho^P = \frac{1}{Q_P^2}\sum_{i, j\in \text{heavy}}(|\vec p_i||\vec p_j| - \vec p_i\cdot\vec p_j)$
\item $\rho^P_q(y,z) = \frac{Q^2}{2 Q_P^2} y \bigl[\text{mod}(y,1-z)+y(1-z)+2 z-1\bigr]$
\item $\rho^P_\text{min} = 0$
\item $\rho^P_\text{max} = 1/3$
\item $f_{\rho^P}(z) = \frac{2z + v - 1}{2v^2}$
\item $ \begin{aligned}[t]
I_{\rho^P}(\hat m) ={}& (1-2 \hat m^2) \biggl[\text{Li}_2(1-v)+\text{Li}_2\biggl(\frac{1}{v+1}\biggr)\biggr]\\
&+\frac{1}{2} \log \biggl(\frac{1+v}{1-v}\biggr) [1-2(1-2 \hat m^2) \log (2 v^2)]-\frac{1}{2} (1-2 \hat m^2) \log^2(1-v)\\
&-\frac{1}{3} (1-2 \hat m^2) [\pi ^2-3 \log ^2(v+1)]-v \log (\hat m)+ v \log (v)
\end{aligned}$
\end{itemize}

\noindent\underline{E-scheme}
\begin{itemize}
\item $\rho^E = \frac{1}{Q^2}\sum_{i, j\in \text{heavy}}\frac{E_i E_j}{|\vec p_i||\vec p_j|}(|\vec p_i||\vec p_j| - \vec p_i\cdot\vec p_j)$
\item $\rho^E_q(y,z) = \frac{y (1 - y (1-z)) [\text{mod}(y,1-z)+y(1-z)-1+2 z]}{2\, \text{mod}(y,1-z)}$
\item $\rho^E_\text{min} = 0$
\item $\rho^E_\text{max} = \frac{(2 - r)^2}{3}$
\item $f_{\rho^E}(z) = v f_{\rho^P}(z)$
\item $I_{\rho^E}(\hat m) = I_{\rho^P}(\hat m) + \log(v) \,\mathcal I(\hat m)$
\end{itemize}

\subsection*{Thrust / 2-Jettiness}

\noindent\underline{P-scheme, original definition}
\begin{itemize}
\item $\tau = \frac{1}{Q_P} \sum_i p_{i,\bot} \mathrm e^{-|\eta_i|} = \frac{1}{Q_P} \min_{\hat t} \sum_i(|\vec p_i|-|\hat t\cdot\vec p_i|)$
\item $\tau_q(y, z) = \text{mod}(y,1-z) - \text{mod}(y,z) + y$
\item $\tau_\text{min} = 0$
\item $\tau_\text{max} = \rho^P_\text{max}$
\item $f_\tau(z) = f_{\rho^P}(z)$
\item $	I_{\tau}(\hat m) = I_{\rho^P}(\hat m)$
\end{itemize}

\noindent\underline{E-scheme}
\begin{itemize}
\item $\tau^E = \frac{1}{Q_P} \min_{\hat t} \sum_i \frac{E_i}{|\vec p_i|}(|\vec p_i|-|\hat t\cdot\vec p_i|)$
\item $\tau^E_q(y, z) = \frac{1}{2} \biggl(\frac{[1 - y (1-z)] [4 \hat m^2+y^2 (1-z) z-1+y]}{\text{mod}(y,z)\,\text{mod}(y,1-z)}-
\frac{y [1 + (y-2) z]}{\text{mod}(y,z)}-y (1-z)+ 1 + y\biggr)$
\item $\tau^E_\text{min} = 0$
\item $\tau^E_\text{max} =\rho^E_\text{max}$
\item $f_{\tau^E}(z) = f_{\rho^E}(z)$
\item $I_{\tau^E}(\hat m) = I_{\rho^E}(\hat m)$
\end{itemize}

\noindent\underline{2-jettiness}
\begin{itemize}
\item $\tau_J = \frac{1}{Q} \sum_i m_{i,\bot} \mathrm e^{-|\eta_i|} = \frac{1}{Q} \min_{\hat t} \sum_i (E_i-|\hat t\cdot\vec p_i|)$
\item $\tau_{J,q}(y, z) = 1 - \text{mod}(y,z)$
\item $\tau_{J,\,\text{min}} = 1 - v$
\item $\tau_{J,\,\text{max}} = \rho_\text{max}$
\item $f_{\tau_J}(z) = z/v$
\item $I_{\tau_J}(\hat m) = I_\rho(\hat m) - \log(v) \,\mathcal I(\hat m)$
\end{itemize}

\subsection*{C-Parameter / C-Jettiness}

Here we provide results for reduced C-parameter $\widetilde C = C/6$. The relation to the original version for the event shape dependent
integral is simply
\begin{equation}
I_C(\hat m) = I_{\widetilde C}(\hat m) - \log(6)\, \mathcal I(\hat m)\,.
\end{equation}

\noindent\underline{P-scheme, original definition}
\begin{itemize}
\item $\widetilde C = \frac{1}{4}\Bigl[1 - \frac{1}{Q_P^2} \sum_{i, j} \frac{(\vec p_i\cdot \vec p_j)^2}{|\vec p_i||\vec p_j|}\Bigr]$
\item $\widetilde C_q(y, z) = \frac{2 y \,[(1-y) (1-z) z-\hat m^2]}{\text{mod}(y,1-z)\, \text{mod}(y,z)
[\text{mod}(y,1-z)+\text{mod}(y,z)+y]}$
\item $\widetilde C_\text{min} = 0$
\item $\widetilde C_\text{max} = 1/8$
\item $f_{\widetilde C}(z) = \frac{(1-z) z - \hat m^2}{v^3}$
\item $\begin{aligned}[t]
I_{\widetilde C}(\hat m)\,={}&(1-2 \hat m^2) \biggl[-2\log^2(2 \hat m)+2\, \text{Li}_2\biggl(\frac{1-v}{1+v}\biggr)+2 \log^2(1+v)-\frac{\pi ^2}{3}\\
&+3 \log(v) \log\biggl(\frac{1-v}{1+v}\biggr)+2 \log (2) \log \biggl(\frac{1-v}{1+v}\biggr)\biggr]+v \log (v)-\log\biggl(\frac{1-v}{1+v}\biggr)
\end{aligned}$
\end{itemize}

\noindent\underline{E-scheme}
\begin{itemize}
\item $\widetilde C^E = \frac{1}{4}\Bigl[1 - \frac{1}{Q^2} \sum_{i, j} \frac{E_i E_j (\vec p_i\cdot \vec p_j)^2}{|\vec p_i|^2|\vec p_j|^2}\Bigr]$
\item $\widetilde C^E_q(y, z)\! =\! \frac{y \{\hat m^2 [y (6 (1-z) z+1)-3 y^2 (1-z) z-4 (1-z) z-1]-2 \hat m^4
(y-2)+(1-y) (1-z) z [1-y (1-z)] (1-y z)\}}{\text{mod}^2(y,z)\,\,\text{mod}^2(y,1-z)}$
\item $\widetilde C^E_\text{min} = 0$
\item $\widetilde C^E_\text{max} = 1/8$
\item $f_{\widetilde C^E}(z) = v f_{\widetilde C}(z)$
\item $	I_{\widetilde C^E}(\hat m) = I_{\widetilde C}(\hat m) + \log(v) \,\mathcal I(\hat m)$
\end{itemize}
\vspace{1em}

\noindent\underline{C-jettiness}
\begin{itemize}
\item $\widetilde C_J = \frac{1}{4}\Bigl[2 - \frac{1}{Q^2} \sum_{i\neq j} \frac{(p_i\cdot p_j)^2}{E_i E_j}\Bigr]$
\item $\widetilde C_{J,q}(y, z) = \frac{z(1-z)y(1-y)+2\hat m^2(1-y)-2\hat m^4}{(1-zy)[1-(1-z)y]}$
\item $\widetilde C_{J,\,\text{min}} = 2 \hat m^2 (1 - \hat m^2)$
\item $\widetilde C_{J,\,\text{max}} =
\begin{cases}
(1 + 16 \hat m^2 + 32 \hat m^4) / (1 + 2 \hat m^2)^2 / 8\quad& (\hat m < 0.39307568887871164)\\
4 \hat m^2 (1 + 2 \hat m^2)/(1 + 4 \hat m^2)^2\quad& (\hat m > 0.39307568887871164)
\end{cases}$
\item $f_{\widetilde C_J}(z) = z(1-z) - 2 \hat m^4$
\item $\begin{aligned}[t]
I_{\widetilde C_J}(\hat m)={}&\frac{1}{2} (2 \hat m^2-1) \biggl\{-2 \text{Li}_2\biggl(\frac{1-v}{1+u}\biggr)+
2\, \text{Li}_2\biggl(\frac{1-u}{1-v}\biggr)-2\, \text{Li}_2\biggl(\frac{1-u}{1+v}\biggr)\nonumber\\
&+2\, \text{Li}_2\biggl(\frac{1+v}{1+u}\biggr)+\log^2\biggl(\frac{1-v}{1+v}\biggr)+2 \log \biggl(\frac{1-v}{1+v}\biggr)
\log\biggl[\frac{(1+u) (1+v)}{4}\biggr]\biggl\}\nonumber\\
&-\frac{1}{2\hat m^2}\,\biggl[\log \biggl(\frac{1-v}{1+v}\biggr)-u \log\biggl(\frac{u-v}{u+v}\biggr)\biggr]- v \log\bigl[\hat m^2(1-2\hat m^2)\bigr]
\end{aligned}$
\end{itemize}
with $u\equiv\sqrt{1-8\hat m^4}$.

\subsection*{Broadening}

\noindent\underline{P-scheme, original definition}

\begin{itemize}
\item $B_T = \frac{1}{2Q_P}\sum_i p_{i,\bot} = \frac{1}{2 Q_P} \sum_i (|\vec p_i| - |\hat t\cdot\vec p_i|)^{1/2}(|\vec p_i| + |\hat t\cdot\vec p_i|)^{1/2}$
\item $B_{T, q}(y, z)\! =\! \frac{Q}{Q_P}\Biggl\{\frac{\sqrt{\text{mod}(y, z)\, \text{mod}(y, 1-z)-v^2+y^2 (1-z) z+y}
\sqrt{\text{mod}(y, z)\, \text{mod}(y, 1-z)+v^2-y^2 (1-z) z-y)}}{4 \text{mod}(y, z)}\\
+\frac{y \sqrt{[\text{mod}(y, z)+(2-y) z-1] [\text{mod}(y, z)+1-(2-y) z]}}{4\, \text{mod}(y, z)}\Biggl\}$
\item $B_{T, \text{min}} = 0$
\item $B_{T, \text{max}} = \frac{1}{2 \sqrt{3}}$
\item $f_{B_T}(z) = \sqrt{\frac{f_{\widetilde C}(z)}{v}}$
\item $I_{B_T}(\hat m) = \frac{1}{2} \bigl[I_{\widetilde C}(\hat m) - \log(v) \,\mathcal I(\hat m)\bigr]$
\end{itemize}

\noindent\underline{E-scheme}
\begin{itemize}
\item $B_T^E = \frac{1}{2 Q} \sum_i \frac{E_i}{|\vec p_i|}(|\vec p_i| - |\hat t\cdot\vec p_i|)^{1/2}(|\vec p_i| + |\hat t\cdot\vec p_i|)^{1/2}$
\item $B^E_{T,q}(y, z) = \frac{y \text{mod}(y, 1-z) \sqrt{(1-y) (1-z) z-\hat m^2}}{2\, \text{mod}(y, z)\, \text{mod}(y, 1-z)}\\
+ \frac{(1-y (1-z)) \sqrt{\text{mod}(y, z)\, \text{mod}(y, 1-z)-v^2+y^2 (1-z) z+y}
\sqrt{\text{mod}(y, z)\, \text{mod}(y, 1-z)+v^2-y^2 (1-z) z-y}}{4\, \text{mod}(y, z)\, \text{mod}(y, 1-z)}$
\item $B^E_{T, \text{min}} = 0$
\item $B^E_{T, \text{max}} = \frac{2 - r}{2 \sqrt{3}}$
\item $f_{B_T^E}(z) = \frac{1+v}{2} f_{B_T}(z)$
\item $I_{B_T^E}(\hat m)=I_{B_T}(\hat m)+\log\bigl(\frac{v+1}{2}\bigr) \,\mathcal I(\hat m)$
\end{itemize}

\subsection*{Angularities}

Here, we consider angularities in the parameter region $a < 1$ only. The soft expansion of the event shape contains a term proportional to $y^{2-a}$,
which therefore only contributes to $f_{\tau_a}$ if $a = 1$. This case has already been considered separately since $\tau_a \to 2 B_T$ $(a \to 1)$.

\vspace{1em}

\noindent\underline{E-scheme, original definition}
\begin{itemize}
\item $\tau_a = \frac{1}{Q} \sum_i \frac{E_i}{|\vec p_i|} p_{i, \bot}\mathrm e^{-|\eta|(1-a)} = \frac{1}{2 Q} \sum_i \frac{E_i}{|\vec p_i|}(|\vec p_i| - |\hat t\cdot\vec p_i|)^{1-\frac{a}{2}}(|\vec p_i| + |\hat t\cdot\vec p_i|)^{\frac{a}{2}}$
\item $\tau_{a,q}(y, z) = \frac{1}{2} \,\biggl\{\!\Bigl[\frac{y [\text{mod}(y, z)-(2-y) z+1]}{\text{mod}(y, z)}\Bigr]^{\!\frac{a}{2}}
\Bigl[y-\frac{y [1-(2-y) z]}{\text{mod}(y, z)}\Bigr]^{\!1-\frac{a}{2}}\\
+ \frac{[1-y (1-z)] [\text{mod}(y, z)\, \text{mod}(y, 1-z)-v^2+y^2 (1-z) z+y]^{1-\frac{a}{2}}
[\text{mod}(y, z)\, \text{mod}(y, 1-z)+v^2-y^2 (1-z) z-y]^{\frac{a}{2}}}{\text{mod}(y, z)\text{mod}(y, 1-z)}\biggl\}$
\item $\tau_{a, \text{min}} = 0$
\item $\tau_{a, \text{max}} = 3^{\frac{a}{2} - 1} (2 - r)$
\item $f_{\tau_{a<1}}(z) = \frac{(1+v-2 z)^{\frac{a}{2}} (v+2 z-1)^{1-\frac{a}{2}}}{2 v}$
\item $\begin{aligned}[t]
I_{\tau_{a<1}}(\hat m) ={} & \frac{1-a}{4} \Biggl\{(1+v^2) \biggl[2\, \text{Li}_2(1-v)+
2\, \text{Li}_2\biggl(\frac{1}{1+v}\biggr)-2 \,\text{Li}_2\biggl(\frac{1-v}{1+v}\biggr)-\frac{\pi ^2}{3}\\
&-\frac{1}{2} \log^2\biggl(\frac{1-v}{1+v}\biggr)+\log ^2(1+v)\biggr]-4 v \log(\hat m)\Biggl\}-2 \log\biggl(\frac{1-v}{1+v}\biggr)\\
&+\frac{1+v^2}{4} \biggl[2\, \text{Li}_2\biggl(\frac{1-v}{1+v}\biggr)-\frac{1}{2} \log^2\biggl(\frac{1-v}{1+v}\biggr)-
2 \log \biggl(\frac{1+v}{2v}\biggr) \log \biggl(\frac{1-v}{1+v}\biggr)-\frac{\pi ^2}{3}\biggr]
\end{aligned}$
\end{itemize}

\noindent\underline{P-scheme}

\begin{itemize}
\item $\tau_a^P = \frac{1}{Q_P} \sum_i (|\vec p_i| - |\hat t\cdot\vec p_i|)^{1-\frac{a}{2}}(|\vec p_i| + |\hat t\cdot\vec p_i|)^{\frac{a}{2}}$
\item $\tau^P_{a, q}(y, z) = \frac{1}{2} \biggl\{\Bigl[\frac{y [\text{mod}(y, z)-(2-y) z+1]}{\text{mod}(y, z)}\Bigr]^{\!\frac{a}{2}}
\Bigl[y-\frac{y [1-(2-y) z]}{\text{mod}(y, z)}\Bigr]^{\!1-\frac{a}{2}}\\
+\frac{[\text{mod}(y, z)\, \text{mod}(y, 1-z)-v^2+y^2 (1-z) z+y]^{1-\frac{a}{2}}
[\text{mod}(y, z)\, \text{mod}(y, 1-z)+v^2-y^2 (1-z) z-y]^{\frac{a}{2}}}{\text{mod}(y, z)}\biggl\}$
\item $\tau^P_{a, \text{min}} = 0$
\item $\tau^P_{a, \text{max}} = 3^{\frac{a}{2} - 1}$
\item $f_{\tau^P_{a<1}}(z) = \frac{f_{\tau_{a<1}}(z)}{v}$
\item $I_{\tau_a^P} = I_{\tau_a} - \log(v) \,\mathcal I(\hat m)$
\end{itemize}

\bibliography{thrust3}
\bibliographystyle{JHEP}

\end{document}